\documentclass{article}

\usepackage{arxiv}

\usepackage[utf8]{inputenc} % allow utf-8 input
\usepackage[T1]{fontenc}    % use 8-bit T1 fonts

\usepackage[colorlinks = true, citecolor=red, linkcolor = blue, urlcolor = black]{hyperref}  
\usepackage{url}            % simple URL typesetting
\usepackage{booktabs}       % professional-quality tables
\usepackage{amsfonts}       % blackboard math symbols
\usepackage{nicefrac}       % compact symbols for 1/2, etc.
\usepackage{microtype}      % microtypography
\usepackage{lipsum}		% Can be removed after putting your text content
\usepackage{graphicx}
\usepackage{doi}
\usepackage[thinc]{esdiff}

\usepackage{amsmath}
\usepackage{color}
\usepackage{overpic}
\usepackage{stmaryrd}
\usepackage{ulem}
\usepackage{caption}
\usepackage{subcaption}
\usepackage{cases}
\usepackage{appendix}
\usepackage{setspace}

\usepackage[linesnumbered,ruled,vlined]{algorithm2e}

\SetCommentSty{mycommfont}
\SetKwInput{KwInput}{Input}                % Set the Input
\SetKwInput{KwOutput}{Output}  

\makeatletter
\def\widebreve#1{\mathop{\vbox{\m@th\ialign{##\crcr\noalign{\kern3\p@}%
      \brevefill\crcr\noalign{\kern3\p@\nointerlineskip}%
      $\hfil\displaystyle{#1}\hfil$\crcr}}}\limits}

\def\brevefill{$\m@th \setbox\z@\hbox{$\braceld$}%
  \bracelu\leaders\vrule \@height\ht\z@ \@depth\z@\hfill\braceru$}
\makeatletter

\title{An adaptive ensemble filter for heavy-tailed distributions:\\
tuning-free inflation and localization}

\date{}

\author{ \href{https://orcid.org/0000-0003-0396-5740}{\includegraphics[scale=0.06]{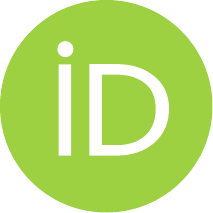}\hspace{1mm}Mathieu Le Provost}\\
Department of Aeronautics and Astronautics\\
	Massachusetts Institute of Technology\\
	Cambridge, MA, 02139, USA\\
	\code{mleprovo@mit.edu} \\
	\And
	\href{https://orcid.org/0000-0002-0421-890X}{\includegraphics[scale=0.06]{orcid.pdf}\hspace{1mm}Ricardo Baptista} \\
	Computing and Mathematical Sciences\\
	California Institute of Technology\\
	Pasadena, CA, 91125, USA\\
	\code{rsb@caltech.edu} \\
 	\And
	\href{https://orcid.org/0000-0002-2672-706X}{\includegraphics[scale=0.06]{orcid.pdf}\hspace{1mm}Jeff D. Eldredge}\\
	Mechanical \& Aerospace Engineering Department\\
	University of California, Los Angeles\\
	Los Angeles, CA, 90095, USA\\
	\code{jdeldre@ucla.edu} 
 	\And
    \href{https://orcid.org/0000-0001-8242-3290}{\includegraphics[scale=0.06]{orcid.pdf}
	\hspace{1mm}Youssef Marzouk}\\
	Department of Aeronautics and Astronautics\\
	Massachusetts Institute of Technology\\
	Cambridge, MA, 02139, USA\\
	\code{ymarz@mit.edu} \\
}

\renewcommand{\headeright}{}
\renewcommand{\undertitle}{}
\renewcommand{\shorttitle}{\textit{Le Provost, Baptista, Eldredge, Marzouk / An adaptive ensemble filter for heavy-tailed distributions}}

\hypersetup{
pdftitle={LePBapMarEld22robustfilter},
pdfsubject={q-bio.NC, q-bio.QM},
pdfauthor={Mathieu Le Provost, Ricardo Baptista, Jeff D. Eldredge, Youssef Marzouk},
pdfkeywords={ensemble Kalman filter, \textit{t}--distributions, transport maps, data-informed dimension reduction.},
}

\begin{document}

\definecolor{forestgreen}{RGB}{1, 93, 45}
\definecolor{pigmentgreen}{RGB}{0, 165, 80}
\definecolor{custompurple}{RGB}{129, 19, 239}
\definecolor{customred}{RGB}{208, 2, 27}
\definecolor{customyelloworange}{RGB}{255, 137, 0}
\definecolor{customturquoise}{RGB}{7, 143, 181}
\definecolor{palegreen}{rgb}{0.263,0.804,0.502}
\definecolor{blue(pigment)}{rgb}{0.2, 0.2, 0.6}
\definecolor{cadet}{rgb}{0.33, 0.41, 0.47}
\definecolor{amaranth}{rgb}{0.9, 0.17, 0.31}

\newcommand{\jeffcomment}[1]{\textcolor{amaranth}{\bf{[Jeff: #1]}}}
\newcommand{\ricomment}[1]{\textcolor{blue}{\bf{[Ricardo: #1]}}}
\newcommand{\matcomment}[1]{\textcolor{forestgreen}{\bf{[Mathieu: #1]}}}
\newcommand{\youcomment}[1]{\textcolor{blue(pigment)}{\bf{[Youssef: #1]}}}

\newcommand{\diag}[1]{\text{diag}\left(#1\right)}
\newcommand{\epde}{ePDE}
\newcommand{\pde}{PDE}

\newcommand{\jeff}[1]{\textcolor{amaranth}{#1}}
\newcommand{\ric}[1]{\textcolor{blue}{#1}}
\newcommand{\mat}[1]{\textcolor{forestgreen}{#1}}
\newcommand{\youssef}[1]{\textcolor{blue(pigment)}{#1}}

\newcommand{\eqn}{equation}
\newcommand{\Eqn}{Equation}

\newcommand{\argmin}{argmin\;}
\newcommand{\argmax}{argmax\;}

\newcommand{\atm}{\texttt{AdaptiveTransportMap.jl}}

\newcommand{\eg}{e.g.,\xspace}
\newcommand{\ie}{i.e.,\xspace}
\newcommand{\st}{s.t.\xspace}
\newcommand{\etal}{et \textit{al.}\xspace}

\newcommand{\lo}[1]{Lorenz-#1}
\newcommand{\ks}{Kuramoto-Sivashinsky}
\newcommand{\mx}{Maxey-Riley}

\newcommand{\BB}{\boldsymbol}
\newcommand{\be}{\begin{equation}}
\newcommand{\ee}{\end{equation}}

\newcommand{\ba}{\begin{align}}
\newcommand{\ea}{\end{align}}

% Numerical analysis notations
\newcommand{\real}[1]{\mathbb{R}^{#1}}
\newcommand{\complex}[1]{\mathbb{C}^{#1}}

\newcommand{\code}[1]{\texttt{#1}}

\newcommand{\id}[1]{\BB{I}_{#1}}
\newcommand{\zero}[1]{\BB{0}_{#1}}
\newcommand{\one}{\BB{1}}
\newcommand{\trace}[1]{\text{tr}(#1)}

% Define colors
\definecolor{forecast}{RGB}{19, 24, 143}
\definecolor{analysis}{RGB}{208, 2, 27}

\definecolor{posterior}{rgb}{0.263,0.804,0.502}
\definecolor{truth}{rgb}{0.125,0.698,0.667}
\definecolor{enkf}{rgb}{0.933,0.251,0.0}

% \definecolor{enkf}{rgb}{0.4,0.804,0.0}
\definecolor{smap0}{rgb}{0.686, 0.624, 0.11}
\definecolor{smap1}{rgb}{0.275,0.51,0.706}  
\definecolor{smap2}{rgb}{0.545, 0.0, 0.0}

\definecolor{ccfd}{rgb}{0.0,0.0,0.502}
\definecolor{cenkf}{rgb}{0.0,0.502,0.502}
\definecolor{cetkf}{rgb}{0.933,0.251,0.0}  
\definecolor{cetkfconf}{RGB}{246, 157, 127}
\definecolor{cgust}{rgb}{0.196,0.804,0.196}
\definecolor{ctmap}{rgb}{0.502,0.0,0.502}
\definecolor{cgold}{rgb}{0.8, 0.61, 0.11}

\newcommand{\commentcode}[1]{\textcolor{cenkf}{\footnotesize{\texttt{\% #1}}}}

\definecolor{customblue}{RGB}{19, 24, 143}

\definecolor{orangered}{RGB}{168,58,8}
\definecolor{customyellow}{RGB}{252,227,3}
\definecolor{customgreen}{RGB}{0,158,115}

\newcommand\crule[3][black]{\textcolor{#1}{\rule{#2}{#3}}}
\newcommand\lcolor[1]{\crule[#1]{2mm}{2mm}}
\newcommand\squarecolor[1]{\crule[#1]{2mm}{2mm}}
\newcommand\linecolor[1]{\textcolor{#1}{\rule[1pt]{4mm}{0.8mm}}}

% Notations for vortex models
\newcommand{\lesp}{LESP}
\newcommand{\lespc}{$\mbox{LESPc}$}
\newcommand{\cn}{$C_n$}
\newcommand{\ts}{t^\star}

% Notations for EM and EMq algorithms
\newcommand{\Lq}{\mathcal{L}_q}

\newcommand{\indep}{\rotatebox[origin=c]{90}{$\models$}}

% Notations for data assimilation 

\newcommand{\xf}{{\x^f}}
\newcommand{\xfi}{{\x^{f,i}}}
\newcommand{\xa}{{\x^a}}
\newcommand{\xai}{{\x^{a,i}}}
\newcommand{\xbar}{\overline{\x}}
\newcommand{\xbara}{\xbar^a}
\newcommand{\xbarf}{\xbar^f}

\newcommand{\covm}{\BB{Q}}
\newcommand{\covo}{\BB{R}}
\newcommand{\score}{\BB{\Omega}}

\newcommand{\Ne}{N_e}
\newcommand{\Nx}{N_x}
\newcommand{\Ny}{N_y}

\newcommand{\K}{\BB{K}}

\newcommand{\KH}{\BB{KH}}

\newcommand{\Pa}{\BB{P}^a}
\newcommand{\Pf}{\BB{P}^f}

\newcommand{\Pbar}{\overline{\BB{P}}}
\newcommand{\Pbara}{\overline{\BB{P}}^a}
\newcommand{\Pbarf}{\overline{\BB{P}}^f}

% Anomalies and ensemble matrices

\newcommand{\Ens}{\BB{E}}
\newcommand{\Ensa}{\BB{E}^a}
\newcommand{\Ensf}{\BB{E}^f}

\newcommand{\ensemble}[1]{\left\{#1\right\}}

\newcommand{\U}{\BB{U}}
\newcommand{\J}{\BB{J}}
\newcommand{\V}{\BB{V}}
\newcommand{\T}{\BB{T}}
\newcommand{\G}{\BB{G}}
\newcommand{\sqG}{{\BB{G}^{1/2}}}

% Localisation tools
\newcommand{\Rho}{\BB{\rho}}

% Notations for probability and transport maps

\newcommand\given[1][]{\:#1\vert\:}

\newcommand{\N}[2]{\mathcal{N}\left(#1, #2\right)}
\newcommand{\chitwo}[1]{\mathcal{\chi}^2_{#1}}
\newcommand{\St}[3]{\mathit{St}\left( #1, #2, #3 \right)}
\newcommand{\reftdist}[2]{\eta_{#1, #2}}
\newcommand{\tdist}{\textit{t}--distribution\xspace}
\newcommand{\tdisted}{\textit{t}--distributed\xspace}
\newcommand{\tdists}{\textit{t}--distributions\xspace}
\newcommand{\gauss}{Gaussian\xspace}
\newcommand{\dof}[1]{\nu_{#1}}
\newcommand{\sdof}[1]{\hat{\nu}_{#1}}
\newcommand{\emq}{\texttt{EMq}\xspace}
\newcommand{\tlasso}{\texttt{tlasso}\xspace}
\newcommand{\glasso}{\texttt{glasso}\xspace}

\newcommand{\qform}[2]{\left(#1 \right)^\top #2 \left(#1 \right)}
\newcommand{\deltaform}[2]{\delta_{#1}\bigl(#2 \bigr)}
\newcommand{\sdeltaform}[2]{\hat{\delta}_{#1}\bigl(#2 \bigr)}
\newcommand{\maha}{Mahalanobis squared distance\xspace}

\newcommand{\scaling}[2]{\alpha_{#1}\bigl(#2 \bigr)}
\newcommand{\invscaling}[2]{\alpha^{-1}_{#1}\bigl(#2 \bigr)}

\newcommand{\iup}[1]{#1^{(i)}}
\newcommand{\zi}{\iup{\z}}
\newcommand{\tauiJ}{\tau^{(i), \Jidx}}

\newcommand{\E}[2]{\mathrm{E}_{#1}\left[#2\right]}
\newcommand{\cov}[1]{\BB{\Sigma}_{#1}}
\newcommand{\scov}[1]{\widehat{\BB{\Sigma}}_{#1}}
\newcommand{\mean}[1]{\BB{\mu}_{#1}}
\newcommand{\smean}[1]{\widehat{\mu}_{#1}}
\newcommand{\scale}[1]{\BB{C}_{#1}}
\newcommand{\anom}[1]{\BB{A}_{#1}}
\newcommand{\sscale}[1]{\widehat{\BB{C}}_{#1}}
\newcommand{\logdet}{\text{logdet}\;}
\newcommand{\preci}[1]{\BB{\Theta}_{#1}}
\newcommand{\spreci}[1]{\widehat{\BB{\Theta}}_{#1}}

\newcommand{\pdfx}{\pdf{\X}}
\newcommand{\meanx}{\mean{\X}}
\newcommand{\covx}{\cov{\X}}
\newcommand{\scovx}{\scov{\X}}
\newcommand{\scalex}{\scale{\X}}
\newcommand{\dofx}{\dof{\X}}

\newcommand{\Stx}{\St{\meanx}{\scalex}{\dofx}}

\newcommand{\pdfy}{\pdf{\Y}}
\newcommand{\meany}{\mean{\Y}}
\newcommand{\covy}{\cov{\Y}}
\newcommand{\scovy}{\scov{\Y}}
\newcommand{\scaley}{\scale{\Y}}
\newcommand{\dofy}{\dof{\Y}}
\newcommand{\Sty}{\St{\meany}{\scaley}{\dofy}}

\newcommand{\meanxy}{{\mean{\X, \Y}}}
\newcommand{\scalexy}{{\scale{\X, \Y}}}
\newcommand{\dofxy}{{\dof{\X, \Y}}}
\newcommand{\Stxy}{\St{\meanxy}{\scalexy}{\dofxy}}

\newcommand{\meanyx}{{\mean{\Y, \X}}}
\newcommand{\scaleyx}{{\scale{\Y, \X}}}
\newcommand{\schurxy}{{\scale{\X} - \scale{\X, \Y} \scale{\Y}^{-1} \scale{\X, \Y}^\top }}

\newcommand{\dofyx}{{\dof{\Y, \X}}}
\newcommand{\Styx}{\St{\meanyx}{\scaleyx}{\dofyx}}

\newcommand{\pdfz}{\pdf{\Z}}
\newcommand{\meanz}{\mean{\Z}}
\newcommand{\covz}{\cov{\Z}}
\newcommand{\scovz}{\scov{\Z}}
\newcommand{\scalez}{\scale{\Z}}
\newcommand{\preciz}{\preci{\Z}}
\newcommand{\spreciz}{\spreci{\Z}}
\newcommand{\dofz}{\dof{\Z}}
\newcommand{\Stz}{\St{\meanz}{\scalez}{\dofz}}
\newcommand{\smeanz}{\widehat{\BB{\mu}}_{\Z}}
\newcommand{\sscalez}{\widehat{\BB{C}}_{\Z}}
\newcommand{\anomz}{\BB{A}_{\Z}}
\newcommand{\sdofz}{\widehat{\nu}_{\Z}}

\newcommand{\stauz}{\BB{S}_{\tau \Z}}

\newcommand{\swonei}{\hat{w}^{(i),\Jidx}_1}
\newcommand{\swtwoi}{\hat{w}^{(i),\Jidx}_2}

\newcommand{\swqi}{\hat{w}^{(i),\Jidx}_{q}}
\newcommand{\sWqi}{\hat{W}^{(i),\Jidx}_{q}}
\newcommand{\swoneqi}{\hat{w}^{(i),\Jidx}_{1,q}}
\newcommand{\swtwoqi}{\hat{w}^{(i),\Jidx}_{2,q}}
\newcommand{\szetaqi}{\hat{\zeta}^{(i),\Jidx}_{q}}

\newcommand{\zeroidx}{\mathsf{0}}
\newcommand{\Jidx}{\mathsf{J}}

\newcommand{\kf}{\text{KF}\xspace}
\newcommand{\enkf}{\text{EnKF}\xspace}
\newcommand{\senkf}{\text{sEnKF}\xspace}
\newcommand{\lrenkf}{\text{LREnKF}\xspace}
\newcommand{\etkf}{ETKF\xspace}
\newcommand{\smf}{SMF\xspace}
\newcommand{\rf}{\text{RF}\xspace}
\newcommand{\enrf}{\text{EnRF}\xspace}
\newcommand{\adaptenrf}{\code{AdaptEnRF}\xspace}
\newcommand{\fixedenrf}{\code{FixedEnRF}\xspace}
\newcommand{\refreshenrf}{\code{RefreshEnRF}\xspace}
\newcommand{\senkfglasso}{\text{sEnKF-\glasso}\xspace}

\newcommand{\enrfdof}{$\text{EnRF}_{\nu = 100}$\xspace}

\newcommand{\obsstate}{\BB{Hx}}
\newcommand{\Obsstate}{\BB{\mathsf{HX}}}

\newcommand{\x}{\BB{x}}
\newcommand{\y}{\BB{y}}
\newcommand{\ystar}{\BB{y}^\star}

\newcommand{\ybarf}{\overline{\y}^f}
\newcommand{\dyn}{\BB{f}}
\newcommand{\Dyn}{\BB{F}}
\newcommand{\obs}{\BB{h}}
\newcommand{\Obs}{\BB{H}}
\newcommand{\tObs}{\tilde{\BB{H}}}
\newcommand{\Noisedyn}{\BB{\mathsf{W}}}
\newcommand{\Noiseobs}{\BB{\mathcal{E}}}
\newcommand{\noisedyn}{\BB{w}}
\newcommand{\noiseobs}{{\BB{\epsilon}}}
\newcommand{\covdyn}{\BB{W}}
\newcommand{\covobs}{\BB{V}}

%%%%%%%%%%%%%%%%%%%%%%%%%%%%%%%%%%%%%%%
%%%%%%% Notations for the densities
\newcommand{\pdf}[1]{\pi_{#1}}
\newcommand{\pdfprior}{\pdf{\X}}
\newcommand{\pdflik}{\pdf{\Y \given \X}}
\newcommand{\pdfjoint}{\pdf{\Y, \X}}
\newcommand{\pdfpost}{\pdf{\X \given \Y}}

\newcommand{\stack}[2]{\begin{bmatrix} #1 \\ #2 \end{bmatrix}}
\newcommand{\X}{\BB{\mathsf{X}}}
\newcommand{\Y}{\BB{\mathsf{Y}}}

\newcommand{\YX}{\stack{\Y}{\X}}
\newcommand{\XY}{\stack{\X}{\Y}}

\newcommand{\Xup}{\boldsymbol{\mathcal{X}}}
\renewcommand{\Yup}{\boldsymbol{\mathcal{Y}}}

\newcommand{\z}{\BB{z}}
\newcommand{\Z}{\BB{\mathsf{Z}}}

\newcommand{\proj}{\BB{P}}

\newcommand{\push}[1]{{#1}_\sharp}
\newcommand{\pull}[1]{{#1}^\sharp}

\newcommand{\tmap}{\BB{T}}
\newcommand{\stmap}{\widehat{\BB{T}}}
\newcommand{\tmappush}{{\push{\tmap}}}
\newcommand{\tmappull}{{\pull{\tmap}}}
\newcommand{\tmapdof}{\BB{T}_{\dof{}}}
\newcommand{\stmapdof}{\widehat{\BB{T}}_{\dof{}}}
\newcommand{\tmapkf}{\BB{T}_{\kf}}

\newcommand{\smap}{\BB{S}}
\newcommand{\smappush}{{\push{\smap}}}
\newcommand{\smappull}{{\pull{\smap}}}

\newcommand{\umap}{\BB{U}}

\newcommand{\lmap}{\BB{L}}
\newcommand{\qmap}{\BB{Q}}

\newcommand{\rmap}{\BB{R}}

\newcommand{\Hmap}{\BB{\mathcal{H}}}
\newcommand{\Hmapdelta}{\BB{\mathcal{H}}_\Delta}
\newcommand{\Hmapk}{\mathcal{H}_k}
\newcommand{\Hmapkh}{\mathcal{H}_{k,a}}

\newcommand{\ui}{\BB{\mathfrak{u}}_i}
\newcommand{\uk}{\BB{\mathfrak{u}}_k}
\newcommand{\psij}{\psi_j}
\newcommand{\Uk}{U^k}

\newcommand{\Cx}{\BB{C}_{\X}}
\newcommand{\Cy}{\BB{C}_{\Y}}
\newcommand{\sCx}{\widehat{\BB{C}}_{\X}}
\newcommand{\sCy}{\widehat{\BB{C}}_{\Y}}
\newcommand{\Sx}{\BB{S}_{\X}}
\newcommand{\Sy}{\BB{S}_{\Y}}

\newcommand{\rx}{r_{\X}}
\newcommand{\ry}{r_{\Y}}
\newcommand{\Kbreve}{\breve{\BB{K}}}

\newcommand{\Q}{\BB{\mathsf{Q}}}
\newcommand{\q}{\mathcal{\BB{q}}}

\def\dkl#1#2{\mathcal{D}_{\mathrm{KL}}\left(#1 || #2\right)}

% Metric
\newcommand{\rmse}{\mbox{RMSE}}

%%%%%%%%%%%%%%%%%%%%%%%%%%%%%%%%%%%%%%%%%%%%%%%%%%%%%%
%%%%%%%%%%% Notations for the adaptive transport map % 

\newcommand{\recti}[2]{\mathcal{R}_{#1}(#2)}
\newcommand{\lossJ}[2]{\mathcal{J}_{#1}(#2)}
\newcommand{\lossL}[2]{\mathcal{L}_{#1}(#2)}
\newcommand{\coeff}{c_{\BB{\alpha}}}

\newcommand{\sketch}{\BB{\Omega}}
\newcommand{\sketchbis}{\BB{\Gamma}}
\newcommand{\tbi}{\texttt{TransportBasedInference.jl}}
\newcommand{\dist}{distribution\xspace}
\newcommand{\kr}{Knothe-Rosenblatt rearrangement\xspace}
\newcommand{\rea}{rearrangement\xspace}

\maketitle

\begin{abstract}
Heavy tails is a common feature of filtering distributions that results from the nonlinear dynamical and observation processes as well as the uncertainty from physical sensors. In these settings, the Kalman filter and its ensemble version --- the ensemble Kalman filter (\enkf) --- that have been designed under Gaussian assumptions result in  degraded performance. \tdists are a parametric family of distributions whose tail-heaviness is modulated by a degree of freedom $\dof{}$. Interestingly, Cauchy and Gaussian distributions correspond to the extreme cases of a \tdist for $\dof{} = 1$ and $\dof{} = \infty$, respectively. Leveraging tools from measure transport (Spantini \etal, SIAM Review, 2022), we present a generalization of the \enkf whose prior-to-posterior update leads to exact inference for \tdists. We demonstrate that this filter is less sensitive to outlying synthetic observations generated by the observation model for small $\nu$. Moreover, it recovers the Kalman filter for $\dof{} = \infty$. For nonlinear state-space models with heavy-tailed noise, we propose an algorithm to estimate the prior-to-posterior update from samples of joint forecast distribution of the states and observations. We rely on a regularized expectation-maximization (EM) algorithm to estimate the mean, scale matrix, and degree of freedom of heavy-tailed \tdists from limited samples (Finegold and Drton, arXiv preprint, 2014). Leveraging the conditional independence of the joint forecast distribution, we regularize the scale matrix with an $l1$ sparsity-promoting penalization of the log-likelihood at each iteration of the EM algorithm. This $l1$ regularization draws upon the graphical lasso algorithm (Friedman \etal, Biostatistics,  2008) to estimate sparse covariance matrix in the Gaussian setting. By sequentially estimating the degree of freedom at each analysis step, our filter has the appealing feature of  adapting the prior-to-posterior update to the tail-heaviness of the data. This new filter intrinsically embeds an adaptive and data-dependent multiplicative inflation mechanism complemented with an adaptive localization through the $l1$-penalization of the estimated scale matrix. We demonstrate the benefits of this new ensemble filter on challenging filtering problems with heavy-tailed noise. 
\end{abstract}

% \textbf{Data availability:} All the computational results are reproducible and code is available at: \href{https://github.com/mleprovost/Paper-Ensemble-Robust-Filter.jl.git}{https://github.com/mleprovost/Paper-Ensemble-Robust-Filter.jl.git}

\section{Introduction \label{sec:intro}}

%Heavy-tailedness is a recurring feature of geophysical systems. 
%This result from the nonlinearity.

 %Heavy-tailedness is formally defined as a distribution whose tails are not exponentially bounded \cite{nair2022fundamentals}. 

 Departure from Gaussian tails is a common feature of the distributions for the states and observations in geophysical systems. Sardeshmukh \etal \cite{sardeshmukh2015understanding} argue that the observed tail-heaviness is fundamentally related to the quadratic nonlinearity of the dynamical processes in geosciences. Tail-heaviness can also be caused by extreme events resulting from couplings of the stochasticity of certain parameters with the nonlinearity of the dynamical process \cite{farazmand2019extreme, sapsis2021statistics}. Some examples include extreme climate patterns (\eg tornadoes, thunderstorms, extreme precipitations), earthquakes, or rogue waves. Those last ones are large amplitude and unpredictable waves resulting from the constructive interference of multiple waves \cite{dysthe2008oceanic} with potentially disastrous consequences on ships. In 1984, an oceanic rogue wave of $11$m height was recorded near the Grom oil platform in the central North sea  \cite{alimohammadi2022recognize}. Tail-heaviness can also result from the nonlinearity of the observation operator and the uncertainty of the physical sensors. Tracers transported in a fluid such as drifters in the ocean or balloon sondes in the atmosphere are common techniques to obtain Lagrangian measurements of the fluid velocity field. Studies consistently report the heavy-tailedness of the probability distribution for the position of these tracers \cite{bracco2003lagrangian, elipot2016global}.  Data assimilation is a mathematical paradigm to improve the state estimate of a dynamical system by combining two complementary sources of information: a state prediction from a numerical model and observations collected from the real system \cite{asch2016data, evensen2009ensemble, law2015data}. In the filtering problem, we seek to estimate the filtering distribution, namely the conditional distribution of the state  $\x_t \in \real{n}$ at time $t$ given all the realizations of the observation variable up to that time \cite{asch2016data, spantini_coupling_2022, leprovost2022low}. This paper introduces a novel filtering algorithm for heavy-tailed filtering distributions.

% \cite{qi2016predicting} paper that discusses intermittency as a motivation for heavy tails behavior. 

% \cite{apte2013impact} paper that discusses nonlninearity of Lagrangian data assimilation

The ensemble Kalman filter (\enkf) \cite{evensen1994sequential} is usually considered as the state of the art filtering algorithm for high-dimensional and nonlinear problems \cite{asch2016data,carrassi2018data}. The \enkf propagates over time a set of $M$ particles $\{\x^{(1)}, \ldots, \x^{(M)}  \}$ with $\iup{\x} \in \real{n}$ to form an empirical approximation of the filtering distribution. At each assimilation cycle, the \enkf repeats a two-step procedure: a forecast step followed by an analysis step. In the forecast step, filtering samples from the previous time step $\{\x^{(1)}, \ldots, \x^{(M)}\}$ are propagated through the dynamical model. In the analysis step, synthetic observation samples $\{ \y^{(1)}, \ldots, \y^{(M)} \}$ are generated from the observation model and the resulting joint forecast samples $\{(\iup{\y}, \iup{\x})\}$ are used to estimate the Kalman gain matrix that maps observation discrepancies to state updates \cite{evensen2009ensemble}.  The ensemble Kalman filter belongs to a broader class of algorithms called \textit{ensemble filtering algorithms}. Many of these methods utilize the same forecast step as the \enkf, but differ in the forecast transformation that is applied in the analysis step to generate the filtering samples\cite{spantini_coupling_2022, leprovost2021low}. Following Spantini \etal \cite{spantini_coupling_2022}, we view the analysis step of an ensemble filter as the application of a transformation, called the \textit{prior-to-posterior transformation} or \textit{analysis map}, that maps the samples from forecast distribution (prior) to the samples from filtering distribution (posterior).

It is enlightening to look at ensemble filters through the lens of measure transport theory. Measure transport is a branch of mathematics aiming at identifying transformations that map random variables with a target density $\pdf{}$ to a reference density $\rho$, \eg a standard normal density \cite{villani2008optimal, marzouk2016sampling}. A map that verifies this condition is called a \textit{transport map} and we say that $\smap$ \textit{pushes forward} $\pdf{}$ to $\rho{}$, denoted $\smappush \pdf{} = \rho{}$. Among numerous candidate transport maps, and under mild assumptions on the target and reference densities, the \kr is defined as the unique lower triangular and monotonically increasing transformation $\smap$ that pushes forward $\pdf{}$ to $\rho$ \cite{rosenblatt1952remarks, marzouk2016sampling}. Spantini \etal \cite{spantini_coupling_2022} derived a generic formula to build a prior-to-posterior transformation from the \kr that pushes forward the joint forecast distribution to the reference density. Thus, the analysis step of an ensemble filter is fully characterized by the choice of three elements: the reference density, the class of functions to represent the \kr or the analysis map, and its estimation from samples \cite{spantini_coupling_2022, baptista_probabilistic_2022}. By choosing the standard normal density as the reference density and restricting to linear approximations to the \kr, we recover the analysis map of stochastic \enkf \cite{spantini_coupling_2022}. To capture non-Gaussianity that arises from the nonlinearity of the state-space model, Spantini \etal \cite{spantini_coupling_2022} introduced a new ensemble filter called the stochastic map filter (\smf) that relies on  interpretable and parsimonious nonlinear transformations to gradually depart from the linear setting of the \enkf. In practice, the estimated rearrangement must revert to a linear behavior in the tails to avoid additional variance due the limited number of samples in these low-density regions. By restricting to a linear asymptotic behavior and choosing a Gaussian reference density, the \enkf and nonlinear extensions such as the \smf assume at least that the tails of the joint forecast distribution have a Gaussian decay. Thus for filtering problems with heavy-tailed distributions, these algorithms are fundamentally biased and do not lead to consistent inference. The nonlinear transformation of the \smf targets can only capture non-Gaussian behaviors in the bulk of the joint forecast distribution such as skewness and multimodal behaviors.

Rather than changing the class of functions as in the \smf, we use a parametric \tdisted reference (instead of the Gaussian reference) and restrict to the class of linear transformations to estimate the \kr. The \tdists are a parametric family of distributions whose tail-heaviness is modulated by a degree of freedom $\dof{} \in [1, \infty]$ \cite{kotz_multivariate_2004}. By using a \tdisted reference density with tunable degree of freedom, we can adapt the analysis map to the tail-heaviness of the joint forecast distribution. A multivariate \tdisted random variable $\X \in \real{n}$  is characterized by a mean vector $\mean{\X} \in \real{n}$, a positive definite matrix called the scale matrix $\scale{\X} \in \real{n \times n}$, and a degree of freedom $\dof{\X} \in [1, \infty]$ \cite{kotz_multivariate_2004}. Interestingly, \tdists revert to Cauchy distributions and Gaussian distributions for $\dof{} = 1$, $\dof{} = \infty$, respectively. Using the methodology presented in \cite{spantini_coupling_2022}, we derive in closed form the linear analysis map, denoted $\tmapdof$, for a joint \tdist for the states and observations  with degree of freedom $\dof{}$.  We show that the analysis map $\tmapdof$ reverts to the transformation of the Kalman filter when $\dof{} = \infty$.  We introduce the \textit{ensemble robust filter (\enrf)} that estimates this analysis map from joint forecast samples. We interpret the \enrf as a generalization of the \enkf with the  appealing feature of adapting the prior-to-posterior update to the tail-heaviness of the joint forecast distribution. Roth \etal \cite{roth2013student} previously derived closed-form update equations for the filtering mean, covariance, and degree of freedom. They considered a \tdisted linear state-space model with a \tdisted initial distribution. They argue that the applicability of \tdist-based filters is limited as the filtering distribution becomes lighter over time and rapidly reverts to a Gaussian distribution. The advantages of the \enrf become apparent when dealing with nonlinear dynamical models. In such systems, the tail-heaviness of the state distribution, represented by the empirical degree of freedom, evolves over time, depending on the local expansive or contractive nature of the dynamical model. In our numerical experiments, we observed that the empirical degree of freedom of the state distribution remains bounded over time. This observation indicates a balance between the state distribution gaining heavier tails during the forecast step (resulting in a decrease of the empirical degree of freedom) and becoming lighter-tailed during the analysis step (leading to a decrease of the empirical degree of freedom).

%For the sake of conciseness, the degree of freedom of a distribution which is not \tdisted are not closed under nonlinear transformations, , we implicitly the notion of degree of freedom o to refer to an empirical estimate. 
% The novelty of this work is to derive a linear ensemble filter which is applicable to nonlinear state-space models. 

The estimation of high-dimensional and heavy-tailed distributions from limited samples is a long standing challenge in Statistics \cite{hampel1986robust, huber2011robust}. Empirical estimators for mean vectors and covariance matrices derived under light-tails assumptions are unbiased with infinite samples, but are highly sensitive to samples with large norm. Thus, these estimators suffer from large variance with limited samples \cite{hampel1986robust, resnick2007heavy, nair2022fundamentals}. To reduce the variance in the estimated parameters, we will use empirical estimators for the mean and covariance based on heavy-tails assumptions. Before to present our estimation methodology, let us consider the following scenario that highlights some undesired effects of light-tailed estimators: we apply the \enkf over several assimilation cycles with a light-tailed initial distribution. Each analysis step will produce a more condensed light-tailed filtering distribution. The only option to increase the spread of the forecast distribution is to rely on the nonlinearity and stochasticity of the dynamical model, or some \textit{ad-hoc} inflation \cite{asch2016data}. Over many assimilation cycles, the forecast distribution becomes so condensed that we have \textit{over-confidence} in the forecast distribution: observations collected from the real system are no longer assimilated in the state estimate. This lead to poor inference further exacerbated for heavy-tailed systems. 

% We recall that we perform inference with $M \sim 100$ for states of  dimension $n \sim 10^3-10^5$ and observations of  dimension $d \sim 10^2-10^3$. 
% We note that inflation is \textit{ad-hoc} both in its theoretical justification and practical implementation.

% \matcomment{Rewrite this paragraph}
% We present our methodology to estimate parameters of heavy-tailed \tdists from limited samples. 
The maximum likelihood (ML) method is a popular method to estimate parameters that maximizes the log-likelihood of a probability distribution given observation samples \cite{hastie2009elements}. The estimation equations for the parameters are derived by setting the gradient of the empirical log-likelihood with respect to the parameters to zero. For Gaussian distributions, these equations can be solved in closed form leading to the classical formulas for the sample mean and sample covariance matrix. Unfortunately for \tdists, the ML method gives coupled estimation equations for the mean, scale matrix, and degree of freedom. Liu \etal \cite{liu1995ml, liu1997ml} proposed an iterative expectation-maximization (EM) algorithm to solve these coupled estimation equations. The EM algorithm leverages the characterization of a \tdist as a the mixture of a Gaussian distribution and a chi-squared distribution \cite{kotz_multivariate_2004, mclachlan2007algorithm}. The resulting estimators for the mean and scale matrix adaptively weight the importance of the samples based on the \maha. In the context of chaotic dynamical systems, we expect the degree of freedom of the joint forecast distribution to vary as the system evolves in the state space. Thus, the empirical degree of freedom should be updated over time rather than fixed \textit{a priori}. In practice, we are often in the low-data regime: the number of samples $M$ is small compared to the dimension of the observation and state spaces. In this regime, we will show that it is essential to regularize the empirical scale matrix to maintain the adaptivity of the \enrf. In data assimilation, there are usually weak statistical dependencies between state and observation variables at long distance. Distance localization leverages this fact to systematically remove all long-range correlations in the empirical covariance matrices \cite{asch2016data}. Instead of this \textit{ad-hoc} treatment, we apply the \tlasso algorithm: an $l1$ regularized EM algorithm introduced by Finegold \etal  \cite{finegold2014robust} to estimate \tdists from samples. The \tlasso draws upon the graphical lasso  (\glasso) algorithm \cite{friedman2008sparse} to solve an $l1$ sparsity-promoting penalized optimization problem for the  empirical inverse scale matrix. In the context of data assimilation, we interpret this $l1$ regularization of the empirical scale matrix as an adaptive localization of the scale matrix \cite{asch2016data}. 

% The \enrf has the appealing feature of adapting its prior-to-posterior update to the (empirical) tail-heaviness of the joint forecast distribution. 

To avoid over-confidence in the forecast distribution, existing ensemble filters like the \enkf and the \smf rely on covariance inflation to artificially inflate the forecast covariance \cite{evensen2009ensemble, carrassi2018data, spantini_coupling_2022}. For instance, multiplicative inflation rescales the forecast covariance by a factor $\alpha > 0$. However, the manual tuning of this extra parameter is time-consuming. In contrast, our principled filter incorporates a data-dependent and tuning-free multiplicative inflation mechanism that complements the adaptive localization of the \tlasso. For a joint \tdist for the states and observations, the posterior distribution is also \tdisted and known in closed form \cite{kotz_multivariate_2004}. The posterior scale matrix is given by the Schur complement of scale matrix for the states $\scale{\X}$ with the scale matrix for the observations $\scale{\Y}$ (similar to the posterior covariance matrix for Gaussian distributions) scaled by a factor $\scaling{\Y}{\y} > 0$. The scaling factor $\scaling{\Y}{\y}$ grows linearly with the \maha of the realization $\y$ of the observation variable $\Y$. We interpret this scaling of the posterior scale matrix as an adaptive and data-dependent multiplicative inflation. Our numerical experiments shows that the \enrf without any tuning outperforms the \enkf with optimally tuned multiplicative inflation and distance localization. 

The remainder of the paper is organized as follows. Section \ref{sec:nomenclature} presents our notation conventions. Section \ref{sec:tdist} reviews some useful properties of \tdists. Section \ref{sec:filtering_measure} provides a basic outline for the filtering problem and introduces the necessary tools from measure transport to derive the analysis map. Section \ref{sec:enrf} specializes the results of Section \ref{sec:filtering_measure} to derive the analysis map for \tdists. We follow by discussing its estimation from joint forecast samples. Example problems are treated in Section \ref{sec:examples}. Concluding remarks follow in Section \ref{sec:conclusion}. For the sake of 
completeness, details on the derivation of the analysis map $\tmapdof$ are provided in Appendix \ref{apx:tmapdof} as well as  pseudo-codes for the \tlasso and the \enrf, see Algorithms \ref{algo:tlasso_algo} and \ref{algo:enrf} in Appendices \ref{apx:tlasso} and \ref{apx:enrf}.

\section{Nomenclature \label{sec:nomenclature}}

In the rest of this manuscript, we use the following conventions. Serif fonts refer to random variables, \eg $\BB{\mathsf{Q}}$ on $\real{n}$ or $\mathsf{Q}$ on $\real{}$. Lowercase roman fonts refer to realizations of random variables, \eg $\BB{q}$ on $\real{n}$ or $q$ on $\real{}$. $\pdf{\BB{\mathsf{Q}}}$ denotes the probability density function for the random variable $\BB{\mathsf{Q}}$, and $\BB{q} \sim \pdf{\BB{\mathsf{Q}}}$ means that $\BB{q}$ is a realization of $\BB{\mathsf{Q}}$.  $\mean{\BB{\mathsf{Q}}}$ and $\cov{\BB{\mathsf{Q}}}$ denote the mean and the covariance matrix of the random variable $\BB{\mathsf{Q}}$, respectively. $\cov{\BB{\mathsf{Q}}, \BB{\mathsf{R}}}$ denotes the cross-covariance matrix of the random variables $\BB{\mathsf{Q}}$ and $\BB{\mathsf{R}}$. Similarly for \tdists, $\scale{\BB{\mathsf{Q}}}$ denote the scale matrix of the random variable $\BB{\mathsf{Q}}$. $\scale{\BB{\mathsf{Q}}, \BB{\mathsf{R}}}$ denotes the cross-scale matrix of the \tdisted random variables $\BB{\mathsf{Q}}$ and $\BB{\mathsf{R}}$.  Empirical quantities are differentiated from their asymptotic counterparts using carets above each symbol. 

\section{Useful properties of \tdists \label{sec:tdist}}

\begin{figure}
    \centering
    \includegraphics[width = 0.45\linewidth]{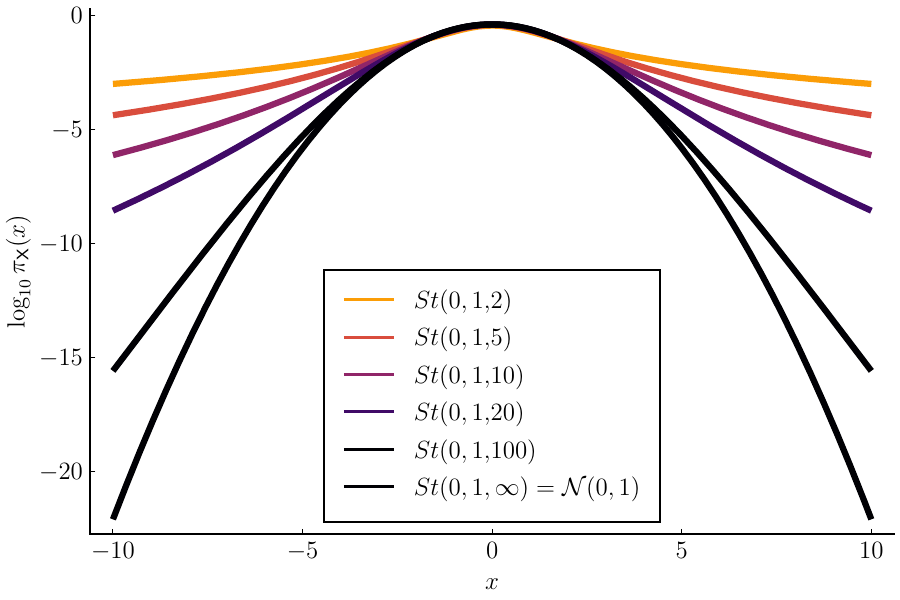}
        \includegraphics[width = 0.45\linewidth]{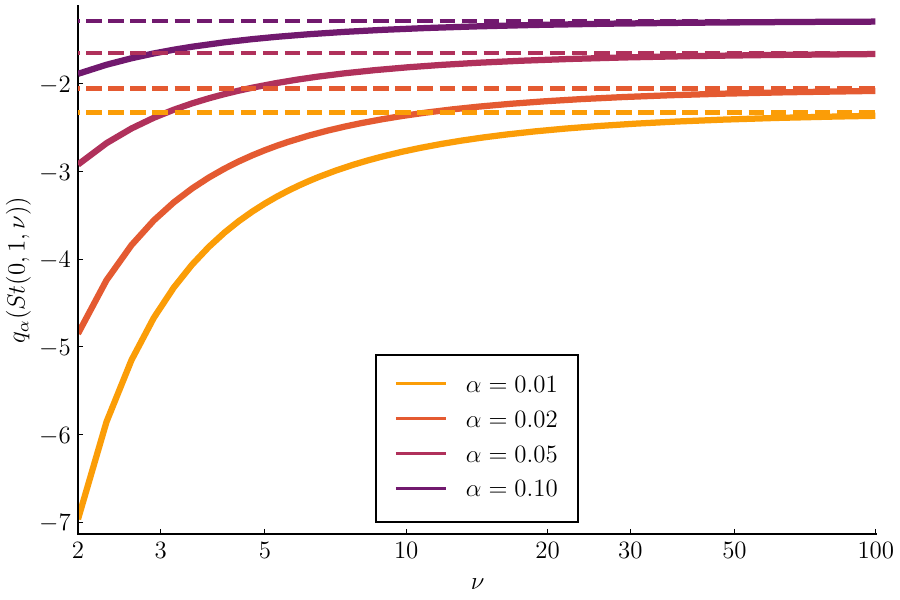}
    \caption{Left panel: Logarithm (with base $10$) of the probability density function of the univariate standard \tdist $\St{0}{1}{\dof{}}$  for $\dof{} = 2, 5, 10, 20, 100$ and the univariate standard Gaussian distribution $\N{0}{1} = \St{0}{1}{\infty}$ (black). The $y$ axis is in log scale. Right panel: Evolution of $\alpha$-quantile $q_{\alpha}$ for $\alpha = 1\%, 2\%, 5\%, 10\%$ with the degree of freedom $\dof{}$ of the univariate standard \tdist. The $x$ axis is in log scale. The corresponding $\alpha$-quantiles of the standard Gaussian distribution $\N{0}{1}$ --- corresponding to $\St{0}{1}{\infty}$ --- are given in dashed lines.}
    \label{fig:logpdf}
\end{figure}

In this section, we present some useful properties of \tdists for the derivation of the analysis map $\tmapdof$ in Sec. \ref{sec:enrf}. A random variable $\X \in \real{n}$ following a \tdist $\St{\meanx}{\scalex}{\dofx}$ is characterized by three parameters: a mean $\meanx \in \real{n}$, a positive definite matrix $\scalex \in \real{n \times n}$ called the scale matrix, and a degree of freedom $\dofx \in [1, \infty)$. Its probability density function $\pdf{\X}$ is given by 
\begin{equation}
\label{eqn:pdf_tdist}
\pdf{\X}(\x) = \frac{\Gamma((\dofx + n)/2)}{\Gamma(\dofx/2)} \frac{1}{(\dofx \pi)^{n/2} \sqrt{\det \scalex}} \left(1 + \frac{\qform{\x - \meanx}{{\scalex}^{-1}}}{\dofx} \right)^{-\frac{\dofx + n}{2}},
\end{equation}
where $\Gamma$ is the Gamma function. The family of \tdists for different values of $\dofx$ %form a family of distributions 
interpolates between a Cauchy distribution for $\dofx = 1$ and a Gaussian distribution for $\dofx = \infty$. We refer to the $n$-dimensional \tdist with zero mean, identity scale matrix and $\dof{}$ degree of freedom as standard \tdist, denoted by $\reftdist{\dof{}}{n}$. To connect the degree of freedom with the tail-heaviness of a \tdist, Fig. \ref{fig:logpdf} depicts the logarithm of the density function and quantiles of the univariate standard \tdist for increasing degrees of freedom.
%\ricomment{It would be great to add a sentence with brief observations from the plot}\matcomment{
%What about: 
A \tdist $\St{0}{1} {\dof{}}$ with finite degree of freedom $\dof{}$ %smaller than $O(50)$ 
has heavier tails than the standard Gaussian distribution. For $\dof{} = 2$, the $1\%, 2\%, 5\%, 10\%$ quantiles are approximately $-7.0, -4.8, -2.9, -1.9$ versus $-2.3, -2.1, -1.6, -1.3$  for the standard Gaussian distribution, respectively. For $\dof{} > O(50)$, the differences in tail behavior are only marginal. %''} %We stress than density values smaller than machine precision cannot be distinguished if we use single-precision float numbers.}

Given the mean and scale matrix, the moments of the t-distribution can be computed as: % have the following relations between its moments and its mean and scale matrix: 
\begin{equation}
    \label{eqn:moments}
    \begin{aligned}
        &\E{\pdf{\X}}{\x} = \meanx, \mbox{ for } \dofx > 1\\
        &\E{\pdf{\X}}{(\x - \meanx)(\x - \meanx)^\top} = \cov{\X} =  \frac{\dofx}{\dofx -2}\scalex \mbox{ for } \dofx > 2.
    \end{aligned}
\end{equation}
We note that the first and second order moments (\ie covariance matrices) are not defined outside of the specified ranges. In this work, we will only consider \tdists with degree of freedom $\dof{\X} > 2$ to ensure the existence of the two first moments.  

An appealing property of \tdists is their stability under affine transformations: if $\Z = \BB{A} \X + \BB{b}$, then $\Z \sim \St{\BB{A} \meanx + \BB{b}}{\BB{A} \scalex \BB{A}^\top}{\dofx}$ for appropriate $\BB{A}$ and $\BB{b}$. We emphasize that the degree of freedom $\dofx$ remains unchanged by an affine transformation. 

We define $\deltaform{\X}{\x} = \qform{\x - \meanx}{\scalex^{-1}} $ to be the \textit{\maha} (square of the Mahalanobis distance) of $\x$ from the distribution $\pdf{\X}$. We emphasize that the \maha is invariant (\ie conserved) under invertible affine transformations of $\x \mapsto \BB{A} \x + \BB{b}$, where $\BB{A} \in \real{n \times n}$ is an invertible matrix and $\BB{b} \in \real{n}$.

Let's consider two random variables $\X \in \real{n}$ and $\Y \in \real{d}$ such that $(\Y, \X)$ is jointly \tdisted with degree of freedom $\dof{}$, \ie 
\begin{equation}
    \stack{\Y}{\X} \sim \St{\stack{\meanx}{\mean{\Y}}}{
    \begin{bmatrix}
    \scale{\Y} & \scale{\X, \Y}^\top \\
    \scale{\X, \Y} &  \scalex
    \end{bmatrix}}{\dof{}}.
\end{equation}
Remarkably, the marginal and conditional distributions are known in closed form \cite{kotz_multivariate_2004}. By applying the linear transformation $\BB{A} = [\zero{d}, \id{n}]$ to the joint random variable $(\Y, \X)$, we obtain the marginal distribution for $\X$ is $\Stx$. The conditional distribution $\pdf{\X \given \Y= \y}$ is also \tdisted with mean $\mean{\X \given \Y = \y}$, scale matrix $\scale{\X \given \Y = \y}$, and degree of freedom $\dof{\X \given \Y = \y}$ given by \cite{kotz_multivariate_2004}:
\begin{equation}
\label{eqn:conditional_tdist}
\begin{aligned}
    \mean{\X \given \Y = \y} & = \mean{\X} + \scale{\X, \Y} \scale{\Y}^{-1}(\y - \meany),\\
    \scale{\X \given \Y = \y} & = \scaling{\Y}{\y}
    \left( \scalex - \scalexy {\scaley}^{-1} \scalexy^\top \right),\\
    \scaling{\Y}{\y} & = \frac{\dof{} + \qform{\y - \meany}{{\scaley}^{-1}}}{ \dof{} + d } = \frac{\dof{} + \deltaform{\Y}{\y}}{ \dof{} + d },\\
    \dof{\X \given \Y = \y} & = \dof{} + d,
\end{aligned}
\end{equation}

Several comments are in order regarding this conditional \tdist, in particular how it compares with the Gaussian case. By noting that $\scalexy {\scaley}^{-1} = \cov{\X, \Y} {\cov{\Y}}^{-1}$ for \tdists, we conclude that the conditional mean is identical to the Gaussian case. The conditional scale matrix $\scale{\X \given \Y = \y}$ of a \tdist is given by the Schur complement $\scalex$ with $\scaley$, namely $\scalex - \scalexy {\scaley}^{-1} \scalexy$, multiplied by a scalar $\scaling{\Y}{\y} >0$. The key difference with the conditional Gaussian covariance is the additional scaling factor $\scaling{\Y}{\y} >0$. Interestingly, $\scaling{\Y}{\y}$ depends linearly on the \maha of the realization $\y \sim \pdf{\Y}$ and thus grows quadratically with the normalized deviation of $\y$ with respect to $\meany$. As a result, \tdists capture that the uncertainty in the conditional distribution is larger when we condition on an outlier realization $\y \sim \Y$ than on a realization close to the mean $\meany$. Finally, the degree of freedom of the conditional \tdist $\X \given \Y = \y$ increases (relative to the prior $\X$) by the dimension $d$ of the observation variable $\Y$. We will later comment on the evolution of the degree of freedom of the state in the context of sequential data assimilation.

We finish this discussion on the properties of \tdists by noting that unlike standard Gaussian distributions,  standard \tdists cannot be factorized as the product of their marginals \cite{kotz_multivariate_2004}, even for \tdists with isotropic scale matrices (\ie $\scale{} = \alpha \id{}$ with $\alpha > 0$). We will show in Sec. \ref{subsec:analysis_map} how to bypass this limitation of \tdists in the construction of the prior-to-posterior transformation $\tmapdof$.

\section{Filtering problem, ensemble filter, and measure transport \label{sec:filtering_measure}}

This section provides an outline of the filtering problem (see \cite{asch2016data, spantini_coupling_2022, law2015data} for further details), and motivates the use of ensemble methods. We follow by presenting the \kr between two distributions \cite{rosenblatt1952remarks, marzouk2016sampling}. Finally, we present the derivation of the analysis map from the \kr that maps the joint forecast distribution to a reference distribution \cite{spantini_coupling_2022}.

\subsection{Filtering problem and ensemble filtering methods \label{subsec:filtering}}

The filtering problem is a classical data assimilation problem in which we seek to sequentially estimate the state $\X_t$ at time $t$ based on all the observations $\y_{1:t}$ available up to that time. More precisely, we seek to estimate the filtering density $\pdf{\X \given \Y_{1:t} = \y_{1:t}}$. In this work, we consider generic nonlinear state-space models given by the pair of a dynamical model and an observation model. The evolution of the state $(\X_t)_{t \geq 0}$ is fully described by the initial density $\pdf{\X_0}$ and the dynamical model that propagates the state from one time step to the next, \ie 
\begin{equation}
\label{eqn:dynamical_model}
    \X_{t+1} = \dyn(\X_t) + \Noisedyn_t,
\end{equation}
where $\dyn: \real{n} \to \real{n}$ is the dynamical operator
%(flow map of a systems of ordinary or partial differential equations that advanc  
and $\Noisedyn_t$ is an independent process noise. As we propagate the state forward, we collect observations $(\Y_t)_{t \geq 1}$ at every time step according to the observation model:
\begin{equation}
\label{eqn:observation_model}
    \Y_t = \obs(\X_t) + \Noiseobs_t,
\end{equation}
where $\obs: \real{n} \to \real{d}$ is the observation operator and $\Noiseobs_t$ is an independent observation noise. 

In this work, we assume that we do not have analytical access to the dynamical and observation operators. That is, we can only generate samples from the transition kernel $\pdf{\X_{t+1} \given \X_t = \x_t}$ and the likelihood function $\pdf{\Y_{t} \given \X_t = \x_t}$ using the dynamical model and the observation model, respectively.

Ensemble filtering methods are a popular class of methods to approximately solve filtering problems \cite{spantini_coupling_2022, asch2016data}. These methods create a Monte-Carlo approximation of the filtering distribution $\pdf{\X_t \given \Y_{1:t} = \y_{1:t}}$ by propagating a set $M$ samples $\{\x^{(1)}, \ldots, \x^{(M)} \}$ over time. At each assimilation cycle, an ensemble filter operates in two steps: a forecast step followed by an analysis step. In the forecast step, the ensemble filter propagates $M$ samples $\{\x^{(1)}, \ldots, \x^{(M)} \}$ from the filtering density $\pdf{\X_{t-1} \given \Y_{1:t} = \y_{1:t-1}}$ at time $t-1$ through the dynamical equation \eqref{eqn:dynamical_model} to form a particle approximation of the forecast density $\pdf{\X_{t} \given \Y_{1:t-1} = \y_{1:t-1}}$ at time $t$. Using these forecast samples, we generate $M$ samples $\{\y^{(1)}, \ldots, \y^{(M)} \}$ from the likelihood model: $\iup{\y}_t \sim \pdf{\Y_t \given \X_t = \iup{\x}}$. The samples $\{(\iup{\y}, \iup{\x}) \}$ form joint samples from the \textit{joint forecast distribution} $\pdf{\Y_t, \X_t \given \Y_{1:t-1} = \y_{1:t-1}}$. In the analysis step, the joint forecast samples $\{(\iup{\y}, \iup{\x}) \}$ get mapped to samples $\{\iup{\x}\}$ from the filtering density at the next time step $\pdf{\X_t \given \Y_{1:t} = \y_{1:t}}$ by conditioning on the realization $\y_t$ of the observation variable $\Y_t$. Our treatment of the analysis step relies on the existence of an underlying map $\tmap_t:\real{d} \times \real{n} \to \real{n}$, called \textit{analysis map}, that maps the joint forecast density $\pdf{\Y_t, \X_t \given \Y_{1:t-1} = \y_{1:t-1}}$ to the filtering density $\pdf{\X_t \given \Y_{1:t} = \y_{1:t}}$ \cite{spantini_coupling_2022, marzouk2016sampling}. A formal construction of the analysis map is discussed in the next section. We stress that different ensemble filters share the same forecast step but differ in the analysis step by estimating distinct analysis maps $\widehat{\tmap}_t$ from the joint forecast samples \cite{spantini_coupling_2022, leprovost2021low, baptista_probabilistic_2022}. We stress that the analysis step can be treated as a \textit{static} Bayesian inference problem as it does not involve time propagation. Kalman \cite{kalman1960new} derived the exact analysis map $\tmap_{\text{KF},t}$ for linear-Gaussian state-space models:
\begin{equation}
\label{eqn:kf_map}
\tmap_{\text{KF},t}(\y_t, \x_t) = \x_t - \cov{\X_t, \Y_t} \cov{\Y_t}^{-1}(\y_t - \y_t^\star),
\end{equation}
where $\ystar_t$ is the observation to assimilate, $\cov{\X_t, \Y_t}$ is the cross-covariance matrix between the state and observation variables, and $\cov{\Y_t}^{-1}$ is the precision matrix (inverse of the covariance matrix) of the observation variable. The linear operator $\K_t = \cov{\X_t, \Y_t} \cov{\Y_t}^{-1} \in \real{n \times d}$ is called the Kalman gain and maps observation discrepancies to state updates. 
Evensen \cite{evensen1994sequential} introduce the \textit{stochastic ensemble Kalman filter (\senkf)}  that estimates $\tmap_{\text{KF},t}$ from the joint forecast samples. The analysis map of the \senkf replaces covariances with empirical covariances estimated from the joint forecast samples $\{(\iup{\y}_t, \iup{\x}_t)\} \sim \pdf{\Y_t, \X_t \given \Y_{1:t-1} = \y_{1:t-1}}$:
\begin{equation}
\label{eqn:enkf_map}
\widehat{\tmap}_t(\y_t, \x_t) = \x_t - \scov{\X_t, \Y_t} \scov{\Y_t}^{-1}(\y_t - \y_t^\star).
\end{equation}

\textbf{Remark: \textit{In the rest of this paper, we omit the time-dependence subscripts wherever possible since the analysis step can be treated as a static inverse problem and does not involve time propagation.}}

\subsection{The \kr \label{subsec:kr}}
This paper introduces the \textit{ensemble robust filter (\enrf)}: an ensemble filter that estimates the analysis map $\tmapdof$ for \tdisted joint forecast distributions from samples.  To build $\tmapdof$, we rely on tools from measure transport \cite{marzouk2016sampling}. Measure transport theory looks at transformations that map a target density $\pdf{}$ to a reference density $\eta$ defined on $\real{m}$ \cite{villani2008optimal}. A transformation  $\tmap \colon \real{m} \to \real{m}$ that verifies this property  is called a \textit{transport map}, and we say that $\tmap$ \textit{pushes forward} the target density $\pdf{}$  to the reference density $\eta$, denoted $\push{\tmap}\pdf{} = \eta$. This question has been extensively studied in optimal transport where one seeks a map that minimizes some cost function $c \colon \real{m} \to \real{m}$ \cite{villani2008optimal}. For smooth target and reference densities, there is no unique way to build the transport map. In the context of Bayesian inference that underpins this work, we have no need for the optimality of the transport map with respect to a particular cost function $c$. Instead we seek a transformation tailored for conditional sampling and conditional density estimation. One of those is candidates the \textit{\kr (KR rearrangement)} defined as the unique lower triangular and increasing map $\smap \colon \real{m} \to \real{m}$ that pushes forward the target to the reference density (under mild assumptions on the densities) \cite{rosenblatt1952remarks, marzouk2016sampling}:

\begin{equation}
\label{eqn:kr_rearrangement}
     \smap(\z) = \smap(z_1, z_2, \cdots, z_m)=\left[\begin{array}{l}
    S^{1}\left(z_{1}\right) \\
    S^{2}\left(z_{1}, z_{2}\right) \\
    \vdots \\
    S^{m}\left(z_{1}, z_{2}, \ldots, z_{m}\right)
    \end{array}\right],
\end{equation}
where the $k$th component of the map $S^k \colon \real{k} \to \real{}$ only depends on the first $k$ components $(z_1, \ldots, z_k)$ of the input variable $\z \in \real{n}$. The monotonicity of $\smap$ is understood in the following sense:  for $k = 1, \ldots, m$, the induced univariate function of the $k$th component $\xi \mapsto S^k(z_1, \ldots, z_{k-1}, \xi)$ is strictly increasing. As mentioned above, the \kr has many interesting properties for Bayesian inference owing to this lower-triangular structure and monotonicity. First, the \kr fully characterizes the marginal conditionals of the target distribution. Let us consider a random variable $\Z \in \real{m} \sim \pdf{}$ and a reference density $\eta$ that can be factorized as the product of its marginals $\eta(\z) = \prod_{k=1}^m \eta_k(z_k)$, \eg a standard normal. Then, the univariate function $\xi \mapsto S^k(\z_{1:k-1}, \xi)$ pushes forward the marginal conditional density $\pdf{\mathsf{Z}_k \given \Z_{1:k-1}}(\xi | \z_{1:k-1})$ to the $k$th  marginal of the reference density $\eta_k$, \ie $(\push{S^k(\z_{1:k-1}, \cdot)} \pdf{\mathsf{Z}_k \given \Z_{1:k-1} = \z_{1:k-1}})(\xi) = \eta_k(\xi)$ \cite{marzouk2016sampling, baptista2021learning}. We will leverage this property in the construction of the analysis map discussed in the next section. Also, its lower triangular and monotonic structure greatly simplifies computations of the determinant of the Jacobian of the map $\det \nabla_{\z} \smap(\z)$ and reduces the inversion of the map to a sequence of root finding problems \cite{marzouk2016sampling, spantini_coupling_2022, leprovost2021low}. Note that we do not employ these last two properties, as we will derive the analysis map in closed-form.

The choice of the reference density is left unspecified in the rest of this section. Section \ref{sec:enrf} will specialize the present results when the reference density is a well chosen product of independent \tdists. 

\subsection{Construction of the analysis map \label{subsec:analysis_map}}

\begin{figure}
    \centering
    \vskip 0.5cm
    \begin{overpic}[width = 0.8\linewidth]{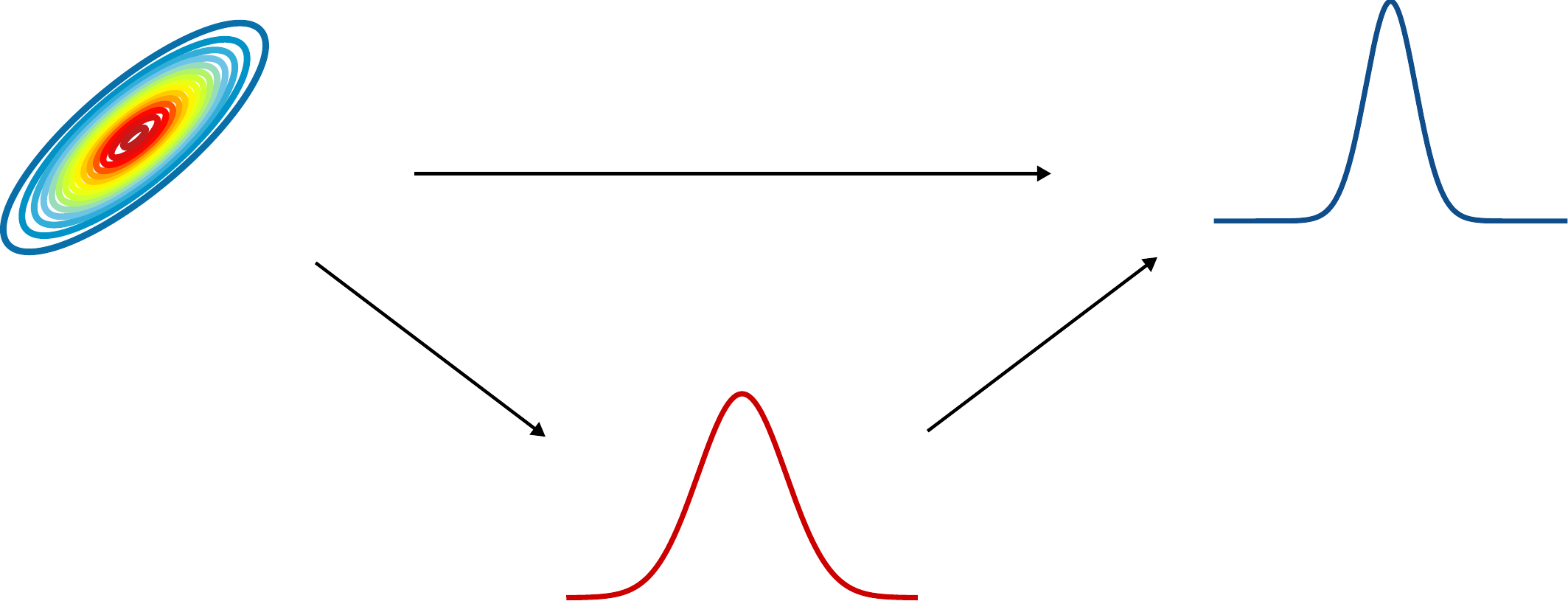}
    \put(10,  22){\large{$\pdfjoint$}}
    \put(46, -4){\large{$\eta_{\X}$}}
    \put(84, 20){\large{$\pdf{\X \given \Y = \ystar}$}}
    
    \put(18, 10){\large{$\smap^{\Xup}(\y, \x)$}}
    
    \put(29, 30){\large{$\smap^{\Xup}(\ystar, \cdot)^{-1} \circ \smap^{\Xup}(\y, \x)$}}
    \put(40, 22){\large{$\tmap(\y, \x)$}}
    \put(63, 10){\large{$\smap^{\Xup}(\ystar, \cdot)^{-1}$}}
    \end{overpic}
    \vskip 1.0cm 
    \caption{Construction of the analysis map $\tmap$ for a univariate state and observation (\ie $n = 1$ and $d = 1$), adapted from Le Provost \etal \cite{leprovost2021low}.} 
    \label{fig:construction_analysismap}
\end{figure}
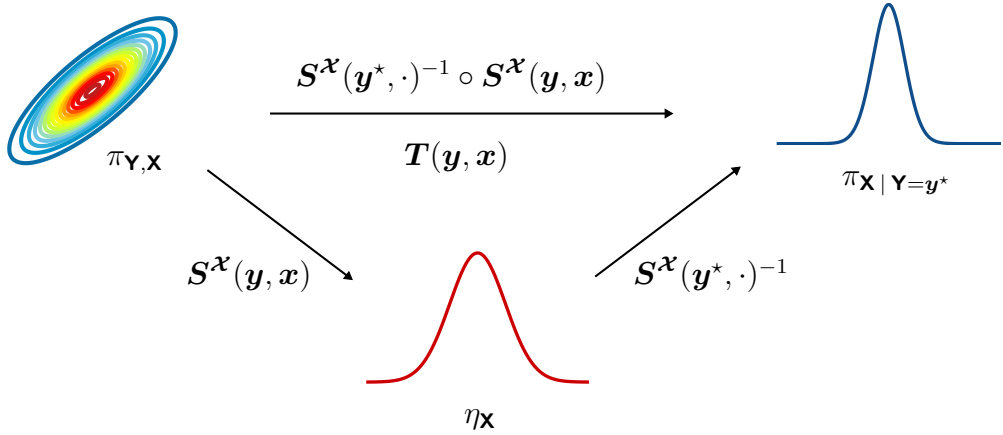

Let us consider a pair of random variables $(\Y, \X) \sim \pdfjoint$ with $ \Y \in \real{d}$ and $\X \in \real{n}$. In the rest of this paper, we will frequently identify $\real{d + n}$ with $\real{d} \times  \real{n}$. This section presents the construction of the analysis map $\tmap \colon \real{d + n} \to \real{n}$ that pushes forward the prior distribution $\pdf{\X}$ to  the posterior distribution $\pdf{\X \given \Y = \ystar}$, where $\ystar \sim \pdf{\Y}$. Let $\smap: \real{d + n} \to \real{d + n}$ denote the \kr that pushes forward $\pdfjoint$ to some reference density $\eta$ defined on $\real{d + n}$, \ie $\smappush \pdfjoint = \eta$. From its lower triangular structure, $\smap$ can be partitioned as follows:

\begin{equation}
\label{eqn:split}
\smap(\y, \x)=\left[\begin{array}{c}
\begin{aligned}
& \smap^{\Yup}(\y) \\
& \smap^{\Xup}(\y, \x)
\end{aligned}
\end{array}\right],
\end{equation}
where $\smap^{\Yup}\colon\real{d} \xrightarrow{} \real{d}$ and $\smap^{\Xup}\colon\real{d} \times \real{n} \xrightarrow{} \real{n}$. Let us assume that the reference density can be factorized as $\eta(\y, \x) = \eta_{\Y}(\y) \eta_{\X}(\x)$. By definition of $\smap$, we have $\smap^{\Xup}(\Y, \X) \sim \eta_{\X}(\x)$.  Also, from the monotonicity of the \kr, the map $\BB{\xi} \mapsto \smap^{\Xup}(\ystar, \BB{\xi})$ is a bijection on $\real{n}$ and pushes forward $\pdf{\X \given \Y = \ystar}$ to the reference density $\eta_{\X}(\x)$. Thus for any pair of samples $(\y, \x) \sim \pdfjoint$, there exists a unique element $\x_a \in \real{n}$ such that $\smap^{\Xup}(\y, \x) = \smap^{\Xup}(\ystar, \x_a)$. The solution $\x_a$ is exactly the posterior update of the prior estimate $\x$ given the realization $\ystar$ of the observation variable \cite{leprovost2021low}. Spantini \etal \cite{spantini_coupling_2022} define the prior-to-posterior transformation, or analysis map, $\tmap \colon \real{d} \times \real{n} \to \real{n}$ that pushes forward the prior $\pdf{\X}$ to the posterior $\pdf{\X \given \Y = \ystar}$ as:
\begin{equation}
\label{eqn:tmapdef}
    \tmap(\y, \x) = \smap^{\Xup}(\ystar, \cdot)^{-1}\circ \smap^{\Xup}(\y, \x),
\end{equation}
where $\smap^{\Xup}(\ystar, \cdot)^{-1}$ denotes the inverse of the map $\BB{\xi} \mapsto \smap^{\Xup}(\ystar, \BB{\xi})$ that comes from setting the first $d$ entries of $\smap^{\Xup}$ to $\ystar$, see Fig. \ref{fig:construction_analysismap}. From its lower triangular structure, the partial inversion of $\smap^{\Xup}$ can be reduced to solving a sequence of one dimensional root finding problems \cite{marzouk2016sampling}. The next section will show that the analysis map can be computed in closed form for a joint \tdist and a reference density given by a particular product of independent \tdists. We stress that $\smap^{\Yup}$ is only an artifact in the construction of the analysis map  \eqref{eqn:tmapdef}. Baptista \cite{baptista_probabilistic_2022} showed that the analysis map defined by \eqref{eqn:tmapdef} is broadly applicable and only requires that the map $\smap^{\Xup}$ removes the dependence of the output on the observation variable $\Y$.

\section{The ensemble robust filter (\enrf): an ensemble filter for \tdists \label{sec:enrf}}

In this section, we present the ensemble robust filter (\enrf), a new ensemble filter for heavy-tailed filtering problems. We discuss the derivation of the analysis map for a joint \tdist $\pdfjoint$. First, we construct the \kr $\widetilde{\smap}_{\dof{}}$ that pushes forward $\pdfjoint$ to a well chosen reference density $\eta$. The analysis map $\tmapdof$ follows naturally  by applying the partial inversion of the \kr $\widetilde{\smap}_{\dof{}}$ given in \eqref{eqn:tmapdef}. Then, we discuss the estimation of the map $\tmapdof$ from joint forecast samples. Finally, we present two strategies to reduce estimation occurrences for the degree of freedom in the \enrf. %to reduce the computational cost of estimating the degree of freedom regularization techniques for the \enrf and different variants of the \enrf based on the variability of the empirical degree of freedom of the joint forecast distribution over time.

\subsection{Construction of the analysis map for \tdists}

Let $(\Y, \X)$ be a pair of random variables with $ \Y \in \real{d}$ and $\X \in \real{n}$ that are jointly \tdisted with degree of freedom $\dof{}$, \ie
\begin{equation}
    \stack{\Y}{\X} \sim \St{\stack{\meanx}{\mean{\Y}}}{
    \begin{bmatrix}
    \scale{\Y} & \scale{\X, \Y}^\top \\
    \scale{\X, \Y} &  \scalex
    \end{bmatrix}}{\dof{}}.
\end{equation}
By marginal conditional factorization, we have $\pdf{\Y, \X}(\y, \x) = \pdf{\Y}(\y) \pdf{\X \given \Y}(\x \given \y)$  where the marginal distribution is given by $\pdf{\Y} = \St{\mean{\Y}}{\scale{\Y}}{\dof{}}$ and the conditional distribution is given by $\pdf{\X \given \Y = \y} = \St{\mean{\X \given \Y = \y}}{\scale{\X \given \Y = \y}}{\dof{}  + d}$.
Let $\Z = [\Z_1; \Z_2] \in \real{d + n}$ with $\Z_1 \in \real{d}$ and  $\Z_2 \in \real{n}$, be a random variable  following a $d + n$ standard \tdist $\reftdist{\nu}{d + n}$ with degree of freedom $\dof{}$:
\begin{equation}
    \Z = \stack{\Z_1}{\Z_2} \sim \reftdist{\nu}{d + n} = \St{\zero{d + n}}{\begin{bmatrix}\id{d} &  \zero{}\\
    \zero{} & \id{n}
    \end{bmatrix}}{\dof{}}.
\end{equation}
Similarly, we have $\pdf{\Z_1, \Z_2}(\z_1, \z_2) = \pdf{\Z_1}(\z_1) \pdf{\Z_2 \given \Z_1 = \z_1}(\z_2)$ with $\pdf{\Z_1}(\z_1)  = \St{\zero{d}}{\id{d}}{\dof{}}$ and $\pdf{\Z_2 \given \Z_1  = \z_1} = \St{\zero{n}}{\scaling{\Z_1}{\z_1} \id{n}}{\dof{} + d}$. 

To build the analysis map, it is natural to consider the \kr $\smap_{\dof{}}$ that pushes forward the joint distribution to a reference density given by the standard \tdist with the same degree of freedom $\dof{}$. Unlike standard Gaussian distributions,  standard \tdists cannot be factorized as the product of their marginals \cite{kotz_multivariate_2004}. For two realizations $\y$ and $\ystar$ of $\Y$, the partials maps $\x \mapsto \smap_{\dof{}}^{\Xup}(\y, \x)$ and $\x \mapsto \smap_{\dof{}}^{\Xup}(\ystar, \x)$ push forward the posterior distribution $\pdf{\X \given \Y= \y}$ and $\pdf{\X \given \Y= \ystar}$ to two different reference densities $\pdf{\Z_2 \given \Z_1 = \z_1(\y)}$ and $\pdf{\Z_2 \given \Z_1 = \z_1(\ystar)}$. Baptista \cite{baptista_probabilistic_2022} showed that it is sufficient for the lower-block of the \kr to remove the dependence of $\X$ on $\Y$ to build the analysis map \eqref{eqn:tmapdef}. Using the independence of $\Z_1$ and $\Z_2/\sqrt{\scaling{\Z_1}{\z_1}}$ \cite{ding2016conditional}, we define the rescaled variable $\tilde{\Z} = [\tilde{\Z}_1, \tilde{\Z}_2] = [\Z_1, \Z_2/\sqrt{\scaling{\Z_1}{\z_1}}]$, such that $\pdf{\tilde{\Z}}$ can be factorized as the products of its marginals, \ie $\pdf{\tilde{\Z}}(\tilde{\z}) = \pdf{\tilde{\Z}_1}(\tilde{\z}_1) \pdf{\tilde{\Z}_2}(\tilde{\z}_2)$ with $\pdf{\tilde{\Z}_1} =\pdf{\Z_1} = \reftdist{\nu}{d}$ and $\pdf{\tilde{\Z}_2} = \reftdist{\nu + d}{n}$. We note that $\tilde{\Z}$ is not \tdisted.

% Let $\widetilde{\smap}_{\dof{}}$ be the \kr  that pushes forward $\pdfjoint$ to the reference density $\pdf{\tilde{\Z}}$. 
Let $\lmap_{\Y} \in \real{d \times d}, \lmap_{\X \given \Y = \y} \in \real{n \times n}$ be the Cholesky factors of the inverse observation scale matrix $\scale{\Y}^{-1}$ and the inverse posterior scale matrix $\scale{\X \given \Y = \y}^{-1}$, respectively:
\begin{equation}
    \begin{aligned}
        & \scale{\Y}^{-1}   =\lmap_{\Y}^\top \lmap_{\Y}, \\
        & \scale{\X \given \Y = \y}^{-1}  = \lmap_{\X \given \Y = \y}^\top \lmap_{\X \given \Y = \y}.
    \end{aligned}
\end{equation}

Then, the \kr  $\widetilde{\smap}_{\dof{}}$ that pushes forward $\pdfjoint$ to the reference density $\pdf{\tilde{\Z}}$ is given by:
\begin{equation}
\label{eqn:kr_tdist}
\widetilde{\smap}_{\dof{}}(\y, \x)= \left[\begin{array}{c}
\begin{aligned}
& \widetilde{\smap}_{\dof{}}^{\Yup}(\y) \\
& \widetilde{\smap}_{\dof{}}^{\Xup}(\y, \x)
\end{aligned}
\end{array}\right] = \left[\begin{array}{c}
\begin{aligned}
& \lmap_{\Y}(\y - \mean{\Y}) \\
& \lmap_{\X | \Y = \y} \left[ \left(\x - \meanx \right) - \scale{\X, \Y} \scale{\Y}^{-1}(\y -  \mean{\Y})\right]
\end{aligned}
\end{array}\right].
\end{equation}

Following Spantini \etal \cite{spantini_coupling_2022}, we define the analysis map $\tmapdof : \real{d} \times \real{n} \to \real{n}$ by partial inversion of $\widetilde{\smap}_{\dof{}}^{\Xup}$:
\begin{equation}
    \label{eqn:tmapdof}
    \begin{aligned}
    \tmapdof(\y, \x) & = {\widetilde{\smap}}_{\dof{}}^{\Xup}(\ystar, \cdot)^{-1} \circ {\widetilde{\smap}}_{\dof{}}^{\Xup}(\y, \x)\\
    % & = \meanx + \scale{\X, \Y} \scale{\Y}^{-1} (\ystar - \mean{\Y}) + \BB{L}_{\X | \ystar}^{-1} \BB{L}_{\X | \y} \left[ \left(\x - \meanx \right) - \scale{\X, \Y} \scale{\Y}^{-1}(\y -  \mean{\Y})\right]\\
    & = \meanx + \scale{\X, \Y} \scale{\Y}^{-1} (\ystar - \mean{\Y}) + \sqrt{\frac{\scaling{\Y}{\ystar}}{\scaling{\Y}{\y} }} \left[ \left(\x - \meanx \right) - \scale{\X, \Y} \scale{\Y}^{-1}(\y -  \mean{\Y})\right]
    % & = (\meanx + \scale{\X, \Y} \scale{\Y}^{-1} (\ystar - \mean{\Y}))
    % + \sqrt{\frac{\scaling{\Y}{\ystar}}{\scaling{\Y}{\y} }} \left[ \x - \left(\meanx   + \scale{\X, \Y} \scale{\Y}^{-1}(\y -  \mean{\Y}) \right) \right],
    \end{aligned}
\end{equation}
where we recall that $\scaling{\Y}{\y} = (\dof{} + \deltaform{\Y}{\y})/(\dof{} + d)$ with $d$ the dimension of the observation variable $\Y$. Further details on the derivation of $\tmapdof$ are provided in Appendix \ref{apx:tmapdof}. The analysis map $\tmapdof$ exactly pushes forward the prior distribution $\pdf{\X}$ to the posterior distribution $\pdf{\X \given \Y = \ystar}$ with $\ystar \sim \pdf{\Y}$ for a joint \tdist of the observations and states. To the best of our knowledge, this is the first analytical expression for an analysis map based on a non-Gaussian reference density. 
We call \textit{robust filter (\rf)}, the algorithm that applies the analysis map $\tmapdof$ given by \eqref{eqn:tmapdof}. We introduce the \textit{ensemble robust filter (\enrf)}, with analysis map $\stmapdof$, the ensemble version of the \rf that estimates the analysis map $\tmapdof$ from samples $\{(\iup{\y}_t, \iup{\x}_t)\}$ of the forecast distribution $\pdf{\Y_t, \X_t \given \Y_{1:t-1} = \y_{1:t-1}}$. Before presenting the estimation of $\tmapdof$ from samples in Section~\ref{subsec:estimation}, we comment on the robustness to synthetic observations of the analysis map $\tmapdof$ and the connection between the \rf and \kf and their ensemble versions, namely the \enrf and the \enkf.

% \matcomment{It is interesting to think of $\tmapdof$ as an update for the mean (the mean get mapped to the conditional mean) plus a Kalman's update for the deviation from the mean $(\x - \meanx)$ further scaled by $\sqrt{\scaling{\Y}{\ystar}/\scaling{\Y}{\y}}$ }.

\subsubsection{Sensitivity to outlying synthetic observations}

% \mat{The analysis map $\tmapdof(\y, \x)$ balances two contributions: an update for the mean $\meanx$ and a Kalman's like update for deviation $\x - \meanx$. the a shifthe mean $\meanx$ is shifted to the conditional mean $\mean{\X \given \Y = \ystar}$ and a Kalman's 
% balances two contributions  balances two update modelste by  upda$\tmap_{\dof{}, 1} \colon (\y, \x) \mapsto \meanx + \scale{\X, \Y} \scale{\Y}^{-1} (\ystar - \mean{\Y})$
% and the mapping $\tmap_{\dof{}, 2} \colon (\y, \x) \mapsto \sqrt{\scaling{\Y}{\ystar}/\scaling{\Y}{\y}} \left[ (\x - \meanx) - \scale{\X, \Y} \scale{\Y}^{-1} (\ystar - \y) \right]$. In this section, we propose an intuitive interpretation for the analysis map $\tmapdof$. This update balances two}

% \mat{The analysis map $\tmapdof$ can be decomposed in two parts: a translation by the conditional mean --- $\meanx + \scale{\X, \Y} \scale{\Y}^{-1}(\ystar - \meany)$ --- and a Kalman's like update for the deviation $(\x - \meanx)$, scaled by the ratio of Mahalanobis distances $\sqrt{\scaling{\Y}{\ystar}/\scaling{\Y}{\y}}$.} We will leverage this formulation to highlight thThis formulation of $\tmapdof$ will help us to understand the behavior of the ma} 

We note that the analysis map $\tmapdof$ in \eqref{eqn:tmapdof} decomposes in two parts: a constant component corresponding to the posterior mean $\meanx + \scale{\X, \Y} \scale{\Y}^{-1}(\ystar - \meany)$ and a Kalman-like update for the deviation $(\x - \meanx)$ scaled by the ratio of Mahalanobis distances $\sqrt{\scaling{\Y}{\ystar}/\scaling{\Y}{\y}}$. To highlight the robustness of $\tmapdof$ with respect to outlying synthetic observations, we consider a joint forecast sample$(\y^{(j)}, \x^{(j)})$ with an outlying synthetic observation $\y^{(j)}$ generated by the likelihood model $\pdf{\Y \given \X = \x^{(j)}}$, such that $\deltaform{\Y}{\y^{(j)}} \to \infty$, so $\scaling{\Y}{\ystar}/\scaling{\Y}{\y^{(j)}} \to 0$. In this case, the analysis map $\tmapdof$ \eqref{eqn:tmapdof} reduces to
\begin{equation}
    \tmapdof(\y, \x)  = \meanx + \scale{\X, \Y} \scale{\Y}^{-1} (\ystar - \mean{\Y}).
\end{equation}
Thus, the dependence of $\tmapdof$ on the outlying observation $\y^{(j)}$ vanishes, and $\tmapdof$ reduces to its constant component, effectively mapping the prior sample $\x^{(j)}$ to the posterior mean/median/mode. This behavior reveals the reduced sensitivity of the analysis map $\tmapdof$ \eqref{eqn:tmapdof} to outlying synthetic observations. This situation is frequently encountered in real world problems that deviate from the twin experiment setting. Indeed, the observation model usually represents a simplified physics and potentially suffers from mis-specifications of the observation operator and observation noise \cite{leprovost2021ensemble}. 

The analysis map $\tmapdof$ balances two update models that both depend on the true observation $\ystar$. As a result, the behavior of the analysis map with respect to an outlying true observation $\ystar$, such that $\scaling{\Y}{\ystar} \to \infty$, is difficult to interpret. 
%The analysis map $\tmapdof$ balances two update models that both depend on the true observation $\ystar$.}

% \sout{The behavior of $\tmapdof$ for a true observation $\ystar$ with large \maha is difficult to interpret, due to the balance of the second and third right hand-side terms in the right-hand-side of \eqref{eqn:tmapdof}. The last formulation of $\tmapdof$ in \eqref{eqn:tmapdof} suggests that the update balances two posterior means: one by conditioning on  $\ystar$ (first term) and one by conditioning on  $\y$. The desired behavior of the map is unclear to us as $\ystar$ is the only knowledge of the true physical system.
% There is no expectation on the \maha of a true observation $\ystar$  to be large or small as the computed distance inherently depends on the choice of dynamical and observation models to represent the true system. Also, there is a choice in the distance used. For certain dynamical systems, one could consider the distance of the observation from the manifold of the attractor of the dynamical system to determine if this is an outlying observation. We leave this consideration for future work.} 

We conclude this discussion by noting that the adaptivity of the map does not depend on the individual values of $\scaling{\Y}{\ystar}$ and $\scaling{\Y}{\y}$, but on their ratio $\scaling{\Y}{\ystar}/\scaling{\Y}{\y}$. If $\ystar$ and $\y$ have the same \maha, \ie they are on the same hypersphere of $\real{d}$, then the analysis map reduces to the Kalman filter's update.

%However, we believe that the ability to interpolate between heavy-tailed densities and light-tailed densities is extremely valuable, while using only linear transformations. 

\subsubsection{Connection with the Kalman filter \label{subsubsec:connection_kf}}

Spantini \etal \cite{spantini_coupling_2022} re-derived the analysis map of the Kalman filter $\tmapkf$ from \eqref{eqn:tmapdef} by assuming the Gaussianity of the joint distribution of the observations and states and the reference density. To highlight the connection between the map $\tmapdof$  and $\tmapkf$, we perform an asymptotic expansion of $\tmapdof$ \eqref{eqn:tmapdof} for large $\dof{}$:

\begin{equation}
\label{eqn:expansion_tmapdof}
\begin{aligned}
    \tmapdof = & \;  \x - \scale{\X, \Y} \scale{\Y}^{-1} (\y- \ystar) + \frac{1}{\dof{}} \frac{ \deltaform{\Y}{\ystar} - \deltaform{\Y}{\y} }{2} \left[ \left(\x - \meanx \right) - \scale{\X, \Y} \scale{\Y}^{-1}(\y -  \mean{\Y})\right] + O \left(\frac{1}{\dof{}^2} \right)\\
     = & \;  \tmap_{\infty} + \frac{1}{\dof{}} \frac{ \deltaform{\Y}{\ystar} - \deltaform{\Y}{\y} }{2} \left[ \left(\x - \meanx \right) - \scale{\X, \Y} \scale{\Y}^{-1}(\y -  \mean{\Y})\right] + O \left(\frac{1}{\dof{}^2} \right).\\
    \end{aligned}
\end{equation}
Several comments are in order regarding this result. First, the zeroth order term of the expansion $\tmap_{\infty}$ (case $\dof{} = \infty$) is exactly the analysis map of the Kalman filter $\tmapkf$ of \eqref{eqn:kf_map} by noting that $\cov{\X, \Y} \cov{\Y}^{-1} = \scale{\X, \Y} \scale{\Y}^{-1}$. The higher order terms of the expansion correct the analysis map $\tmapkf$ to account for the finite degree of freedom. We interpret the \enrf as a generalization of the \enkf{} to joint \tdists for the observations and states.

Let us consider a joint \tdist $(\Y, \X)$ with degree of freedom $\dof{}$ and compare the statistics of the pushforward of the prior density with the two maps $\tmapdof$ and $ \tmapkf$. By definition, $\tmapdof$ exactly pushes forward the prior $\pdfprior$ to the posterior $\pdfpost$, \ie $\push{\tmapdof}\pdfprior = \pdfpost = \St{\mean{\X \given \ystar}}{\scale{\X \given \ystar}}{\dof{\X \given \ystar}}$:
\begin{equation}
\label{eqn:tmapdof_stats}
\begin{aligned}
    \mean{\push{\tmapdof}\pdfprior} = \mean{\X \given \ystar} = \; & \mean{\X} + \cov{\X, \Y} \cov{\Y}^{-1} (\ystar - \mean{\Y})  = \mean{\X} + \scale{\X, \Y} \scale{\Y}^{-1} (\ystar - \mean{\Y}),\\
    \scale{\push{\tmapdof}\pdfprior} = \scale{\X \given \ystar} = \;  &  \scaling{\Y}{\ystar} \left(\schurxy \right),\\
    \dof{\push{\tmapdof}\pdfprior} = \dof{\X \given \ystar} = \;  & \dof{} + d,
\end{aligned}
\end{equation}
where $d$ is the dimension of the observation variable. By linearity of $\tmapkf$, the pushforward density of the prior $\pdfprior$ by  $\tmapkf$ is also \textit{t-}distributed: $\push{\tmapkf} \pdfprior = \St{\mean{\push{\tmapkf}\pdfprior}}{\scale{\push{\tmapkf}\pdfprior}}{\dof{\push{\tmapkf}\pdfprior}}$ with
\begin{equation}
\label{eqn:tmapkf_stats}
\begin{aligned}
    \mean{\push{\tmapkf}\pdfprior} = \; & \mean{\X} + \cov{\X, \Y} \cov{\Y}^{-1} (\ystar - \mean{\Y})  = \mean{\X} + \scale{\X, \Y} \scale{\Y}^{-1} (\ystar - \mean{\Y}),\\
     \scale{\push{\tmapkf}\pdfprior} = \;  &  \schurxy,\\
    \dof{\push{\tmapkf}\pdfprior} = \;  & \dof{}.
\end{aligned}
\end{equation}
We note that the map of the Kalman filter $\tmapkf$ correctly estimates the posterior mean $\mean{\X \given \ystar}$ for \tdists. However, the estimator for posterior scale matrix does not include the scaling factor $\scaling{\Y}{\ystar}$ nor update the degree of freedom. The update of degree of freedom by the maps $\tmapdof$ and $\tmapkf$ suggest that the \rf creates a posterior distribution with lighter tails than the Kalman filter. However, it is important to stress the uncertainty in the \rf is modulated by the scaling factor $\scaling{\Y}{\ystar}$ that depends on the \maha of $\ystar$. To better understand the action of this scaling, we compute the expected value of $\scaling{\Y}{\ystar}$ for $\ystar \sim \pdfy = \St{\meany}{\scaley}{\dof{}}$. Using the cyclic property of the trace operator and the relation between the scale and covariance of a \tdist, see \eqref{eqn:moments}, we get:
\begin{equation}
      \E{\pdfy}{\scaling{\Y}{\ystar}}   =  %\E{\pdfy}{\frac{\dof{} + \qform{\ystar - \meany}{\scaley^{-1}}}{\dof{} + d}} 
       \E{\pdfy}{\frac{\dof{} + \text{tr}(\scaley^{-1} (\ystar - \meany)(\ystar - \meany)^\top)}{\dof{} + d} } 
      =  \frac{\dof{} + \text{tr}(\scaley^{-1} \covy)}{\dof{} + d}  = \frac{\dof{} + \frac{ \dof{} d}{\dof{} - 2}}{\dof{} + d} > 1.
\end{equation}
Over expectation, this scaling factor inflates the posterior Schur complement $(\schurxy)$. To summarize, the \rf balances three sources of information in the posterior uncertainty: the degree of freedom of the joint distribution, the dimension of the observation, and the \maha distance of the realization $\ystar \sim \pdf{\Y}$.

\begin{figure}
    \centering
    \includegraphics[width = 0.7
    \linewidth]{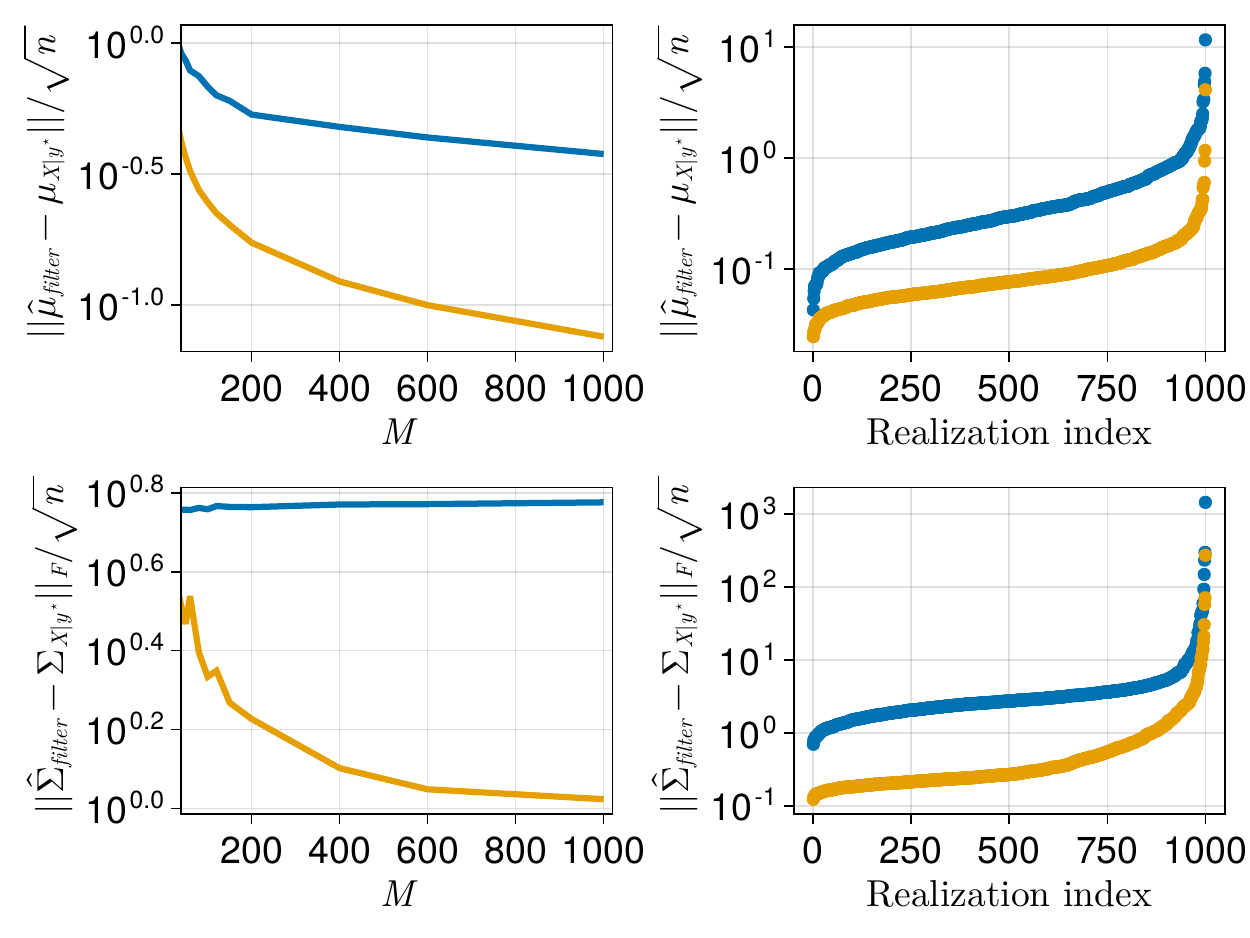}

    \caption{Panel (a): Evolution of the error $||\widehat{\mean{}}_{filter} - \mean{\X \given \ystar}||_2/\sqrt{n}$ with the ensemble size $M$ (averaged over $1000$ realizations), where $\widehat{\mean{}}$ denotes the sample mean with the \enkf (blue) and the \enrf (orange), respectively.
    Panel (b): Empirical distribution of the error $||\widehat{\mean{}}_{filter} - \mean{\X \given \ystar}||_2/\sqrt{n}$ over $1000$ realizations of $\ystar$ using $M = 600$ samples with the \enkf (blue) and the \enrf (orange), respectively.
    Panel (c): Evolution of the error $||\widehat{\cov{}}_{filter} - \cov{\X \given \ystar}||_F/\sqrt{n}$ with the ensemble size $M$ (averaged over $1000$ realizations), where $||\cdot||_F$ denotes the Frobenius norm and $\widehat{\cov{}}$ denotes the sample covariance with the \enkf (blue) and the \enrf (orange), respectively.
    Panel (d): Empirical distribution of the error $||\widehat{\cov{}}_{filter} - \cov{\X \given \ystar}||_F/\sqrt{n}$ over $1000$ realizations of $\ystar$ using $M = 600$ samples with the \enkf (blue) and the \enrf (orange), respectively.}
    \vspace{-0.5cm}
    \label{fig:convergence_analysis}
\end{figure}

It is crucial to emphasize that the previous conclusions regarding the Kalman filter's performance only hold for the asymptotic case, where the number of samples is infinite. In practical scenarios, estimating the mean and covariance matrix of a heavy-tailed distribution from limited samples poses challenges. For instance, the sample mean and sample covariance utilized in the \enkf{} are derived from an $l2$ loss, making them sensitive to outliers. To assess the performance of the sample mean and covariance estimators of the \enkf and \enrf, we conduct the following numerical experiment. We use a standard \tdist  $\pdfjoint = \St{\zero{d +n}}{\id{d + n}}{\dof{}}$ for the joint distribution for the observations and states,  and generate $M$ samples $\{(\iup{\y}, \iup{\x}) \}$ from it, where $n = 10$, $d = 5$, and $\dof{} = 2.5$. Then, we apply the analysis map of the \enkf and the \enrf for a realization $\ystar \sim \pdf{\Y} = \St{\zero{d}}{\id{d}}{\dof{}}$ to assimilate, and obtain posterior samples $\{\iup{\x}_a \} \sim \pdf{\X \given \Y =\ystar}$ for the two ensemble filters. We compute the sample mean and sample covariance for these two posterior ensembles. The results are reported in Fig. \ref{fig:convergence_analysis}. The posterior sample mean estimate from the \enkf and the \enrf empirically converge with increasing ensemble size to the true posterior mean. This is consistent with the asymptotic results for the mean of the pushforward of the prior by the maps $\tmapdof$ and $\tmapkf$,  see \eqref{eqn:tmapdof_stats} and \eqref{eqn:tmapkf_stats}. The \enrf mean estimate reduces the tracking error of the \enkf by $50\%$. 

To interpret this difference, we recall the central limit theorem \cite{rigollet2015high}. Let us consider $M$ samples $\{\zi\}$ drawn from a distribution $\pdf{\Z}$ with mean $\meanz$ and finite variance. Then the sample mean estimator $\smeanz$ --- used by the \enkf --- verifies $||\smeanz - \meanz|| \leq C_{\Z}/\sqrt{M}$, where the positive constant $C_{\Z}$ depends on the variance of the sample mean estimator $\smeanz$ for $\pdf{\Z}$. For heavy-tailed distributions, the constant $C_{\Z}$ is large resulting in degraded tracking performance of the mean estimator of the \enkf. By adaptively weighting samples, the sample mean estimator of the \enrf has reduced sensitivity to outlying samples, thus better performance.

Interestingly, the sample covariance estimate of the \enkf  does not empirically converge to the true posterior covariance. This is consistent with our previous comment that the posterior estimate of the \enkf  does not include the scaling factor $\scaling{\Y}{\ystar}$ in the posterior covariance matrix.

\textbf{Remark:} \textbf{\textit{The \enrf embeds an adaptive and data-dependent multiplicative inflation mechanism.}}\\
Ensemble filters like the \enkf or the \enrf first estimate the statistics of the joint forecast density $\pdf{\Y_t, \X_t, \given \Y_{1:t-1} = \y_{1:t-1}}$ to build the empirical analysis map. As a result of the limited number of samples, the estimated analysis map suffers from sampling errors, spurious long-range correlations, and underestimation of the true statistics \cite{evensen2009ensemble, carrassi2018data}. Inflation is a common regularization of the \enkf that specifically targets the finite-size sampling errors and the underestimation of the covariance matrices by artificially increasing the spread of the forecast samples \cite{raanes2015improvements, asch2016data}. For instance, multiplicative inflation rescales the deviation of the samples from the mean by a factor $\sqrt{\alpha} >1$:
\begin{equation}
    \iup{\x} \leftarrow \smean{\X} + \sqrt{\alpha} (\iup{\x} - \smean{\X}).
\end{equation}
This corresponds to multiplicative rescaling of the forecast covariance matrix  $\scovx \leftarrow \alpha \scovx$. For a linear-Gaussian observation model, the filtering covariance matrix is also rescaled by $\alpha$: $\scov{\X \given \Y} \leftarrow \alpha \scov{\X \given \Y}$. In the finite-sample setting that motivates this work, we interpret the scaling $\scaling{\Y}{\ystar}$ in the posterior scale matrix \eqref{eqn:tmapdof_stats}
 as an adaptive and data-dependent multiplicative inflation mechanism embedded in  the \enrf. We use the term ``adaptive'' to emphasize that it does not require tuning and ``data-dependent''  to indicate that the multiplicative inflation relies on the \maha of the realization $\ystar$.

% We have found empirically that the sample mean and covariance estimators have a slower convergence to the true posterior mean of \tdists that the ones used by the \enrf presented in the next section.

% \matcomment{Need to move this paragraph around}
We conclude this section by mentioning one limitation of the \enrf. Let us assume that the joint forecast distribution $(\Y_t, \X_t)$ is truly Gaussian, \ie $\dof{(\Y_t, \X_t)} = \infty$. In the analysis step of the \enrf, the \tlasso estimates a \textbf{finite} degree of freedom  $\sdof{(\Y_t, \X_t)} < \infty$ from the joint forecast samples 
$\{ (\iup{\y}_t, \iup{\x}_t)\} \sim \pdf{(\Y_t, \X_t)}$. From \eqref{eqn:expansion_tmapdof}, $\tmap_{\dof{(\Y_t, \X_t)}}$ reverts exactly to the Kalman  update for $\dof{(\Y_t, \X_t)} = \infty$. In practice, the estimated map $\hat{\tmap} _{\dof{(\Y_t, \X_t)}}$ contains the zeroth order term  of \eqref{eqn:expansion_tmapdof}, \ie the map of the Kalman filter $\tmapkf$ as wells as the higher order terms accounting for the spurious finite degree of freedom. Thus, the \enrf can introduce more bias than the \enkf for Gaussian joint forecast distributions. 
% For large degree of freedom, the \enrf estimates a linear transformation (the analysis map of the \enkf) with a quadratic family of estimators: resulting in additional variance with minor bias reduction. 
For large degree of freedom (\ie $\dof{(\Y_t, \X_t)} > 50$), it is reasonable to assume that the underlying distribution is Gaussian and to revert to the \enkf's update. In practice, we could consider a hybrid version of the \enrf and the \enkf. At each assimilation cycle, we would first test if the joint forecast distribution is empirically Gaussian, \ie $\sdof{(\Y_t, \X_t)} > 40$. If the test is verified, the hybrid filter will revert to the \enkf's update, otherwise apply the \enrf.

% \subsection{Estimation of the analysis map \texorpdfstring{$\tmapdof$}{}  from samples \label{subsec:estimation}}
\subsection{Estimation of the analysis map from samples \label{subsec:estimation}}

At each assimilation cycle, the \enrf estimates the analysis map $\tmapdof$ \eqref{eqn:tmapdof} from joint samples of the observations and states $\{(\iup{\y}_t, \iup{\x}_t) \} \sim \pdf{\Y_t, \X_t \given \Y_{1:t-1} = \y_{1:t-1}}$ and applies the estimated map $\stmapdof$
 to generate samples $\{\iup{\x}_{a,t} \}$ from the filtering density at time $t$. The construction of $\stmapdof$ involves the inference of the parameters of \tdists--- namely mean, scale matrix, and degree of freedom --- from samples. As mentioned earlier, the classical sample mean and covariance are not suited for heavy-tailed distributions due to their high sensitivity to outlying samples. The rest of this section is organized as follows. First, we present a methodology to estimate parameters of \tdists from samples. This procedure leverages the conditional independence of the joint forecast distribution to promote sparsity in the scale matrix. Second, we present the algorithm for the analysis step of the \enrf.  Finally, we introduce two variants of the \enrf with symplifying approximations to improve scalability of the \enrf . A pseudo-code for the \enrf algorithm is given in Appendix \ref{apx:enrf}.
 
\subsubsection{Maximum likelihood estimation of \tdists \label{subsubsec:ml-tlasso}}

Let  $\Z \in \real{m}$ be a \tdisted random variable with density $\pdf{\Z}(\cdot; \BB{W}_{\Z}) = \Stz$ parameterized by its mean $\meanz$, scale matrix $\scalez$, and degree of freedom $\dofz$, where $\BB{W}_{\Z}$ denotes the set of parameters $(\meanz, \scalez, \dofz)$. The maximum likelihood estimator $\widehat{\BB{W}}_{\Z}$ is defined as the maximizer of the empirical log-likelihood $l(\BB{W}_{\Z} \given \{\zi \}) = \sum_{i = 1}^M \log \pdf{\Z}(\zi; \BB{W}_{\Z}))$ of the parameters $\BB{W}_{\Z}$ given $M$ samples $\{ \zi \} \sim \pdf{\Z}$:  
\begin{equation}
\label{eqn:ML}
\widehat{\BB{W}}_{\Z} = \text{argmax}_{\BB{W}_{\Z}} l(\BB{W}_{\Z} \given \{\zi \}) = \text{argmax}_{\BB{W}_{\Z}}\sum_{i = 1}^M \log \pdf{\Z}(\zi; \BB{W}_{\Z})),
\end{equation}
 We obtain the estimation equation for $\widehat{\BB{W}}_{\Z}$ by setting the gradient of $l(\BB{W}_{\Z} \given \{\zi \})$ with respect to $\widehat{\BB{W}}_{\Z}$ to $0$:
\begin{equation}
\label{eqn:maxML}
\sum_{i = 1}^M \frac{\partial \log(\pdf{\Z}(\zi; \widehat{\BB{W}}_{\Z}))}{\partial \widehat{\BB{W}}_{\Z}} = 0.
\end{equation}

After some algebraic manipulations, we obtain the following estimation equations for $\widehat{\BB{W}}_{\Z} = (\smeanz, \scalez, \sdofz)$ \cite{katz2001low, liu1995ml, liu1997ml, finegold2014robust}:

\begin{equation}
    \label{eqn:ML_mean_scale}
    \smeanz  =  \frac{\sum_{i = 1}^M \tau^{(i)}  \zi }{\sum_{i = 1}^M \tau^{(i)}},\;
    \sscalez  =  \frac{1}{M} \sum_{i = 1}^M \tau^{(i)} (\zi - \smeanz) (\zi - \smeanz)^\top,
\end{equation}
with weights $\{ \tau^{(i)} \}$ given by:

\begin{equation}
    \label{eqn:ML_weights}
    \tau^{(i)} =  \frac{\sdofz + m}{\sdofz + \hat{\delta}_{\Z}(\zi)},\;  \text{with }
   \deltaform{\Z}{\zi} =  \qform{\zi - \smeanz}{\scalez^{-1}}
\end{equation}

If the degree of freedom $\dofz$ is unknown, the estimate $\sdofz$ is obtained by solving the following one-dimensional root finding problem \cite{dogru2018doubly}:

\begin{equation}
    \label{eqn:ML_dof}
           \sdofz = \text{zero of } \dofz \mapsto \sum_{i = 1}^M  \biggr[ \psi \left(\frac{\dofz + m}{2} \right) -  \psi \left( \frac{\dofz}{2} \right)  +  \log \left( \frac{\dofz}{2} \right) - \log \left(\frac{\dofz + \sdeltaform{\Z}{\zi}}{2} \right) - \frac{\dofz + m}{\dofz + \sdeltaform{\Z}{\zi}} +  1   \biggr] \\
    \end{equation}
where $\psi$ is the Digamma function defined as the derivative of the logarithm of the Gamma function $\Gamma$, \ie $\psi(x) = d(\log \Gamma(x))/dx$.

Several comments are in order concerning these estimation equations. For a finite degree of freedom, these estimators adaptively weight the importance of the samples based on their \maha  $\sdeltaform{\Z}{\zi}$. The weight $\tau(\delta) = (\dofz + m)/(\dofz + \delta)$  is decreasing function of the \maha $\delta$, such that outlying samples are down-weighted in these estimators. For $\dofz{} \to \infty$, these estimators converge to the sample mean and covariance formulas derived for a Gaussian model. However, for finite degree of freedom $\dofz < \infty$, the estimation equations (\ref{eqn:ML_mean_scale}, \ref{eqn:ML_weights}, \ref{eqn:ML_dof}) derived from a maximum likelihood approach for \tdist are coupled and do not have a closed-form solution. A classical solution is to rely on an iterative method such as an expectation-minimization approach (EM) to solve these coupled equations \cite{liu1995ml, liu1997ml, kotz_multivariate_2004}.

The EM approach is a popular technique to find estimators for the parameters of a statistical model (here $\BB{W}_{\Z}$) maximizing the log-likelihood (here $l(\BB{W}_{\Z} \given \{\zi \})$) \cite{mclachlan2007algorithm}. It requires the statistical model to have a representation using latent unobserved variables. The EM algorithm alternates between an expectation (E) step and a maximization (M) step. In the E step, we compute the expectation of the log-likelihood given the current estimate of the parameters. In the M step, we compute the parameters maximizing the expected log-likelihood found in the E step. A \tdisted random variable $\Z \sim \Stz$ can be alternatively represented as a mixture of a Gaussian random variable $\X \sim \N{\meanz}{\scalez}$ and a Gamma random variable $\tau \sim \Gamma(\dofz/2, \dofz/2)$ such that $\Z = \meanz + \X/\sqrt{\tau} \sim \Stz$ \cite{kotz_multivariate_2004}. The rest of this section provides a concise outline of the EM algorithm for \tdists, but we refer readers to \cite{liu1995ml, liu1997ml} for further details. 

From the $\Jidx$th iteration of the EM algorithm, with estimated values $\smeanz^\Jidx$, $\sscalez^\Jidx$, and $\sdofz^\Jidx$ for mean, scale, and degree of freedom, respectively, the $\Jidx+1$th iteration proceeds as follows:

\textbf{E-step:} Compute the \maha $\hat{\delta}_{\Z}^{\Jidx + 1, (i)}$ of the samples $\{ \zi \}$ and the weights $\tau^{\Jidx + 1, (i)}$:
\begin{equation}
    \label{eqn:EM_weights}
    % \begin{aligned}
      \delta_{\Z}^{\Jidx + 1, (i)}   = {(\zi - \smeanz^\Jidx)}^\top \sscalez^{\Jidx,{-1}}(\zi - \smeanz^\Jidx)\; \text{and} \;  
      \tau^{\Jidx + 1, (i)}  = \frac{\sdofz^{\Jidx} + m}{\sdofz + \delta_{\Z}^{\Jidx + 1, (i)}}
    % \end{aligned}
\end{equation}

\textbf{M-step:} Compute the new estimates for the mean $\smeanz^{\Jidx + 1}$, scale $\sscalez^{\Jidx + 1}$, and degree of freedom $\sdofz^{\Jidx + 1}$: 
\begin{equation}
    \label{eqn:EM_mean_scale}
    % \begin{aligned}
        \smeanz^{\Jidx +1 }  =  \frac{\sum_{i = 1}^M \tau^{\Jidx+1, (i)}  \zi }{\sum_{i = 1}^M \tau^{\Jidx +1, (i)}},\;
        \sscalez^{\Jidx + 1}  =  \frac{1}{M} \sum_{i = 1}^M \tau^{(i)} (\zi - \smeanz^{\Jidx}) (\zi - \smeanz^{\Jidx})^\top,
    % \end{aligned}
\end{equation}
and the estimate $\sdofz^{\Jidx + 1}$ is given by the root of the following one-dimensional equation:
\begin{equation}
    \label{eqn:ML_dof2}
           \sdofz^{\Jidx + 1} = \text{zero of } \dofz \mapsto \sum_{i = 1}^M  \biggr[ \psi \left(\frac{\sdofz^{\Jidx + 1} + m}{2} \right) -  \psi \left( \frac{\dofz}{2} \right)  + \log \left(\frac{\dofz}{2} \right) - \log \left( \frac{\sdofz^{\Jidx} + \delta_{\Z}^{\Jidx, (i)}}{2} \right) - \frac{\sdofz^{\Jidx} + m}{\sdofz^{\Jidx} + \delta_{\Z}^{\Jidx, (i)}} +  1   \biggr] \\
\end{equation}

In practice, we are often in the low-data regime: the number of samples $M$ is much smaller than the state dimension $m$, \ie $M < m$. The scale estimate $\sscalez$ of \eqref{eqn:EM_mean_scale} has maximal rank $\min(m, M-1)$ and cannot be positive definite as expected. The sample covariance used in the \enkf  faces similar rank-deficiency issues caused by the limited number of samples and results in sampling errors, spurious long-range correlations, and under-estimation of the true statistics \cite{asch2016data, evensen2009ensemble}. Ultimately, we use the sample scale $\sscalez$ to estimate the analysis map $\tmapdof{}$ \eqref{eqn:tmapdof} from the joint forecast samples $\{ \iup{\y}_t, \iup{\x}_t \}$. Pires and Branco \cite{pires2019high} showed that the Mahalanobis distance where the empirical mean and covariance matrix are estimated with less samples than the state dimension is identical for all the samples. This result does not assume that the samples are drawn from a Gaussian distribution, but relies on algebraic properties of the sample covariance matrix. One can extend this collapse result to the sample scale matrix given by \eqref{eqn:ML_mean_scale}. This collapse implies that the Mahalanobis distance is no longer informative of the ``radial distance'' of the samples in the low-data regime. This is critical as the adaptivity of the map $\tmapdof{}$ relies on the ratio $\scaling{\Y}{\ystar}/\scaling{\Y}{\y}$ with $\scaling{\Y}{\y} = (\nu + \deltaform{\Y}{\y})/(\nu + d)$. Thus, the prior-to-posterior update would revert to a linear transformation, and no longer weights the update based on the \maha of the observation realization $\ystar$ and the observation sample $\iup{\y}$. Thus, to preserve this appealing weighting feature of the \enrf, it is imperative to regularize the scale matrix and ensures its full-rankness. The next section presents a $l1$ sparsity-promoting regularization of the scale matrix based on the conditional independence structure of the joint forecast distribution. 

% Estimate t-distributions from samples
\subsubsection{A \textit{l}1 regularized estimation of \tdists from samples}
This section presents the estimation procedure for the mean, scale matrix, and degree of freedom of a \tdist used in the \enrf. We rely on the \tlasso algorithm introduced by Finegold \etal \cite{finegold2014robust}. This algorithm uses a regularized (in a specific sense) EM-type approach to solve coupled estimation equations for the empirical mean, scale matrix, and degree of freedom of a \tdist derived as first order conditions of a regularized log-likelihood loss. We provide in Appendix \ref{apx:tlasso} a pseudo-code for the \tlasso algorithm to estimate parameters of a \tdist from samples. 

% For the sake of conciseness, we refer readers to Finegold \etal \cite{finegold2014robust} for further details on the \tlasso algorithm. 

Observations to assimilate are often given by local functions of the state, \ie they only depend on the state variables that are close-by in physical distance \cite{leeuwen2019non, leprovost2022low}. As a result, the true correlations between the observation and state variables quickly decay with increasing distance. Distance localization is a popular technique to regularize the sample covariance by removing all long-range correlations using a fixed cut-off radius \cite{asch2016data, evensen1994sequential}. With this technique, the \enkf has successfully tracked the state evolution in high-dimensional problems found in weather prediction or geosciences with as little as $M = O(50)$ samples \cite{asch2016data, evensen2009ensemble}. One could similarly apply distance localization to regularize the sample scale matrix. However, distance localization remains an \textit{ad-hoc} method based on the tuning of a characteristic cut-off length for the correlations. Instead, we aim for an adaptive, \ie tuning-free, method to regularize the scale matrix to complement the adaptive and data-dependent multiplication inflation of the \enrf.

We propose to leverage the conditional independence structure of the joint forecast distribution $\pdf{\Y_t, \X_t \given \Y_{1:t-1}}$ as a means of regularizing the scale matrix \cite{spantini_coupling_2022}. For a random variable $\Z \sim \pdf{\Z}$, we say that two variables $\mathsf{Z}_i$ and $\mathsf{Z}_j$ are conditionally independent given the remaining components $\Z_{-ij}$ if $\pdf{\mathsf{Z}_i, \mathsf{Z}_j \given \Z_{-ij}} = \pdf{\mathsf{Z}_i\given \Z_{-ij}} \pdf{\mathsf{Z}_j \given \Z_{-ij}}$ \cite{koller2009probabilistic, spantini2018inference}. We denote this conditional independence as $\mathsf{Z}_i \indep \mathsf{Z}_j \given \Z_{-ij}$. Conditional independence of the joint forecast distribution is frequently observed in data assimilation settings, and typically arises from the local (even if approximate) nature of the state-space model \cite{spantini_coupling_2022, baptista_probabilistic_2022}. In the case of a Gaussian variable $\Z \sim \N{\meanz}{\covz}$, the variables $\mathsf{z}_i$ and $\mathsf{z}_j$ are conditionally independent if the entry $(i,j)$ of the inverse covariance matrix $\covz^{-1}$ is zero, \ie  $\mathsf{Z}_i \indep \mathsf{Z}_j \given \Z_{-ij}$ if and only if $({\covz^{-1}})_{i,j} = 0$. In the Gaussian case, there is a clear connection between the conditional independence of the distribution of interest (here the joint forecast distribution) and the sparsity of its inverse covariance matrix. This result has motivated the development of $l1$ penalized algorithms that promote sparsity in the inverse covariance matrix $\preciz = \covz^{-1}$. Given  samples $\{ \zi \} \sim \pdf{\Z} = \N{\meanz}{\covz}$, the graphical lasso (\glasso) \cite{friedman2008sparse} estimates the inverse covariance $\spreciz$  by maximizing the log-likelihood of a Gaussian distribution given samples $\{ \zi \}$ with a $l1$ penalization of $\preciz$:
\begin{equation}
\label{eqn:glasso}
    \spreciz = \text{argmax}_{\preciz} \text{logdet} (\preciz) - \text{tr} (\scovz \preciz) - \rho \lVert \preciz \rVert_1, 
\end{equation}
where $\text{tr}(\cdot)$ denotes the trace operator, $\rho >0$ is the $l1$ penalty coefficient (to be estimated), and $\scovz$ is the sample covariance matrix of $\{ \zi \}$. The \glasso uses an iterative procedure to maximize \eqref{eqn:glasso} by leveraging solvers for $l1$-penalized least-square problems \cite{friedman2008sparse, mazumder2012graphical}. If the initial estimate of the inverse covariance is positive definite, Mazumder \etal \cite{mazumder2012graphical} showed that at each iteration of the \glasso algorithm, the inverse covariance estimate remains positive definite.

Unfortunately, the conditional independence structure of a \tdisted random variable $\Z$ does not directly correspond to the sparsity pattern of the inverse scale matrix $\scalez^{-1}$. For a sufficiently smooth distribution $\pdf{\Z}$, Baptista \etal \cite{baptista2021learning} showed that the conditional independence structure is given by the sparsity pattern of the Hessian score matrix $\BB{\Omega}$ where $\BB{\Omega}_{i,j} = \E{\pdfz}{\partial^2_{z_i, z_j} \log \pdfz}$. In other words, we have $\mathsf{Z}_i \indep \mathsf{Z}_j \given \Z_{-ij}$ if and only if $ \E{\pdfz}{\partial^2_{z_i, z_j} \log \pdfz} = 0$. Using this result, Bax \etal \cite{bax2022generalized} derived the corresponding Hessian score for \tdists. Given this result, we make a reasonable approximation by leveraging the sparsity pattern of the inverse scale matrix to regularize its estimation.

Finegold \etal \cite{finegold2014robust} showed that the first-order condition for the inverse scale matrix of a \tdist shares a similar structure with the inverse covariance matrix of a Gaussian distribution. They propose to regularize the scale matrix estimate at the $\Jidx$th iteration of the EM algorithm by adding a similar $l1$ penalization on the inverse of the scale matrix:
\begin{equation}
\label{eqn:tlasso_scale}
    \spreciz^{\Jidx} = \text{argmax}_{\preciz} \text{logdet} (\preciz) - \trace{\sscalez^{\Jidx} \preciz} - \rho \lVert \preciz \rVert_1, 
\end{equation}
where $\sscalez^{\Jidx}$ is computed from \eqref{eqn:EM_mean_scale}. At each iteration of the EM algorithm, they apply the \glasso to estimate the inverse scale matrix $\spreciz^{\Jidx}$. They call \tlasso: the $l1$ regularized EM algorithm that estimates the mean, scale, inverse scale, and degree of freedom of a \tdist. We provide a pseudo-code for the \tlasso, see Algorithm \ref{algo:tlasso_algo} in Appendix \ref{apx:tlasso}, and refer readers to \cite{finegold2014robust} for further details.

We finish this section by presenting some practical considerations in estimating the degree of freedom and the penalty coefficient $\rho$. Jointly estimating the degree of freedom from \eqref{eqn:ML_dof}  with the mean and scale matrix leads to a poor estimate for the degree of freedom. Instead, we estimate the mean and scale matrix over a range of degrees of freedom. Then, we compute the log-likelihood for each triplet formed by the degree of freedom and the associated mean and scale matrix, and keep the triplet $(\smeanz, \sscalez, \sdofz)$ maximizing the empirical log-likelihood $l( (\meanz, \scalez, \dofz) \given \{ \zi \})$. 
Strictly speaking, the penalty coefficient $\rho$ is an hyper-parameter of the \tlasso that requires tuning through cross-validation. We have found it sufficient to use a simple heuristic based on the correlation between two Gaussian variables for the parameter $\rho$. Specifically, we set $\rho = c /\sqrt{M}$, where $c$ is a positive constant and $M$ is the number of samples. We defer the development of more advanced estimation schemes for these parameters for future work.

% The \tlasso algoritmm proposed by Finegold \etal \cite{finegold2014robust} solves iteratively the coupled equations \eqref{eqn:ML_mean_scale} and \eqref{eqn:ML_dof} for the estimate of the mean, scale matrix, and degree of freedom. 

\subsection{Algorithm of the \enrf}

This section summarizes the analysis step of the \enrf, see Fig. \ref{fig:enrf_algorithm} for a schematic. Algorithm \ref{algo:enrf} in Appendix \ref{apx:enrf} provides a companion pseudo-code. At time $t$, the \enrf takes as input $M$ forecast samples $\{ \iup{\x}_t \} \sim \pdf{\X_t}$ and the realization $\ystar_t$ of the observation variable $\Y_t$. We generate $M$ samples $ \{ \iup{\y}_t \}$ from the likelihood density $\pdf{\Y \given \X}$ by evaluating the observation model \eqref{eqn:observation_model} for each forecast sample $\iup{\x}_t$. Then, we apply \tlasso to estimate the mean $\smean{(\Y_t, \X_t)}$, scale $\sscale{(\Y_t, \X_t)}$, and degree of freedom $\sdof{(\Y_t, \X_t)}$ of $\pdf{\Y, \X_t \given \Y_{1:t-1} = \y_{1:t-1}}$ from the joint samples $\{ (\iup{\y}_t, \iup{\x}_t) \}$. We build the estimator $\stmap_{\sdof{(\Y_t, \X_t)}}$ of the analysis map $\tmapdof$ \eqref{eqn:tmapdof} for $\dof{} = \sdof{(\Y_t, \X_t)}$ by replacing the mean vectors and scale matrices by their empirical estimates. The filtering samples $\{ \iup{\x}_{a,t} \} \sim \pdf{\X_t \given \Y_{1:t} = \y_{1:t}}$ are generated by applying $\stmap_{\sdof{(\Y_t, \X_t)}}$ to the joint pairs of observations and states $(\iup{\y}_t, \iup{\x}_t)$ for the true observation $\ystar_t$. It is important to emphasize that the \enrf algorithm is independent of the particular method employed to estimate the parameters of the joint forecast distribution.

\begin{figure}
    \centering
    \vskip 0.5cm
    \begin{overpic}[width = 0.9 \linewidth]{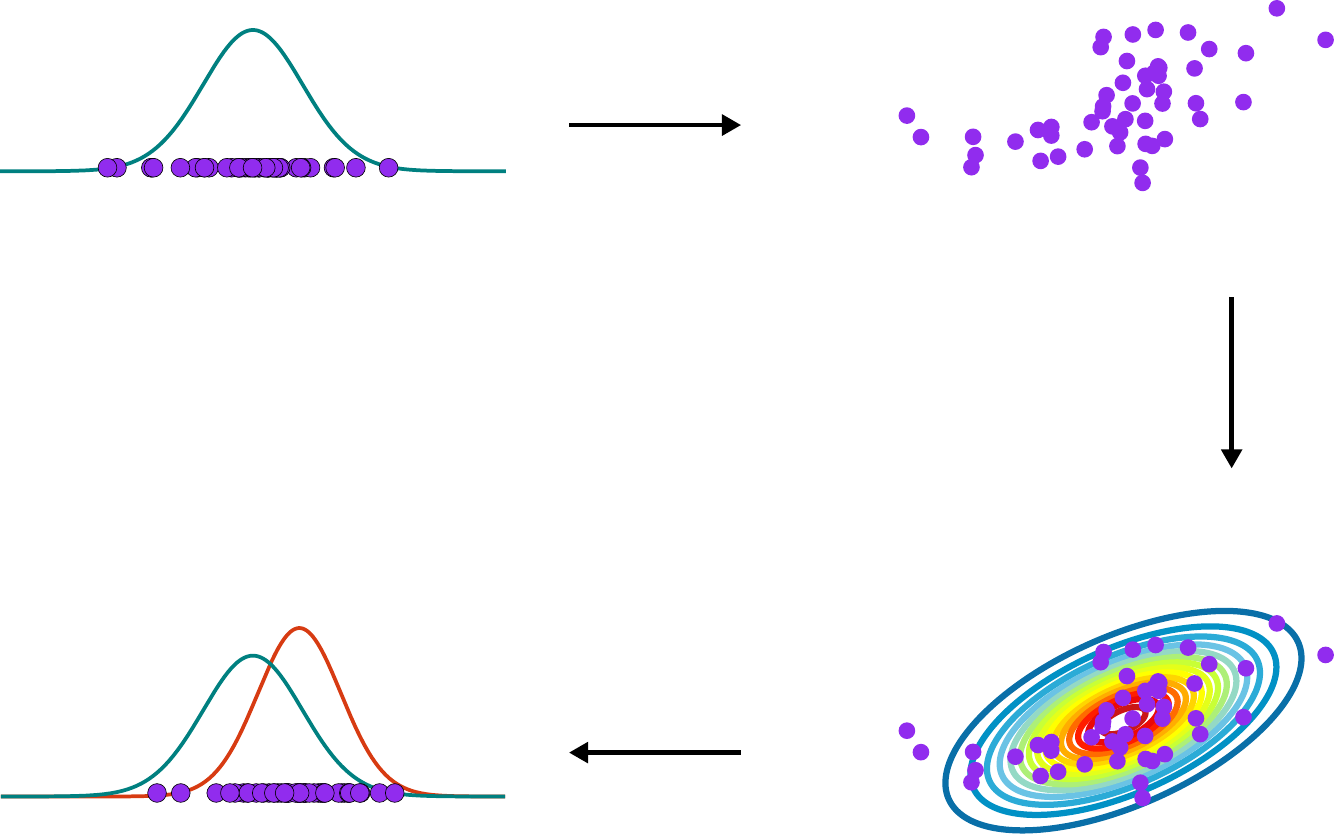}
    % Beginning
    \put(15,  44){$\x^i \sim \pdfprior$}

    % Steps
    % Step 1: Sample the likelihood
    \put(35,62){Sample $\iup{\y}$ from $\pdflik{}(\cdot \given \iup{\x})$:}
    \put(35,57){$\iup{\y} = \obs(\iup{\x}) + \iup{\noiseobs}$, \; eq.~\eqref{eqn:observation_model}}
    
    \put(75, 65){$\{(\iup{\y}, \iup{\x})\} \sim \pdfjoint$}

    % Step 2:
    \put(58, 34){Estimate $\pdfjoint$ from $\{(\iup{\y}, \iup{\x}) \}$}
    \put(58, 30){with \tlasso}

    % Step 3:
    \put(70, 20){Estimated distribution}
    
    \put(38, 15){Estimate and apply $\stmapdof$:}
    \put(38, 10){$\iup{\x}_a = \stmapdof(\iup{\y}, \iup{\x})$}

    % Step 4:
    \put(15, -3){$\iup{\x}_a \sim \pdf{\X \given \Y = \ystar}$}
    \end{overpic}
    \vspace{24pt} 
    \caption{Schematic of the analysis step of the \enrf for a univariate state and observation (\ie $n = 1$ and $d = 1$).}
    \label{fig:enrf_algorithm}
\end{figure}
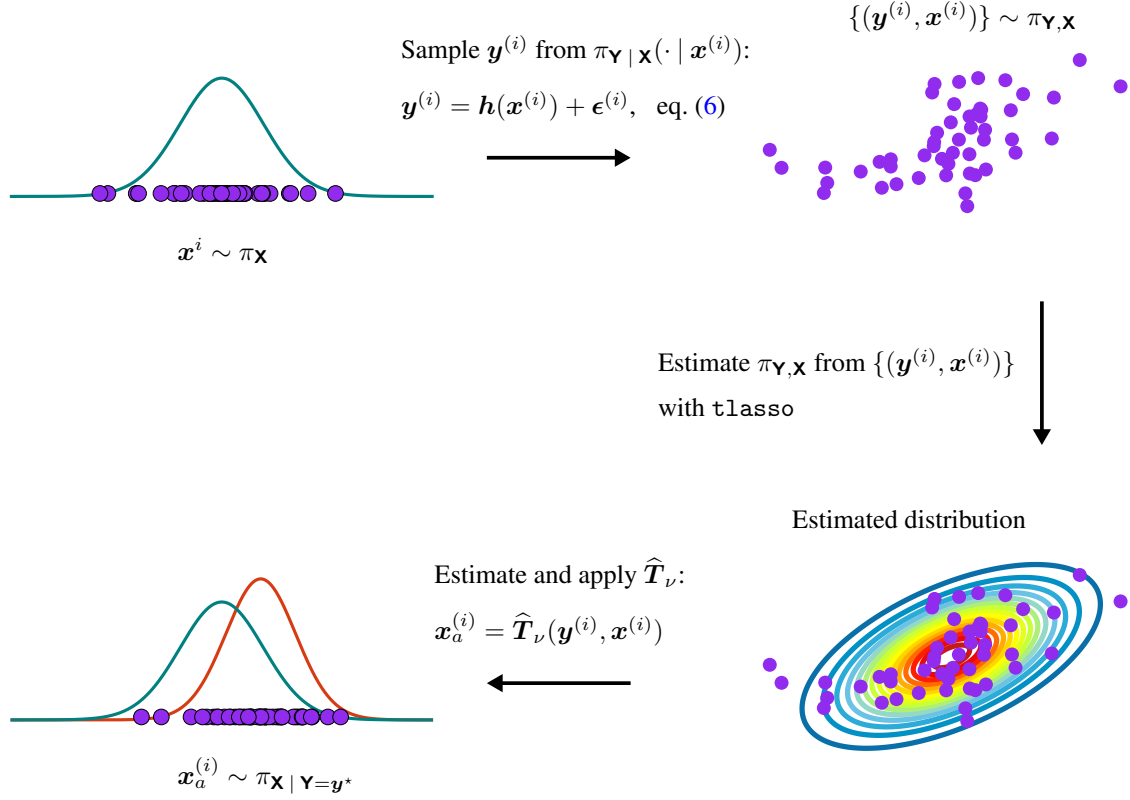

\subsection{Three variants of the \enrf{} algorithm  \label{subsec:variants_enrf}}

% We observe than the estimating the degree of freedom on top of the mean and scale matrix add multiple iterations is more challenging tha
% We observe than the estimation of the degree of freedom renders the 
% In the current state of the art, the estimation of degree of freedom (especially for $\dof{} > 40$) of \tdists from limited samples ($M < 50$) shows 
%  some variability. To alleviate some of these difficulties , we propose three variants of the \enrf{} algorithm: the fixed degree of freedom \enrf (\fixedenrf), the refreshed \enrf (\refreshenrf), and the adaptive \enrf (\adaptenrf). The algorithms differ in the frequency and the samples used to estimate the degrees of freedom. As shown in Fig. \ref{fig:dof_tdist_lorenz63},  

An important practical consideration is the sensitivity of the \enrf algorithm (particularly its tracking performance) to estimation errors of the degree of freedom. In other words, what level of accuracy is required for the empirical degree of freedom to yield meaningful inference? We have observed than estimating the degree of freedom increases the number of iterations of the \tlasso. We explore two different trade-offs to reduce estimation occurrences for the degree of freedom and thus accelerate the \enrf{}. We introduce the three variants of the \enrf{} algorithm: the fixed degree of freedom \enrf (\fixedenrf), the refreshed \enrf (\refreshenrf), and the adaptive \enrf (\adaptenrf). These algorithms differ in the frequency and the samples used to estimate the degrees of freedom.

The \fixedenrf uses a fixed degree of freedom over time. It collects $t_f$ samples of the observation and state variables from a \textit{free-run} (\ie without assimilation of observations) of the state-space model $\{(\y_1, \x_1), \ldots, (\y_{t_f}, \x_{t_f})\}$ for some initial condition $\x_0 \sim \pdf{\X_0}$. These samples are used to estimate the degree of freedom $\hat{\nu}_{\text{fixed}}$ with the \tlasso described in the previous section. At each analysis step, the \fixedenrf estimates the mean and scale matrix of the joint forecast distribution $(\Y_t, \X_t)$ using the \tlasso with the known degree of freedom $\hat{\nu}_{\text{fixed}}$. 

The \adaptenrf estimates the mean, scale matrix, and degree of freedom at every analysis step from the joint forecast samples $\{ (\iup{\y}_t, \iup{\x}_t) \} \sim \pdf{\Y_t, \X_t \given \Y_{1:t-1} = \y_{1:t-1}}$. This is the most demanding version as the estimation of the degree of freedom increases the number of iterations of the \tlasso. 

Finally, the \refreshenrf estimates an initial degree of freedom from a free-run of the state space model. The degree of freedom is then updated every $\Delta t_{refresh}$ (refreshing time step) by using a larger set of $M_{buffer}$ samples of joint forecast samples from the past assimilation cycles. We advise to collect at least $M_{buffer} = 500$ samples. The first estimation of the degree of freedom online occurs when we have collected $M_{buffer}$ joint forecast samples and the assimilation time $t > \Delta t_{refresh}$. We recover the \fixedenrf and the \adaptenrf algorithms by setting $\Delta t_{refresh} = \infty$ and $1$, respectively. The \refreshenrf provides an interesting trade-off between the two previous methods: using a larger ensemble to reduce variance in the empirical degree of freedom, while keeping the flexibility of adapting the degree of freedom as the system evolve in the state space.

% For small ensemble size ($M < 50$), we recommend the \refreshenrf, but we expect improved inference with the \adaptenrf for large ensemble size ($M > 100$). The \fixedenrf is well suited for systems with small variability in the degree of freedom. To make an informed decision, we suggest to estimate the degree of freedom with the \adaptenrf using a large ensemble size. 

\section{Examples \label{sec:examples}}
This section presents the results of our numerical experiments for the stochastic \enkf (\senkf) and the three variants of the \enrf introduced in \ref{subsec:variants_enrf}: the \fixedenrf, the \refreshenrf, and the \adaptenrf. We consider three test cases to assess the filters: the \lo{63} with Gaussian or \tdisted observation noise, and the \lo{96} with \tdisted observation noise. 
The rest of this section is organized as follows. First,  we present the setup of our experiments in \ref{subsec:setup}. Then, we describe the different ensemble filters. Finally, we describe and comment on their performance for these challenging settings of the \lo{63} and \lo{96} systems. 

\subsection{Setup of the numerical experiments \label{subsec:setup}}

The numerical experiments of this paper are performed in the \textit{twin experiment} setting \cite{asch2016data, evensen2009ensemble, spantini_coupling_2022}. We start by sampling the true initial condition $\x^\star_0$ from the initial density $\pdf{\X_0}$  and propagate the state $\x^\star_0$ through the dynamical model \eqref{eqn:dynamical_model} over the time interval $[0, t_f]$ using an adaptive fourth order Runge-Kutta scheme \cite{rackauckas2017differentialequations}. To replicate practical settings, observations are generated with a time step $\Delta t_{\text{obs}}$ larger than a characteristic time scale of the dynamical model. At an observing time $t$ (\ie an integer multiple of $\Delta t_{\text{obs}}$), we collect the true observation $\y_t^\star$ by evaluating the observation model \eqref{eqn:observation_model} for the true state $\x^\star_t$. In this work, we assume that the dynamical observation $\dyn$ and the observation operator $\obs$ are not known. Instead, we can only generate samples from the forecast density $\pdf{\X_t \given \X_{t-1} = \cdot}$ and the likelihood model $\pdf{\Y_t \given \X_t = \cdot}$. We seek to estimate the true state $\x^\star_t$ over time given only the knowledge of the true observations $\{\ystar_t \}$ and the initial state density $\pdf{\X_0}$. In a twin experiment, the dynamical and observation models are identical to generate the truth data and for the filtering. The observation operator will be given by the identity mapping for the \lo{63} system and will select every other component for the \lo{96} system. We set the initial state distribution to the standard Gaussian distribution, \ie $\pdf{\X_0} = \N{\zero{n}}{\id{n}}$. The performance of each filter is assessed with the \textit{root mean square error (RMSE)} and the \textit{spread}. At time $t$, we define the RMSE between the filtering samples $\{ \iup{\x}_{a,t} \}$ and the true state $\x^\star_t$ as $\text{RMSE}_t = || \smean{\X_{a,t}} - \x_t^\star||_2/\sqrt{n}$ where $\smean{\X_{a,t}}$ is the empirical filtering mean at time $t$ and $n$ is the state dimension. We quantify the dispersion of the posterior ensemble with its spread given at time $t$ by $\text{spread}_t = [\trace{ \scov{\X_a}}/n]^{1/2}$, where $\scov{\X_{a,t}}$ is the empirical filtering covariance at time $t$. We compute the RMSE and spread over the last $1000$ time steps of the simulation to remove any influence of the initial ensemble and to ensure the stationarity of the statistics. We optimally tune the multiplicative inflation factor $\alpha$ of the \enkf filters over the range $[0.95, 1.10]$ to obtain the smallest time-averaged RMSE. We report the associated spread of the optimally RMSE-tuned \enkf filters. No multiplicative inflation is applied to the \enrf filters as they embed an adaptive and data-dependent multiplicative inflation mechanism. %The reported RMSE of the \enrf is computed over $20$ realizations of the filter for each ensemble size. 

We assume that we can only sample from the likelihood density via the observation model \eqref{eqn:observation_model}. We have no knowledge of the additive structure of the observation model, nor the underlying distribution for the noise. We perform the inference only given samples from the joint forecast distribution of the states and observations. In particular, we cannot perform sequential assimilation of the observations.

\subsection{Description of the different filters \label{subsec:filters}}

The \enrf embeds three key features: the adaptive and data-dependent multiplicative inflation, the adaptive localization and can account for heavy-tailed distributions. In this study, we assess the performance of different versions of the \senkf and \enrf to isolate the impact of each feature of the \enrf. We use two versions of the \senkf: 
\begin{itemize}
\item The \senkf with optimally tuned multiplicative inflation. For the \lo{96}, we further regularize the \senkf by optimally tuning the distance localization. 
\item The \senkf with optimally tuned multiplicative inflation but we replace distance localization by the \glasso to estimate the covariance matrix of the joint forecast distribution $\pdf{(\Y_t, \X_t)}$ at each assimlation time $t$. We call this filter \senkfglasso in the rest of this paper.
\end{itemize}
By comparing the \senkf with the \senkfglasso, we can assess the benefit of leveraging the conditional independence structure of the problem versus using distance localization. Distance localization regularizes the \senkf by removing all long-range correlations in the empirical covariance of the observations and the empirical cross-covariance of the states and observations. On the other hand, glasso ``discovers'' the conditional independence structure of the joint forecast distribution to regularize the \senkf \cite{spantini_coupling_2022, baptista_probabilistic_2022}. 

We use the following \enrf filters:
\begin{itemize}
\item The \enrf with a large degree of freedom fixed to $\dof{} = 100$. We call this filter \enrfdof in the rest of this paper.
\item The \fixedenrf that uses a fixed freedom estimated from joint samples from a free-run of the state-space model (\ie without data assimilation) with the \tlasso.
\item The \refreshenrf that uses past joint forecast samples to estimate the degree of freedom with a refreshing time $\Delta t_{\text{refresh}}$ set to $20 \Delta t_{\text{obs}}$. We update the degree of freedom of the \refreshenrf every $\Delta t_{\text{refresh}}$ once the buffer reaches $N_{\text{buffer}} = 500$ samples.
\item The \adaptenrf that estimates the degree of freedom at each assimilation cycle from the joint forecast samples. 
\end{itemize}

Comparing the \senkfglasso with the \enrfdof isolates the role of the adaptive and data-dependent multiplicative inflation. 

Additionally, comparing the \fixedenrf, \refreshenrf, and \adaptenrf allows us to determine the accuracy needed for the empirical degree of freedom. Our numerical experiments demonstrate that these three filters yield similar tracking performance. Thus, significantly acceleration of the \enrf by by reducing the frequency of degree of freedom estimations. We stress that the different \enrf filters require no tuning. 

Finally, comparing \enkf filters with \enrf filters helps gauge the significance of considering the tail-heaviness of the joint forecast distribution in designing the prior-to-posterior update. The \rf introduced in this work features a simple and interpretable analysis map that capitalizes on the tail-heaviness of the joint forecast distribution. 

To promote reproducibility, the code for the different filters and the numerical experiments is available at \href{https://github.com/mleprovost/Paper-Ensemble-Robust-Filter.jl}{https://github.com/mleprovost/Paper-Ensemble-Robust-Filter.jl}.

%%%%%%%%%%%%%%%%%%%%%%%%%%%%%%%%%%%%%%%%%%%%%%%%%%%%%%
%%%%%%%%%%%%%%%%%%%%%%%%%%%%%%%%%%%%%%%%%%%%%%%%%%%%%%
%%%%%%%%%%%%%%% Lorenz 63

\subsection{\lo{63} model \label{subsec:lorenz63}}

The  \lo{63}  system is a three-dimensional model for atmospheric convection used as a classical testcase in data assimilation \cite{lorenz1963deterministic, asch2016data}. The state $\x = [x_1, x_2, x_3] \in \real{3}$ is governed by the following set of ordinary differential equations:

\begin{equation}
\label{eqn:lorenz63}
\begin{aligned}
&\frac{\mathrm{d} x_1}{\mathrm{d} t}=\sigma(x_2-x_1),\\
&\frac{\mathrm{d} x_2}{\mathrm{d} t}=x_1(\rho-x_2)-x_2,\\
&\frac{\mathrm{d} x_3}{\mathrm{d} t}=x_1 x_2-\beta x_3,
\end{aligned}
\end{equation}
where we set $\sigma = 10, \beta = 8/3, \rho = 28$ for a chaotic behavior. We integrate \eqref{eqn:lorenz63} over $[0, 200]$ and collect observations every $\Delta t_{\text{obs}} = 0.1$, resulting in $2000$ assimilation cycles. The dynamical operator $\dyn_t$ introduced in \eqref{eqn:dynamical_model} corresponds to the flow generated by \eqref{eqn:lorenz63} over  $\Delta t_{\text{obs}}$. The process noise $\Noisedyn_t$ is Gaussian with zero mean and covariance $\sigma_{\Noisedyn_t}^2\id{3}$ with $\sigma_x^2 = 10^{-4}$. We conduct two experiments for the \lo{63} system, employing either \tdisted observation noise or a Gaussian one.  In the \tdisted setting, the observation noise has zero mean, scale matrix $\scale{\Noiseobs_t} =c_{\Noiseobs_t}^2\id{3}$ with $c_{\Noiseobs_t}^2 = 1.0$, and degree of freedom $\dof{\Noiseobs_t} = 3.0$. In the Gaussian setting, the observation noise has zero mean, covariance matrix $\cov{\Noiseobs_t} =\sigma_{\Noiseobs_t}^2\id{3}$ with $\sigma_{\Noiseobs_t}^2 = 4.0$. We set the $l1$ penalty coefficient in the \tlasso to $0.5/\sqrt{M}$ for the different \enrf filters.

\begin{figure}
    \centering
    \includegraphics[width = 0.45\linewidth]{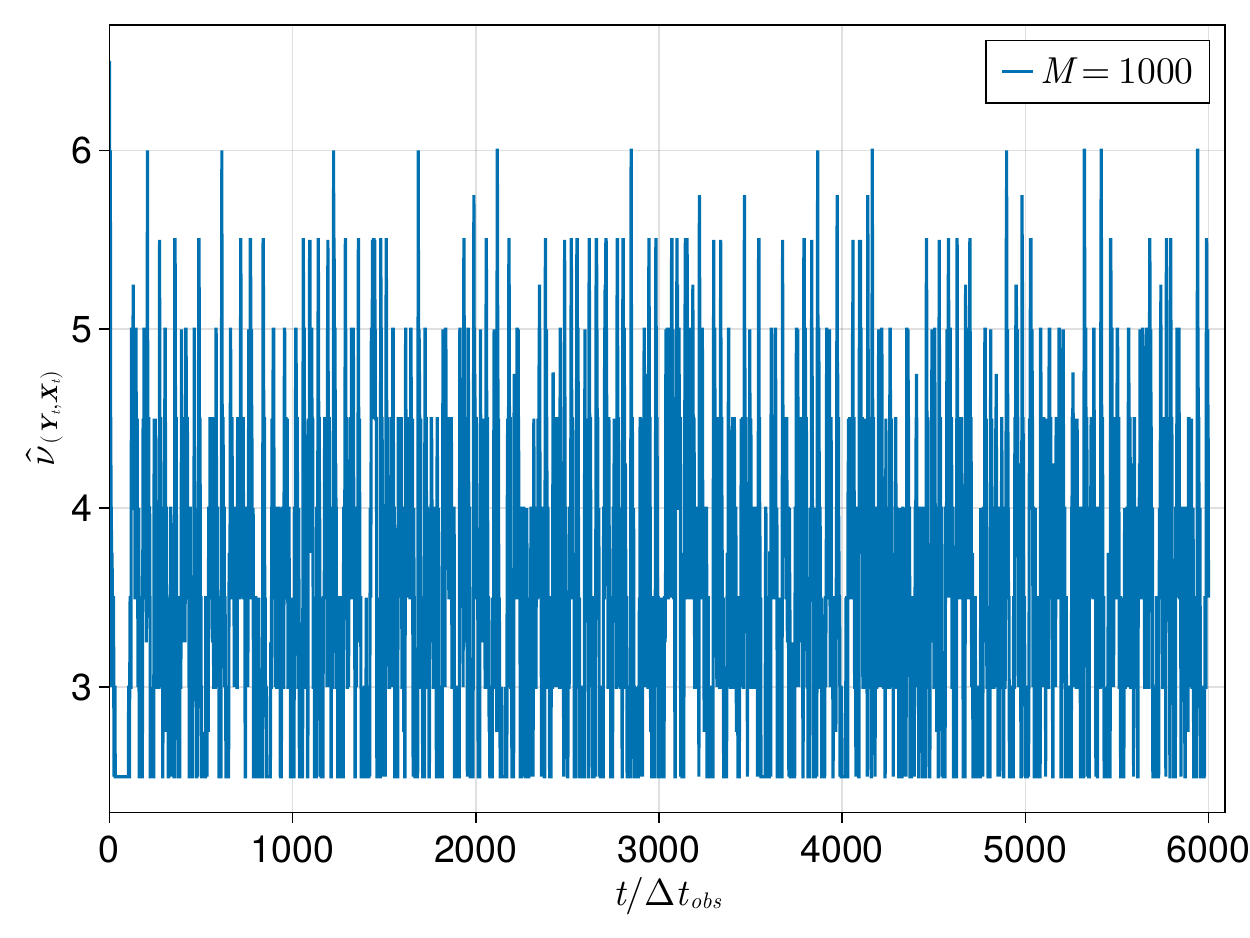}
    \includegraphics[width = 0.45\linewidth]{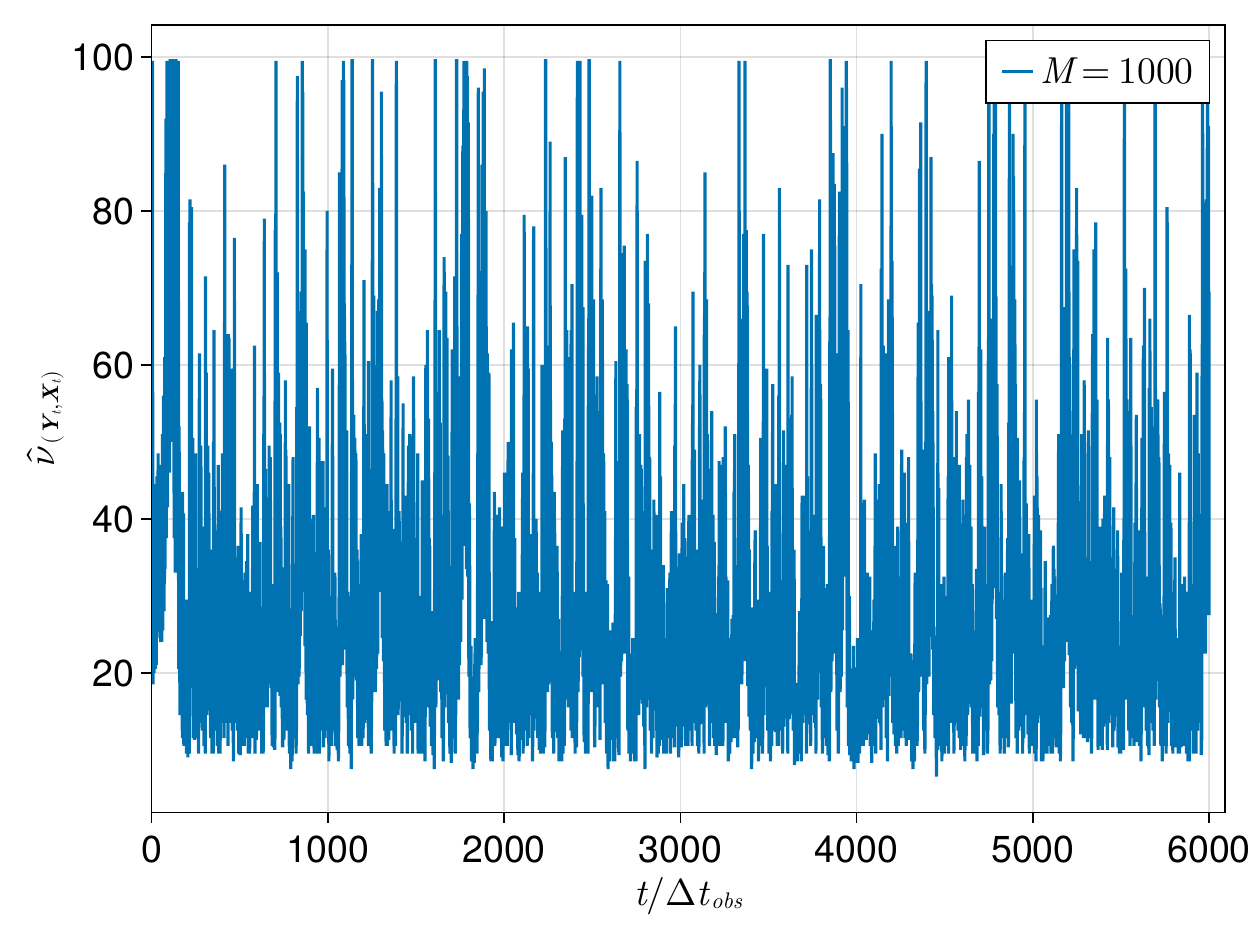}
    \caption{Temporal median of the empirical degree of freedom $\sdof{(\Y_t, \X_t)}$ (computed over $50$ realizations) from $M = 1000$ samples of the joint forecast density $\{(\iup{\y}_t, \iup{\x}_t)\} \sim \pdf{(\Y_t, \X_t) \given \Y_{1:t-1}}$ generated by the \adaptenrf for the \lo{63} problem with \tdisted(left) and Gaussian (right) observation noise.}
    \label{fig:dof_lorenz63}
\end{figure}

Fig .\ref{fig:dof_lorenz63} shows the evolution of the estimated degree of freedom $\dof{(\Y_t, \X_t)}$
of the joint forecast distribution $(\Y_t, \X_t)$ over the time interval $[0, 200]$ in the two settings. We use the adaptive \enrf algorithm with $M = 1000$ samples to estimate  $\dof{(\Y_t, \X_t)}$, choosing a large ensemble size to reduce sampling errors. The results are averaged over $50$ realizations for each case. In the \tdisted case, the estimated degree of freedom is empirically stationary with a median value $5.1$ for $M = 1000$. This estimate confirms the tail-heaviness of the joint forecast distribution.  In the Gaussian case, the degree of freedom shows larger variations with median value $28.9$ and $5\%$ and $95\%$ quantiles $18.7$ and $54.5$. Considering that a distribution with empirical degree of freedom larger than approximately $\sim 50$ exhibits empirically Gaussian tails, we conclude that the joint forecast distribution has slightly heavier tails than a Gaussian distribution. 

It is important to stress that the boundedness of the degree of freedom is connected to the nonlinearity of the dynamical operator. Let us consider an assimilation cycle with a linear-\tdisted state-space model. In the forecast step, the application of the linear dynamical model conserves the degree of freedom. In the analysis step, the conditioning of the state on the new observation increases the degree of freedom of the state by the dimension of the observation \eqref{eqn:conditional_tdist}. Owing to this unbounded arithmetic growth,  the filtering distribution for a \tdisted linear state-space model has lighter tails at each assimilation cycle and converges to a Gaussian distribution ($\dof{} \to \infty$). The application of a nonlinear dynamical model in the forecast step will typically decrease the degree of freedom due to its expansive nature. For sufficiently nonlinear systems,  we conjecture a balance in the degree of freedom over one assimilation cycle resulting from the antagonist behavior of the forecast and analysis steps. We leave for future work to connect the Lyapunov exponents of the dynamical operator with the degree of freedom of the forecast density. 

%%%%%%%%%%%%%%%%%%%%%%%%%%%%%%%%%%%%%%%%%%%%%%%%%%%%%%%
%%%%%%%%%%%%%% Lorenz-63 t-distributed %%%%%%%%%%%%%%%%
%%%%%%%%%%%%%%%%%%%%%%%%%%%%%%%%%%%%%%%%%%%%%%%%%%%%%%%

\subsubsection{Setting with \tdisted observation noise}

\begin{figure}
    \centering  
    \textbf{\small{\lo{63} with \tdisted observation noise}}\\
    \vskip 0.3cm
    \includegraphics[width = 0.45\linewidth]{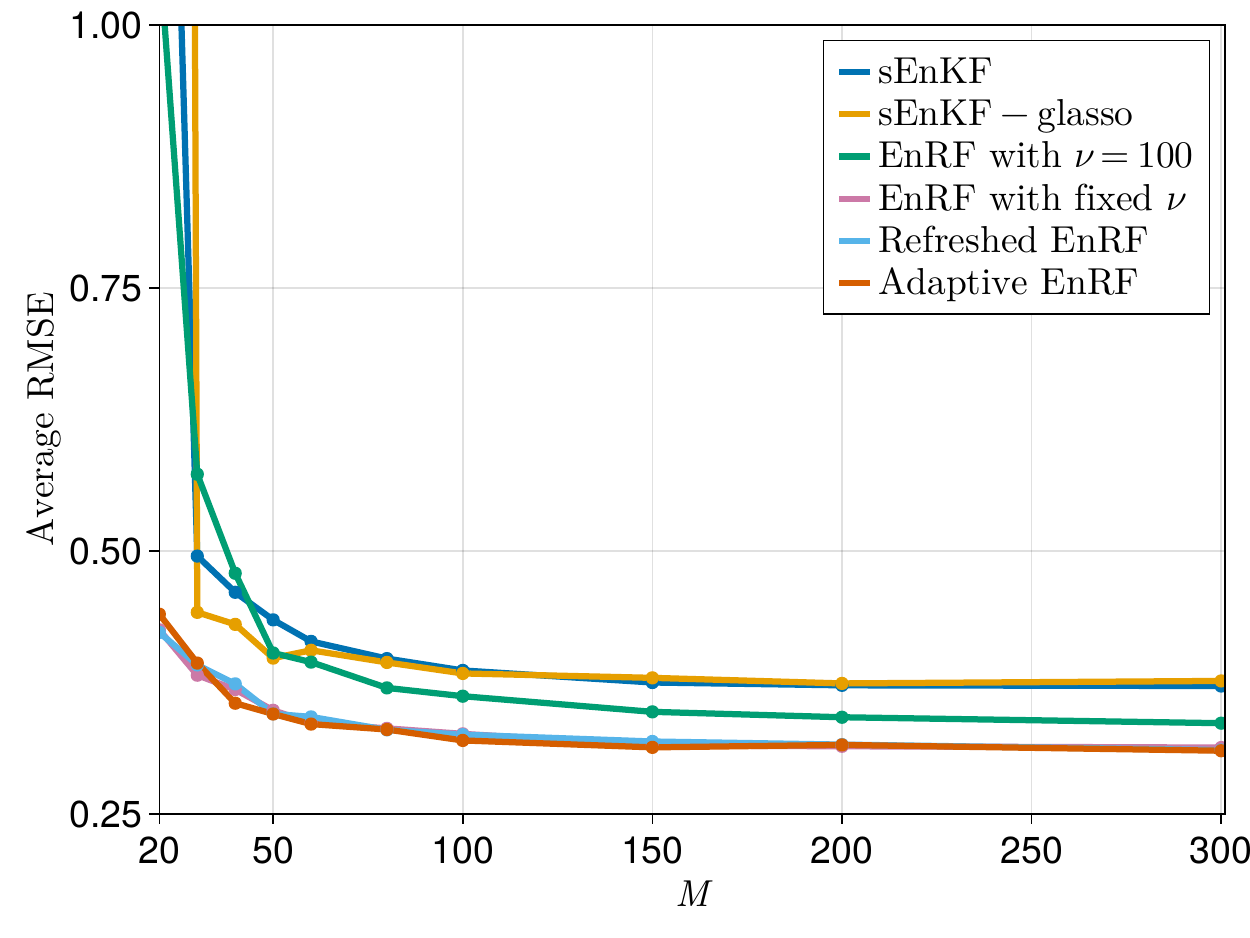}
    \includegraphics[width = 0.45\linewidth]{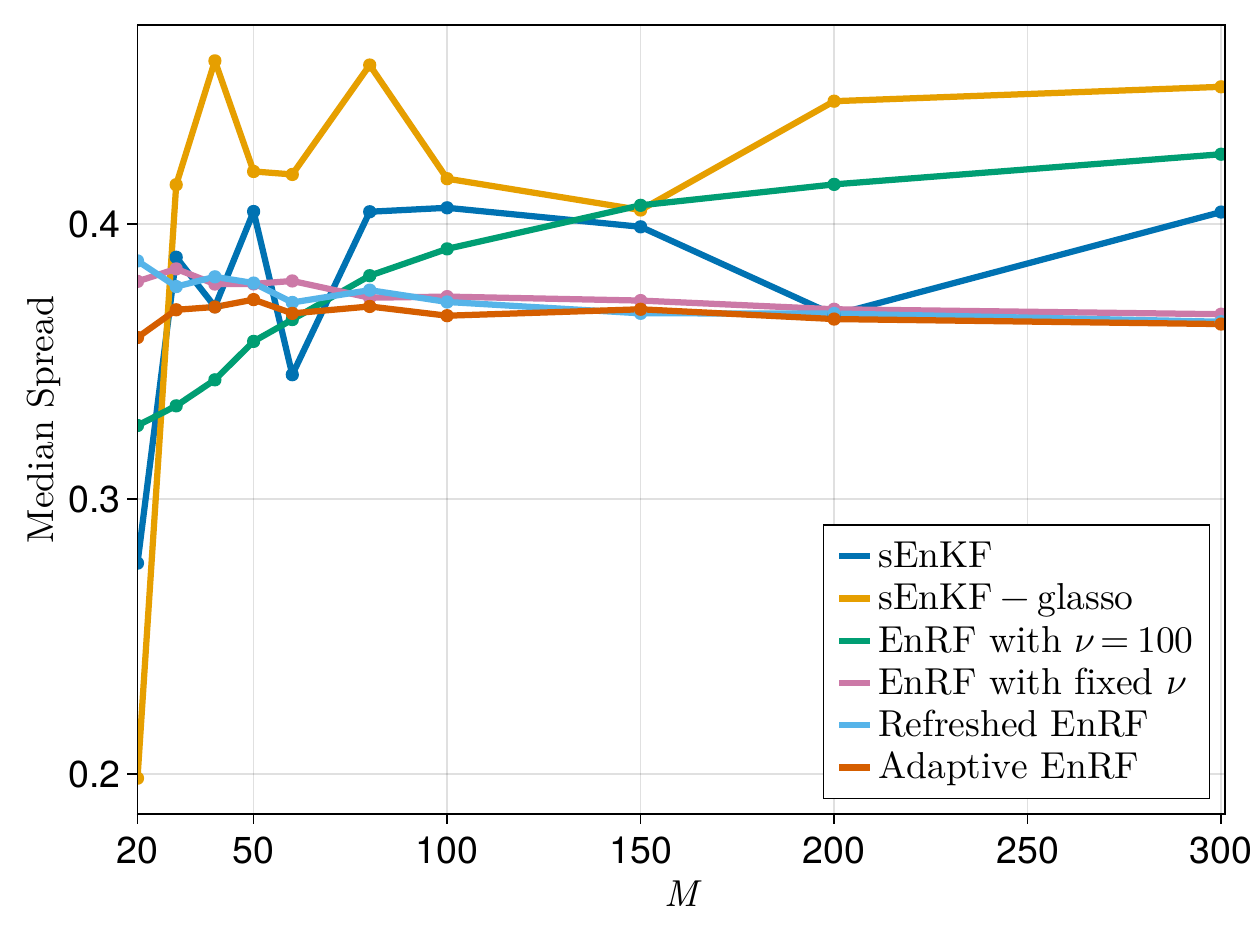}
    \caption{Left panel: Time-averaged evolution of the RMSE with the ensemble size $M$ for the \lo{63} model with \tdisted observation noise using the \senkf  with optimally tuned inflation and localization (blue), the \senkf with optimally tuned inflation and \glasso (orange), the \enrf with fixed $\dof{} = 100$ (green), the \fixedenrf (pink), the \refreshenrf (sky blue) and the \adaptenrf (red). There is no inflation applied to the \enrf filters. Right panel: Median evolution of the spread with the ensemble size $M$ for the same filters.}
    \label{fig:tdist_lorenz63}
\end{figure}

Fig. \ref{fig:tdist_lorenz63} shows the time-averaged RMSE and median spread of the filters for the \lo{63} system with \tdisted noise. We optimally tuned the multiplicative inflation of the \senkf and \senkfglasso, while the \enrf filters require no tuning. We observe that the \senkf and \senkfglasso are unstable for small $M <20$. We recall that the observations to assimilate are given all at once. Additionally, the filters have no prior knowledge of the additive structure of the observation model, and cannot sample from the observation noise. 

We start by commenting on the RMSE performance of the filters. We observe that the \senkf and \senkfglasso have similar performance . The \enrfdof performs asymptotically better than the \senkf filters. We observe that the \fixedenrf, \refreshenrf, and \adaptenrf perform similarly over the range of ensemble size. They perform better than the previous filters and remain stable for small $M$. For $M = 20$, their average RMSE are $0.45, 0.46, 0.52$. For $M > 150$, their tracking error plateau at $0.32, 0.33, 0.33$  corresponding to an improvement of $27\%$ of the RMSE with respect to the optimally tuned \senkf. We attribute the improved performance of the \enrf filters to the adaptive multiplicative inflation mechanism and the intrinsic account for the tail-heaviness of the joint forecast distribution. From this experiment, these two factors contribute equally to the overall performance improvement of the \enrf filter. 
%In this example, the observation operator is the identity. Thus, the \senkfglasso and the \enrf filters further leverage the local likelihood structure \cite{spantini_coupling_2022}. 

The right panel of Fig. \ref{fig:tdist_lorenz63} shows the median spread of the filters. The \fixedenrf, \refreshenrf, and \adaptenrf exhibit minimal spread variation, plateauing at $0.37$ across the range of ensemble sizes. Based on the similarity in RMSE and spread among these three \enrf filters, we conclude that a scattered estimation of the degree of freedom is sufficient. For small ensemble size, the \senkf and the \senkfglasso  have the lowest median spread. For larger ensemble size, they have larger spread than the \enrf filters. We observe that the spread of the \senkfglasso and \enrfdof increase with the ensemble size. We recall the that we report the median spread of the \senkf and \senkfglasso, optimally tuned based on their averaged RMSE. For small ensemble size, the filtering ensembles generated 
by the \senkf and \senkfglasso have large tracking error but small spread. These filters suffer from over-confidence in the forecast distribution, which is caused by the high sensitivity of light-tailed estimators to outlying samples.

%%%%%%%%%%%%%%%%%%%%%%%%%%%%%%%%%%%%%%%%%%%%%%%%%%%%%%%
%%%%%%%%%%%%%%%% Lorenz-63 Gaussian  %%%%%%%%%%%%%%%%%%
%%%%%%%%%%%%%%%%%%%%%%%%%%%%%%%%%%%%%%%%%%%%%%%%%%%%%%%

\subsubsection{Setting with Gaussian observation noise}

\begin{figure}
    \centering
    \textbf{\small{\lo{63} with Gaussian observation noise}}\\
    \vskip 0.3cm
    \includegraphics[width = 0.45\linewidth]{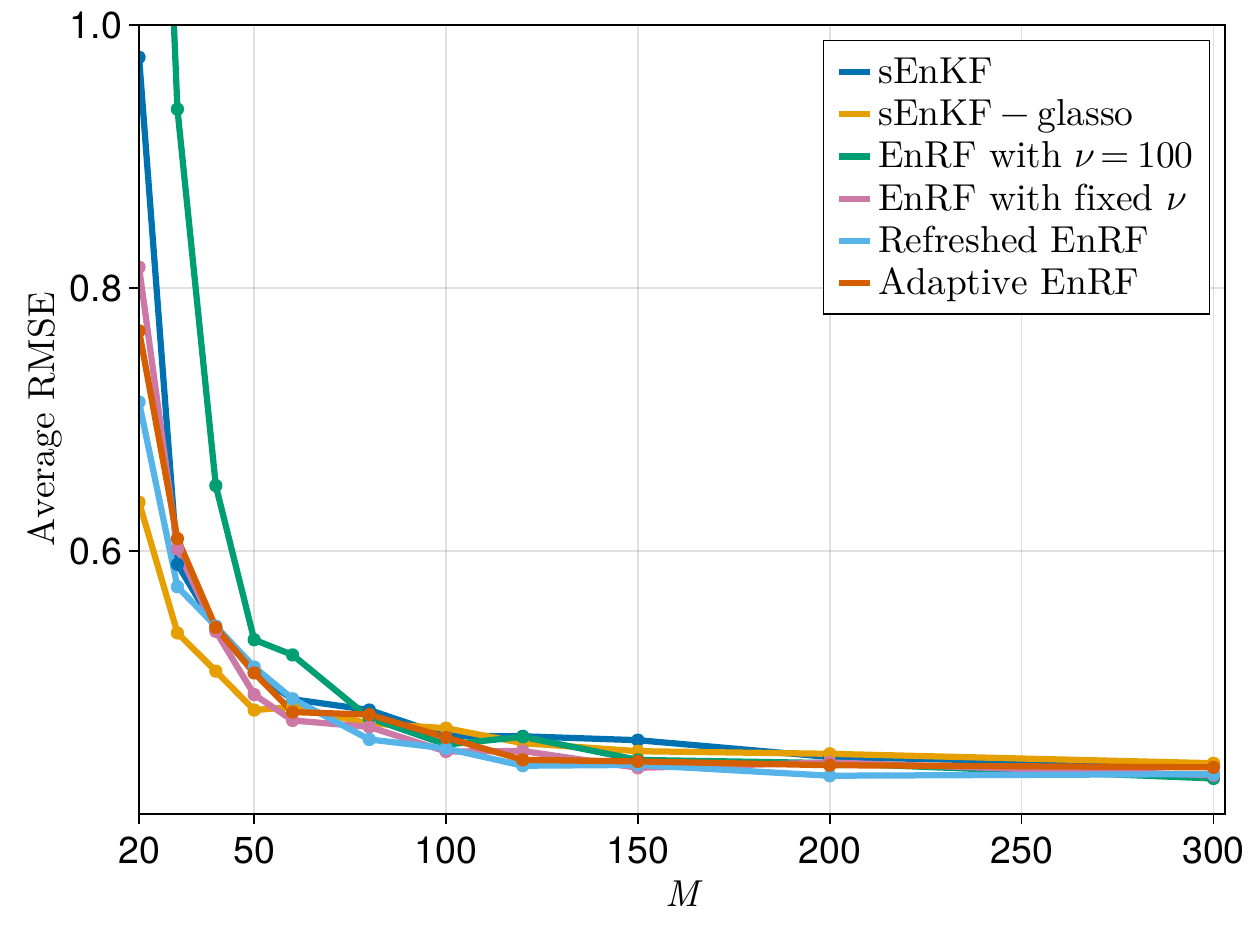}
    \includegraphics[width = 0.45\linewidth]{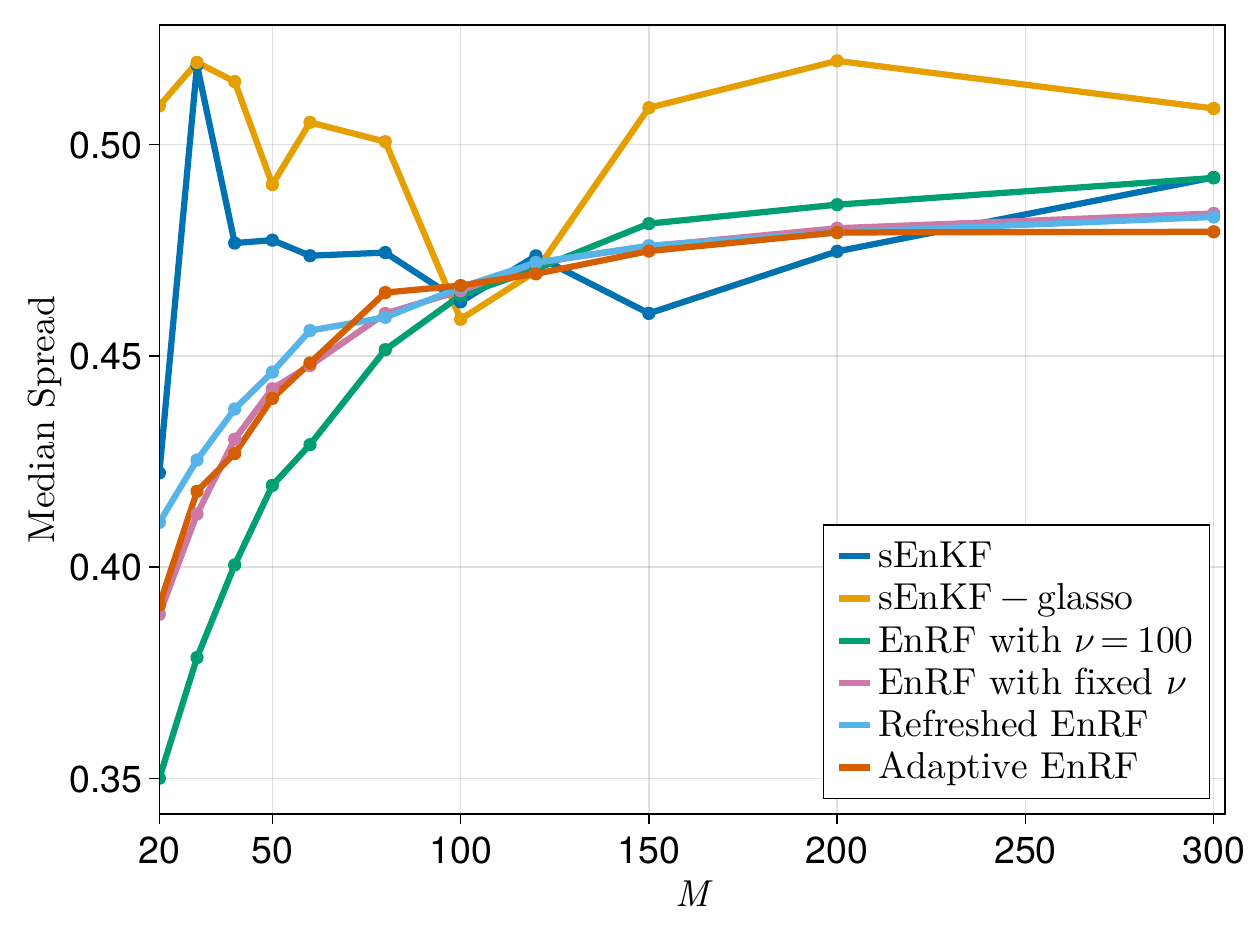}
    \caption{Left panel: Evolution of the temporal median spread with the ensemble size $M$ for the \lo{63} model with Gaussian observation noise using the \senkf  with optimally tuned inflation and localization (blue), the \senkf with optimally tuned inflation and \glasso (orange), the \enrf with fixed $\dof{} = 100$ (green), the \fixedenrf (pink), the \refreshenrf (sky blue) and the \adaptenrf (red). There is no inflation applied to the \enrf filters. Right panel: Median evolution of the spread with the ensemble size $M$ for the same filters.}
    \label{fig:gaussian_lorenz63}
\end{figure}

This section presents the results for the \lo{63} setting with Gaussian observation noise, see Fig. \ref{fig:gaussian_lorenz63}. The time evolution of the empirical degree of freedom, see right panel of \ref{fig:dof_lorenz63}, suggests that the tails of the joint forecast distribution are slightly heavier than Gaussian tails. In this setting, we anticipate similar performance for the \enrf filters and the light-tailed filters, \ie \senkf and \senkfglasso. 

We first comment on the RMSE of these filters, see left panel of Fig. \ref{fig:gaussian_lorenz63}. For large ensemble size, the different filters have similar tracking error. For small ensemble size, the filters can be ranked based on their RMSE performance, from worst to best, as follows: the \enrfdof, the \senkf, the \fixedenrf, the \adaptenrf, the \refreshenrf, and finally the \senkfglasso. We observe that the \enrfdof  is unstable for small ensemble size. Even in this low-dimensional problem ($n = 3$ and $d = 3$), leveraging the conditional independence structure is beneficial. For large ensemble size ($M > 80$), the filters perform similarly.

The right panel of Fig. \ref{fig:gaussian_lorenz63} presents the median spread of the filters. For small ensemble size, the \enrfdof is over-confident in the forecast distribution, resulting in a small spread of the posterior ensemble but a large tracking error. We observe that the spread of the \enrf filters increases with the ensemble size.

%%%%%%%%%%%%%%%%%%%%%%%%%%%%%%%%%%%%%%%%%%%%%%%%%%%
%%%%%%%%%%%%%%%%%%% Lorenz 96 %%%%%%%%%%%%%%%%%%%%%
%%%%%%%%%%%%%%%%%%%%%%%%%%%%%%%%%%%%%%%%%%%%%%%%%%%

\subsection{\lo{96} \label{subsec:lorenz96}}

The \lo{96} model is a popular testbed in data assimilation \cite{lorenz1996predictability, asch2016data}. It is derived from first principles as a one-dimensional model for the response of the mid-latitude atmosphere to forcing input. The $n$-dimensional state $\x = [x_1, \ldots, x_{n}]^\top$ at time $t$ is governed by the following set of ordinary differential equations: 
\begin{equation}
\label{eqn:lorenz96}
\frac{\mathrm{d} x_i}{\mathrm{d} t} = (x_{i+1} - x_{i-2}) x_{i-1} - x_{i} + F, \quad \mbox{for } i = 1, \ldots, n,
\end{equation}
where we enforce periodic boundary conditions and set the forcing to $F = 8.0$. In this experiment, we use $n = 20$. We observe the state with $\Delta t_{obs}=0.4$, every two components ($x_1, x_3, x_5, \ldots, x_{19})$. We set the $l1$ penalty coefficient to $0.5/\sqrt{M}$ for the different \enrf filters. 

\begin{figure}
    \centering
    \includegraphics[width = 0.5\linewidth]{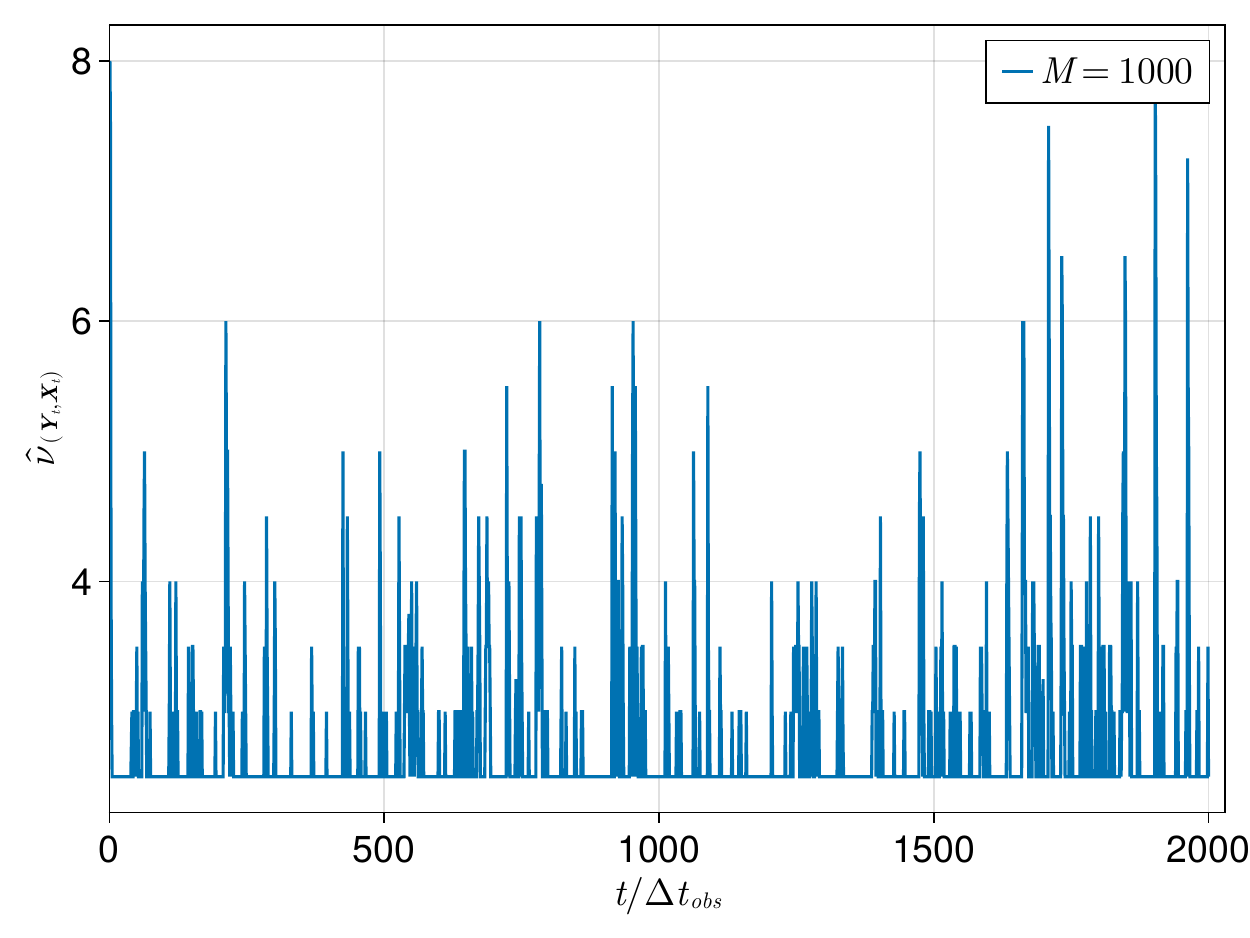}
    \caption{Temporal median of the empirical degree of freedom $\sdof{(\Y_t, \X_t)}$ (computed over $50$ realizations) for $M = 1000$ (blue) samples of the joint forecast density $\{(\iup{\y}_t, \iup{\x}_t)\} \sim \pdf{(\Y_t, \X_t) \given \Y_{1:t-1}}$ generated by the \adaptenrf for the \lo{96} problem with \tdisted observation noise.}
    \label{fig:dof_tdist_lorenz96}
\end{figure}

Fig. \ref{fig:dof_tdist_lorenz96} shows the empirical degree of freedom of the joint forecast distribution $\pdf{(\Y_t, \X_t) \given \Y_{1:t-1}}$ using the estimation procedure described in \ref{subsec:lorenz63}. The joint forecast distribution exhibits heavy tails with a median degree of freedom of $3.6$.

\begin{figure}
    \centering   
    \textbf{\small{\lo{96} with \tdisted observation noise}}\\
    \includegraphics[width = 0.45\linewidth]{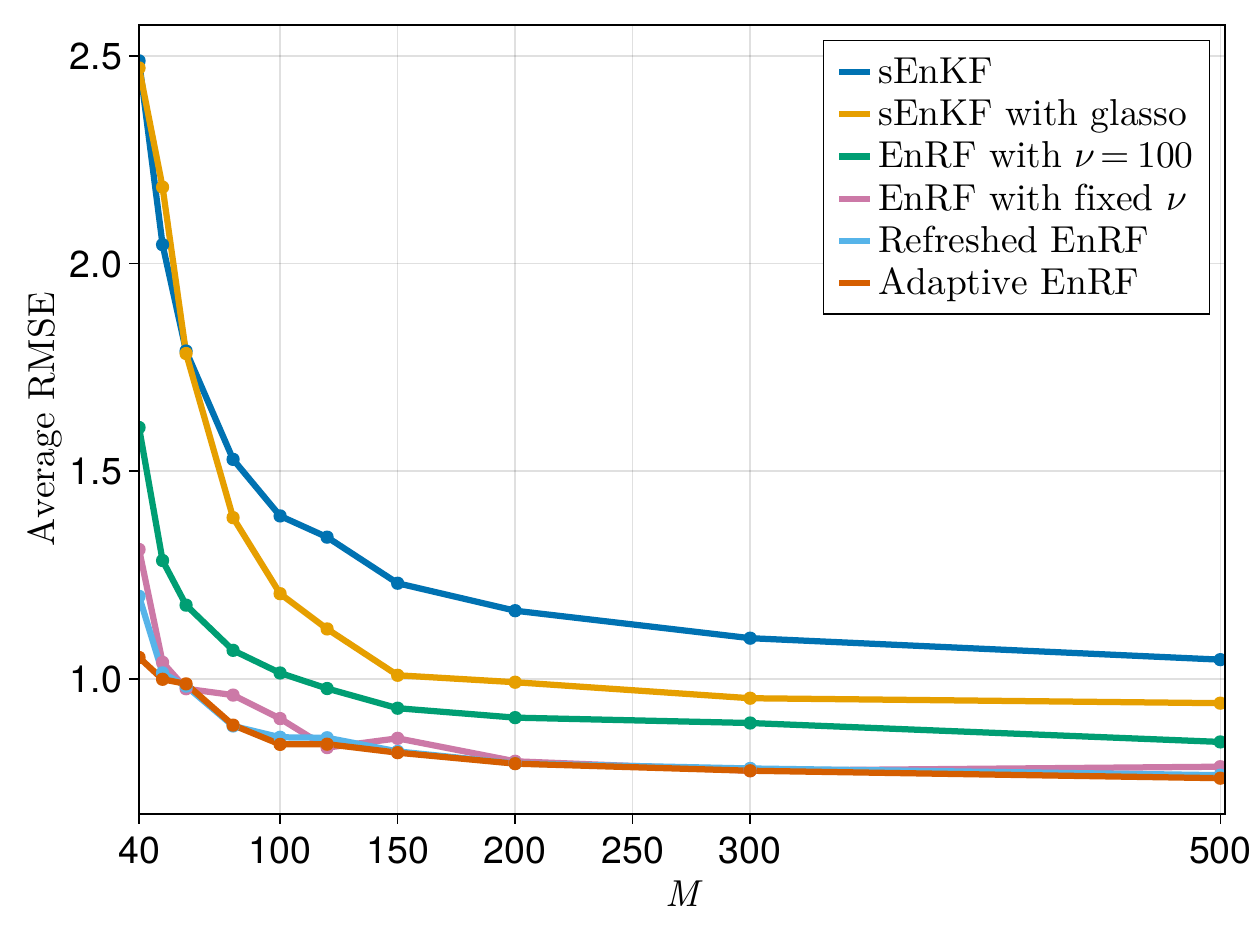}
    \includegraphics[width = 0.45\linewidth]{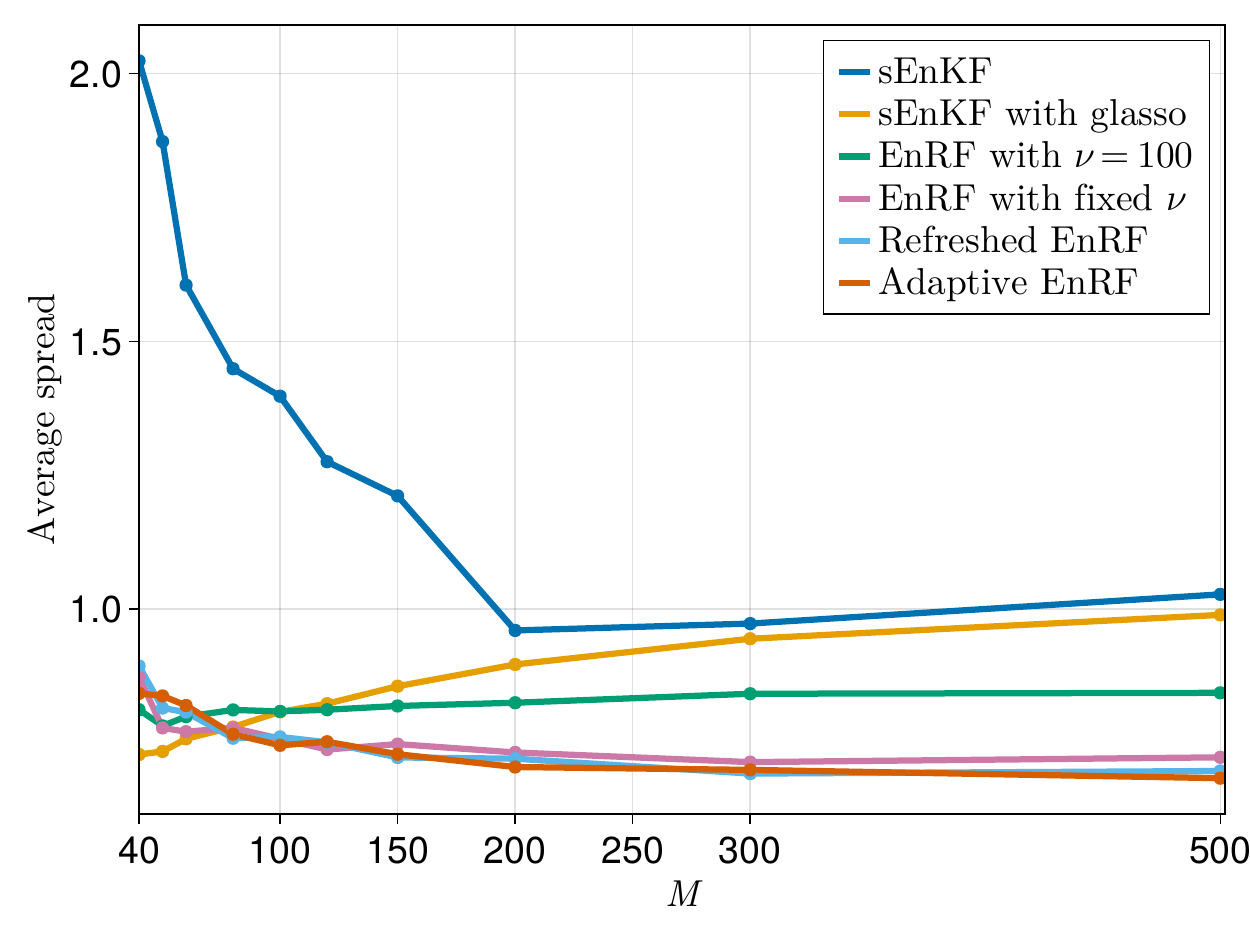}
    \caption{Left panel: Time-averaged evolution of the RMSE with the ensemble size $M$ for the \lo{96} model with \tdisted observation noise with $\dof{} = 3.0$  using the \senkf  with optimally tuned inflation and localization (blue), the \senkf with optimally tuned inflation and \glasso (orange), the \enrf with fixed $\dof{} = 100$ (green), the \fixedenrf (pink), the \refreshenrf (sky blue) and the \adaptenrf (red). There is no inflation applied to the \enrf filters. There is no inflation applied to the \enrf filters. Right panel: Median evolution of the spread with the ensemble size $M$ for the same filters.} 
    \label{fig:rmse_spread_tdist_lorenz96}
\end{figure}

The time-averaged RMSE and time-median spread for the \enkf and \enrf filters are reported on Fig. \ref{fig:rmse_spread_tdist_lorenz96}. The filters can be ranked based on their RMSE performance, from worst to best, as follows: \senkf, \senkfglasso, \enrfdof, \fixedenrf, \refreshenrf, and \adaptenrf. Overall, the \enrf filters perform better than the \senkf filters. The \enrf embeds three key features: the adaptive and data-dependent multiplicative inflation, the adaptive localization and can account for heavy-tailed distributions. 

By comparing the \senkf with the \senkfglasso, we  assess the benefit of leveraging the conditional independence structure of the problem versus using distance localization. For $M < 80$, the \senkf and \senkfglasso have similar RMSE. For larger ensemble size $M$, the \senkfglasso performs better than the optimally tuned \senkf. For $M = 500$, the RMSE of the \senkf and \senkfglasso is $1.05, 0.94$, respectively. Thus, discovering the conditional independence structure can provide better tracking performance than imposing a sparsity pattern on the covariance matrix as distance localization. 

The \senkfglasso and the \enrfdof both leverage the conditional independence structure to sparsify the inverse covariance matrix or the inverse scale matrix. The \enrfdof uses an adaptive and data-dependent multiplicative inflation instead of the fixed multiplicative inflation of the \senkfglasso. The benefits are twofold: the \enrfdof  requires no tuning and better regularizes the \enrf by leveraging the \maha distance of the realization $\ystar$ of the observation variable and the synthetic observation $\iup{\y}$. 
% However, the \enrfdof is not a linear analysis map, and retain the prior-to-posterior update based on the ratio of scaling $\sqrt{\scaling{\Y}{\ystar}/\scaling{\Y}{\y}} = (100 + \deltaform{\Y}{\ystar})/(100 + \deltaform{\Y}{\y})$.  

The \fixedenrf, \refreshenrf, and \adaptenrf have similar RMSE over the range of ensemble size considered with a slightly larger RMSE for the \fixedenrf and the \refreshenrf for $M = 40$. This finding confirms previous results, demonstrating that a scattered approximation of the degree of freedom is adequate for the \enrf. This appealing feature allows for a reduction in the computational cost of the \tlasso by minimizing the occurrences of degree of freedom estimation. For instance, the \fixedenrf estimates the degree of freedom only once in an offline manner. This estimation is carried out using samples from a \textit{free-run} (i.e., without assimilation of observations) of the state-space model ${(\y_1, \x_1), \ldots, (\y_{t_f}, \x_{t_f})}$ over a characteristic time window of the system.

For $M = 500$, we obtain the following RMSE for the \senkf, \senkfglasso, \enrfdof, \fixedenrf, \refreshenrf, and \adaptenrf: $1.05, 0.94, 0.85, 0.79, 0.77, 0.76$. The \enrf filters reduce the RMSE tracking error of the optimally tuned \senkf (inflation and localization) by $27\%$ for $M = 500$. 

Finally, we comment of the spread of the different filters reported in the right panel of Fig. \ref{fig:rmse_spread_tdist_lorenz96}. The \senkf provides an posterior ensemble with the largest spread: the spread is $2.0$ for $M = 40$ and plateaus to $0.96$  for $M > 200$. The spread of the \senkfglasso increases with the ensemble size from $0.73$ for $M = 40$ to $0.99$ for $M = 500$. The spread of \enrf filters slightly decrease over the range of ensemble size from $0.87$ for $M = 40$ to $0.70$ for $M = 500$.  It is interesting to compare these results with our analytical expectations from \ref{subsubsec:connection_kf}.  In \ref{subsubsec:connection_kf}, we show that the posterior scale matrix  of the \rf was scaled by a factor $\scaling{\Y}{\ystar}$ that balances the degree of the freedom and the \maha of the realization $\ystar$ of the observation variable. We show that over expectation this factor is strictly larger than one. One would expect the spread of the \enrf filters to be larger than the spread of the \senkf filters. However, the \senkf relies on empirical estimators based on a $l2$ loss function and thus sensitive to outliers. The spread of the \enrfdof plateaus at $0.82$. Overall, the different \enrf filters provide an improvement of $30\%$ in spread over the optimally tuned \senkf for $M =500$.

\section{Conclusion \label{sec:conclusion}}

In this work, we leveraged tools from measure transport and the properties of \tdists to introduce a new filter tailored for heavy-tailed distributions. We showed that an ensemble filter algorithm is determined (up to the empirical evaluation of its transport map) by a choice of a reference density and a class of functions to estimate $\smap^{\Xup}(\y, \x)$ or $\tmap(\y, \x)$. For instance, the \enkf uses a standard Gaussian reference density and a class of linear transformations for the analysis map $\tmapkf$. The stochastic map filter (\smf) \cite{spantini_coupling_2022} retains a standard Gaussian reference density, but uses sparse and interpretable nonlinear transformations to estimate $\smap^{\Xup}$. These algorithms, among many others, assume at least that the tails of the joint forecast distribution are Gaussian, and thus cannot perform consistent inference for heavy-tailed filtering problems. To improve inference, we proposed a new ensemble filter based on a linear estimator for $\smap^{\Xup}$ and replaced the standard Gaussian reference density by a \tdisted reference density. By tuning the degree of freedom $\dof{}$ of a \tdist, one can model asymptotic tail behavior of densities ranging from algebraic decay (e.g., Cauchy distribution for $\dof{} = 1$) to exponential decay (e.g., Gaussian distribution for $\dof{} = \infty$). Our strategy allowed us to retain the tractability of linear transformations while having the ability to adapt the prior-to-posterior transformation to the tail-heaviness of the joint forecast distribution. The two-step derivation of this new filter exploited properties of \tdists and tools from measure transport presented in Spantini \etal \cite{spantini_coupling_2022}. First, we derived the \kr that pushes forward a joint \tdist with degree of freedom $\dof{}$ to a well chosen product of independent standard \tdists parameterized by $\dof{}$. Then, we constructed the prior-to-posterior transformation/analysis map $\tmapdof$ for a joint \tdist for the observations and states with degree of freedom $\dof{}$. Remarkably, we obtained an interpretable and closed form expression for the resulting transport map $\tmapdof$. To the best of our knowledge, this is the first analytical expression for a \kr and analysis map based on a non-Gaussian reference density.

The map $\tmapdof$ has many appealing features. First, the analysis map $\tmapdof$ shows reduced sensitivity to outlying synthetic observations generated by the likelihood model. For a joint sample $(\iup{\y}, \iup{\x})$ with an outlying synthetic observation, \ie $\scaling{\Y}{\iup{\y}}/\scaling{\Y}{\iup{\y}} \to 0$, $\tmapdof$ discards the synthetic observation $\iup{\y}$ and pushes forward the state $\iup{\x}$ to the posterior mean of the \tdist. Second, $\tmapdof$ reverts to the analysis map of the Kalman filter $\tmapkf = \tmapdof$ for $\dof{} = \infty$. More importantly, the $\tmapdof$ corresponds to an asymptotic expansion about $\tmapdof$ for large degree of freedom $\dof{}$ and thus corrects the Kalman filter's update to account for deviations from Gaussian tails. Interestingly,  the scale matrix of a posterior \tdist  $\pdf{\X \given \Y = \ystar}$ contains a scaling factor $\scaling{\Y}{\ystar}$ that depends linearly on the \maha of $\ystar$. We interpreted this scaling in the posterior scale matrix as an intrinsic  adaptive and data-dependent multiplicative inflation. Our experiments showed that the \enrf filters outperformed the \enkf filters without multiplicative inflation. 

An important challenge with heavy-tailed distributions is to build estimators for their statistics --- and \textit{in fine} for the analysis map --- with reduced sensitivity to outlying samples. In this work, we used a regularized expectation-maximization (EM) algorithm to estimate the mean, scale matrix, and degree of freedom of the joint forecast distribution from samples \cite{liu1995ml, liu1997ml}. These estimators weight the importance of each sample on their \maha such that outlying samples get down-weighted. In the low-data regime, this metric is no longer informative of the ``radial position'' of the samples, and more importantly, the \enrf no longer accounts for the \maha of the $\ystar$ and $\y$ in the update. We regularized the empirical scale matrix by leveraging the conditional independence structure of the joint forecast distribution. We applied the \tlasso algorithm developed in \cite{finegold2014robust} that sparsifies the empirical scale matrix at each iteration of the EM algorithm with a $l1$ penalization of the inverse scale matrix, reminiscent of the \glasso algorithm \cite{friedman2008sparse}. Finally, we note that the algorithm of the \enrf is agnostic to the algorithm used to estimate the statistics of the joint forecast distribution. One can easily replace the \tlasso in the \enrf with a more competitive algorithm to estimate \tdists from samples.

We showed in challenging settings of the \lo{63} and \lo{96} systems, that the \enrf outperformed the \senkf without any tuning. Indeed, the \enrf embeds an adaptive and data-dependent multiplicative inflation as well as an adaptive localization arising from the $l1$ regularization of the inverse scale matrix in the \tlasso. This constitutes a major advantage over existing ensemble filters that require extensive manual parameter tuning. 

We conclude this manuscript by presenting directions for future work. By analogy with state-of-the-art formulations of the \enkf, one can develop a deterministic version of the \enrf based on a square-root factorization of the posterior scale matrix \eqref{eqn:tmapdof_stats} , see \cite{asch2016data, evensen2009ensemble}. We expect this deterministic \enrf to reduce sampling errors from the realizations of the observation noise. 

In many cases, the statistical dependence between the observations and states decay quickly with the distance. In this work, we used the approximate conditional independence structure of the joint forecast distribution to regularize the scale matrix. For non-local functions of the state, given by integral of linear or nonlinear functions of the state, we expect meaningful statistical dependence between the observations and states at long distances. Inverse problems given by non-local observation models usually exhibit the following low-rank structure: low-dimensional projections of the observations strongly inform a low-dimensional subspace of the state space \cite{spantini2015optimal, baptista2022gradient}. Le Provost \etal \cite{leprovost2022low} propose a low-rank factorization of the Kalman gain leveraging the low-rank structure of the inverse problem for nonlinear observation models. Using this factorization, they successfully apply the \senkf to non-local observation models given by elliptic partial differential equations. The analysis step reduces to the estimation of a low-dimensional Kalman gain in the informative subspace of the state and observation spaces. One could derive a similar factorization of $\tmapdof$ and regularize the \enrf by estimating a low-dimensional analysis map $\stmapdof$ in these projected coordinates. 

Finally, a natural extension of this work is to consider nonlinear approximations for   $\smap^{\Xup}$ to capture non-Gaussian behaviors in the bulk of the distribution such as skewness or multi-modality \cite{spantini_coupling_2022, baptista2021learning} combined with an adaptive \tdisted reference to capture the tail behavior. More broadly, this work paves the way for new ensemble filters that tailor features of the joint forecast distribution --- skewness, multi-modal, tail-heaviness --- to improve the inference.

\textbf{Data Accessibility :} All the computational results are reproducible and code is available at: \url{https://github.com/mleprovost/Paper-Ensemble-Robust-Filter.jl.git}

\textbf{Authors’ Contributions:} All authors contributed to the calculations, analysis, and writing and revision of the manuscript. The final version of the manuscript was approved by all authors.

\textbf{Funding:} MLP and YM acknowledge support of the National Science Fundation (award PHY-2028125). RSB gratefully acknowledges support from MURI US Department of Energy AEOLUS center (award DE-SC0019303), the Air Force Office of Scientific Research MURI on “Machine Learning and Physics-Based Modeling and Sim- ulation” (award FA9550-20-1-0358), and a Department of Defense (DoD) Vannevar Bush Faculty Fellowship (award N00014-22-1-2790).

\textbf{Acknowledgements:} The authors would like to thank Nan Chen and  Mathias Drton for insightful discussions.
% MLP and JE gratefully acknowledge support of the Air Force Office of Scientific Research Grant No. FA9550-18-1-0440. RB and YM acknowledge support from the Department of Energy, Office of Advanced Scientific Computing Research, AEOLUS (Advances in Experimental design, Optimal control, and Learning for Uncertain complex Systems) center.

\section*{\begin{center}Appendix \end{center}}
\begin{appendices}

% \section{Derivation of the analysis map \(\tmapdof\) \label{apx:tmapdof}}
\section{Derivation of the analysis map for a joint \tdist \label{apx:tmapdof}}

In this section, we provide further details on the derivation of the analysis map $\tmapdof$ in \eqref{eqn:tmapdof}. We consider a pair of jointly \tdisted random variables $(\Y, \X)$  with degree of freedom $\dof{}$, where $ \Y \in \real{d}$ and $\X \in \real{n}$, \ie
\begin{equation}
    \stack{\Y}{\X} \sim \St{\stack{\meanx}{\mean{\Y}}}{
    \begin{bmatrix}
    \scale{\Y} & \scale{\X, \Y}^\top \\
    \scale{\X, \Y} &  \scalex
    \end{bmatrix}}{\dof{}}.
\end{equation}
We also define the random variable $\Z = [\Z_1; \Z_2] \in \real{d + n}$ with $\Z_1 \in \real{d}$ and  $\Z_2 \in \real{n}$, following a $d + n$ standard \tdist $\reftdist{\nu}{d + n}$ with degree of freedom $\dof{}$:
\begin{equation}
    \Z = \stack{\Z_1}{\Z_2} \sim \reftdist{\nu}{d + n} = \St{\zero{d + n}}{\begin{bmatrix}\id{d} &  \zero{}\\
    \zero{} & \id{n}
    \end{bmatrix}}{\dof{}}.
\end{equation}
As mentioned in Sec. \ref{sec:tdist}, the standard \tdist $\pdf{\Z} = \pdf{\Z_1, \Z_2}$ cannot be factorized as the product of its marginals. To alleviate this, we define the rescaled variable $\tilde{\Z} = [\tilde{\Z}_1, \tilde{\Z}_2] = [\Z_1, \Z_2/\sqrt{\scaling{\Z_1}{\z_1}}]$, such that $\pdf{\tilde{\Z}}$ can be factorized as the products of its marginals, \ie $\pdf{\tilde{\Z}}(\tilde{\z}) = \pdf{\tilde{\Z}_1}(\tilde{\z}_1) \pdf{\tilde{\Z}_2}(\tilde{\z}_2)$ with $\pdf{\tilde{\Z}_1} =\pdf{\Z_1} = \reftdist{\nu}{d}$ and $\pdf{\tilde{\Z}_2} = \reftdist{\nu + d}{n}$.  

Then, we construct the \kr $\widetilde{\smap}_{\dof{}}$ that pushes forward $\pdfjoint$ to the reference density $\pdf{\tilde{\Z}}$. To do so, we introduce the Cholesky factors $\lmap_{\Y} \in \real{d \times d}, \smap_{\X \given \Y} \in \real{n \times n}, \lmap_{\X \given \Y = \y} \in \real{n \times n}$ of the inverse observation scale matrix $\scale{\Y}^{-1}$, the inverse of the Schur complement matrix $\left(\schurxy \right)$, and the inverse posterior scale matrix $\scale{\X \given \Y = \y}^{-1}$, respectively:
\begin{equation}
\label{eqn:apx_cholesky}
    \begin{aligned}
        & \scale{\Y}^{-1}   =\lmap_{\Y}^\top \lmap_{\Y}, \\
        & \schurxy  = \smap_{\X \given \Y}^\top \smap_{\X \given \Y}, \\
        & \scale{\X \given \Y = \y}^{-1}  = \lmap_{\X \given \Y = \y}^\top \lmap_{\X \given \Y = \y}.
    \end{aligned}
\end{equation}

Using properties of \tdists presented in Sec. \ref{sec:tdist}, we obtain the \kr  $\widetilde{\smap}_{\dof{}}$ that pushes forward $\pdfjoint$ to the reference density $\pdf{\tilde{\Z}}$:
\begin{equation}
\widetilde{\smap}_{\dof{}}(\y, \x)= \left[\begin{array}{c}
\begin{aligned}
& \widetilde{\smap}_{\dof{}}^{\Yup}(\y) \\
& \widetilde{\smap}_{\dof{}}^{\Xup}(\y, \x)
\end{aligned}
\end{array}\right] = \left[\begin{array}{c}
\begin{aligned}
& \lmap_{\Y}(\y - \mean{\Y}) \\
& \lmap_{\X | \Y = \y} \left[ \left(\x - \meanx \right) - \scale{\X, \Y} \scale{\Y}^{-1}(\y -  \mean{\Y})\right]
\end{aligned}
\end{array}\right].
\end{equation}

Then, we define the analysis map $\tmapdof : \real{d} \times \real{n} \to \real{n}$ that pushes forward the \tdist $\pdf{\X}$ to the conditional \tdist $\pdf{\X \given \Y = \y}$, by partial inversion of  $\widetilde{\smap}_{\dof{}}^{\Xup}$:  
\begin{equation}
\begin{aligned}
        \tmapdof(\y, \x)  & = {\widetilde{\smap}}_{\dof{}}^{\Xup}(\ystar, \cdot)^{-1} \circ {\widetilde{\smap}}_{\dof{}}^{\Xup}(\y, \x)\\
        & =   \meanx + \scale{\X, \Y} \scale{\Y}^{-1} (\ystar - \mean{\Y}) + \BB{L}_{\X | \ystar}^{-1} \BB{L}_{\X | \y} \left[ \left(\x - \meanx \right) - \scale{\X, \Y} \scale{\Y}^{-1}(\y -  \mean{\Y})\right]
\end{aligned}
\end{equation}
To derive the formula for $\tmapdof$ in eq. \eqref{eqn:tmapdof}, we show that $\BB{L}_{\X | \ystar}^{-1} \BB{L}_{\X | \y} = \sqrt{\scaling{\Y}{\ystar}/\scaling{\Y}{\y}} \id{n}$. From the Cholesky factorization of $\scale{\X \given \Y  = \y}^{-1}$ in \eqref{eqn:apx_cholesky}, we have:
\begin{equation}
\begin{aligned}
    \scale{\X \given \Y  = \y}^{-1} & = \lmap_{\X \given \Y = \y}^\top \lmap_{\X \given \Y = \y}\\
    & = \frac{1}{\scaling{\Y}{\y}} \left( \schurxy \right)^{-1} \\
    & = \frac{1}{\sqrt{\scaling{\Y}{\y}}^2} \smap_{\X \given \Y}^\top \smap_{\X \given \Y} \\
    & = \left( \frac{1}{\sqrt{\scaling{\Y}{\y}}}   \smap_{\X \given \Y}\right)^\top  \left( \frac{1}{\sqrt{\scaling{\Y}{\y}}} \smap_{\X \given \Y} \right).
\end{aligned}
\end{equation}
By uniqueness of the Cholesky factorization of a positive definite matrix (here $\scale{\X \given \Y  = \y}^{-1}$), we conclude that:
\begin{equation}
    \lmap_{\X \given \Y = \y} = \frac{1}{\sqrt{\scaling{\Y}{\y}}}  \smap_{\X \given \Y},
\end{equation}
where $\smap_{\X \given \Y}$ does not depend on the realization $\y$ that we condition on. This allows us to conclude that $\BB{L}_{\X | \ystar}^{-1} \BB{L}_{\X | \y} = \sqrt{\scaling{\Y}{\ystar}/\scaling{\Y}{\y}} \id{n}$, and to derive the expression of eq. \eqref{eqn:tmapdof} for $\tmapdof$.

\section{\tlasso algorithm for \tdists \label{apx:tlasso}}

Algorithm \ref{algo:tlasso_algo} in Appendix~\ref{apx:tlasso} presents pseudo-code of the \tlasso algorithm developed by Finegold and Drton \cite{finegold2014robust} to estimate the mean, scale matrix, inverse scale matrix, and degree of freedom of a \tdist from samples. 

\begin{algorithm}
\DontPrintSemicolon
    \KwInput{$M$ samples $\{\zi\}$ from $\pdf{\Z}$, $\rho$ the penalty coefficient on the $l1$ norm of the off-diagonal entries of the precision matrix, stopping criterion \code{stopping\_rule} and a Boolean \code{is\_dof\_estimated}.}
    \KwOutput{Empirical mean $\smeanz \in \real{m}$, scale matrix $\sscalez \in \real{m \times m}$, precision matrix (inverse of the scale matrix) $\spreciz = \sscalez^{-1} \in \real{m \times m}$, and degree of freedom $\sdofz > 2$ of $\pdf{\Z}$.}
    \vskip 0.2cm
    % \tcc{Tikhonov regularization parameter $\lambda$ for scale matrices}
    % $\lambda = 10^{-8}$\\
    \tcc{Initial estimates for $\meanz, \scalez, \dofz$}
    
    $\smeanz^\zeroidx = \frac{1}{M} \sum_{i = 1}^M \zi,$\\
    
    $\sscalez^\zeroidx = \frac{1}{M-1} \sum_{i = 1}^{M}(\zi - \smeanz^0)(\zi - \smeanz^0)^\top,$\\
    
    $\sdofz^\zeroidx = \dof{0}$

    % \tcc{Form the initial empirical scale matrix $\sscalez^\zeroidx$:}
    % $\sscalez^\zeroidx = \anomz^\zeroidx (\anomz^\zeroidx)^\top + \lambda \id{m}$\\

    % Iteration loop
    \For{$\Jidx= \zeroidx:\mathsf{N}_{\text{iter}}$}{
    
    \tcc{E-Step: Compute the following factors obtained by conditional expectations:}

    \tcc{Extract the Cholesky factor $\lmap_{\Z}^\Jidx \in \real{m \times m}$ of $\sscalez^\Jidx = \lmap_{\Z}^\Jidx {\lmap_{\Z}^\Jidx}^\top$}
    $\lmap_{\Z}^\Jidx  = \code{cholesky}(\sscalez^\Jidx , \code{'lower'})$
    
    \For{$i= 1:M$}{

    \tcc{Compute the \maha of the samples $\{ \zi \}$:}

    % delta equations
    $\delta^{(i),\Jidx} = \qform{\zi - \smeanz^{\Jidx}}{\sscalez^{\Jidx, -1}} = || {\lmap^{\Jidx}_{\Z}} \backslash (\zi - \smeanz^{\Jidx}) ||^2 $

    % tau equations
    $\tau^{(i), \Jidx + 1} = (\sdofz^\Jidx +  m)/(\sdofz^\Jidx +  \delta^{(i),\Jidx})$

    }

    \tcc{M-Step: Update the estimators for the mean, scale, and precision $\smeanz^{\Jidx + 1}, \sscalez^{\Jidx + 1}, \spreciz^{\Jidx +1}$:}
    % Update the mean 
    $\smeanz^{\Jidx + 1} =(\sum_{i = 1}^M \tau^{(i), \Jidx + 1} \zi) /(\sum_{i = 1}^M \tau^{(i), \Jidx + 1} )$\\
    
    \tcc{Compute $\stauz^{\Jidx + 1}$:}
    $\stauz^{\Jidx + 1} = \frac{1}{M-1}\sum_{i = 1}^M \tau^{(i), \Jidx + 1} (\zi - \smeanz^{\Jidx +1 })(\zi - \smeanz^{\Jidx +1 })^\top$\\

    \tcc{Solve graphical lasso problem for  $\spreciz^{\Jidx +1}$, \ie
    $\frac{m}{2} \logdet \spreciz^{\Jidx +1} - \frac{m}{2}\trace{\spreciz^{\Jidx +1}\stauz^{\Jidx + 1}} - \rho ||\spreciz^{\Jidx +1}||_{1,\text{off}}$:}

    $\sscalez^{\Jidx + 1}, \spreciz^{\Jidx + 1} = \code{glasso}(\spreciz^{\Jidx +1}, \rho)$\\

    \tcc{If $\code{is\_dof\_estimated}$ is true,  update  $\sdofz$:}

    $\sdofz^{\Jidx +1} = \text{zero of } \dofz \mapsto \sum_{i=1}^M [ \log(\frac{\dofz}{2}) + 1 - \psi(\frac{\dofz}{2}) + \psi(\frac{\sdofz^{\Jidx} + m }{2}) -\log \left(\frac{\sdofz^{\Jidx} + \delta^{(i), \Jidx}}{2} \right) - \frac{\sdofz^{\Jidx} + m}{\sdofz^{\Jidx} + \delta^{(i), \Jidx}}]$\\

    \vspace{8pt}
    \tcc{Check stopping criterion \code{stopping\_rule}:}
    \If{ \code{stopping\_rule} == \code{true}}{
    \code{break}
    }
    
    }

    return $\smeanz, \sscalez, \spreciz, \sdofz$
\caption{\tlasso($\{\zi \}$, $\rho$, \code{stopping\_rule}, \code{is\_dof\_estimated}) estimates the empirical mean $\smeanz \in \real{m}$, scale matrix $\sscalez \in \real{m \times m}$, precision matrix $\spreciz \in \real{m \times m}$, and degree of freedom $\sdofz > 2$ of the \tdist $\pdf{\Z} = \Stz$ from $M$ samples $\{\z^{(1)}, \ldots, \z^{(M)}\} \sim ~\pdf{\Z}$ with the \code{tlasso} algorithm \cite{finegold2014robust}. The scale and precision matrices are estimated with a graphical lasso procedure \cite{friedman2008sparse}. The degree of freedom $\sdofz$ is only estimated if the Boolean \code{is\_dof\_estimated} is \code{true}. The \tlasso routine keeps iterating until the \code{stopping\_rule} is \code{true}.}
\label{algo:tlasso_algo}
\end{algorithm}

\section{Algorithm for the analysis step of the ensemble robust filter \label{apx:enrf}}
%%%%%%%%%%%%%%%%%%%%%%%%%%%%%%%%%%%%%%%%%%%%%%%%%%%%%%%%%%%%%%%%%%%%%%%%%
%%%%%%%%%%%%%%%%%%%%%%%%%%%% Likelihood model %%%%%%%%%%%%%%%%%%%%%%%%%%%

Algorithm \ref{algo:enrf} presents pseudo-code for one analysis step of the \enrf{}. This algorithm transforms a set of forecast samples to filtering samples by assimilating the realization $\ystar$ of the observation variable.

\begin{algorithm}
\DontPrintSemicolon
    \KwInput{$\ystar \in \real{d}$, likelihood model  $\pdf{\Y \given \X = \cdot}$, $M$ samples $\{\x^i\}$ from $\pdfprior$}
    \KwOutput{$M$ samples $\{\iup{\x_a}\}$ from $\pdfpost(\cdot \given \ystar)$}
    \vskip 0.2cm
    
    \tcc{Generate likelihood samples $\{\iup{\y} \}$}
    \For{$= 1:M$}{
    $\iup{\y} \sim \pdf{\Y \given \X = \iup{\x}}$}

    \tcc{Use the \tlasso procedure to estimate $\stack{\mean{\Y}}{\meanx} \in \real{d + n}, \stack{\scale{\Y} \; \scale{\X, \Y}^\top }{\scale{\X,\Y} \; \scale{\X}} \in \real{(d + n) \times (d + n)}, \dof{\Y, \X}$ from the joint samples $\biggl\{\stack{\iup{\y}}{\iup{\x}} \biggr\}$. $\rho$ refers to the penalty coefficient on the $l1$ norm of the off-diagonal entries of the inverse of $\stack{\scale{\Y} \; \scale{\X, \Y}^\top }{\scale{\X,\Y} \; \scale{\X}}$. }
    $\stack{\mean{\Y}}{\meanx}, \stack{\scale{\Y} \; \scale{\X, \Y}^\top }{\scale{\X,\Y} \; \scale{\X}}, \dof{\Y, \X} = 
    \tlasso \left(\left\{ \stack{\iup{\y}}{\iup{\x}} \right\}, \rho \right)$\\
    
    % \tcc{Assemble the scale of $\Y$}
    % $\scale{\Y} = \anom{\Y} {\anom{\Y}}^\top + \lambda \id{d}$\\

    \tcc{Extract the Cholesky factor $\lmap_{\Y} \in \real{m \times m}$ of $\scale{\Y} = \lmap_{\Y} \lmap_{\Y}^\top$:}
    $\lmap_{\Y} = \code{cholesky}(\scale{\Y}, \code{'lower'})$

    \tcc{Solve the $M$ linear systems for $\{ \iup{\z} \} \in \real{d}$ and compute  $\{ \scaling{\Z}{\iup{\z}} \} \in \real{+}$:}
    \For{$i= 1:M$}{
    $\lmap_{\Y} \iup{\z} =  (\iup{\y} - \mean{\Y})$\\
    $\scaling{\Z}{\iup{\z}} = (\dof{\Y, \X} +  ||\iup{\z}||^2)/(\dof{\Y, \X} + d)$
    }
  \tcc{Solve the linear system for $\z^\star \in  \real{d}$ and compute $\scaling{\Z}{\z^\star} \in \real{+}$:}
    $\lmap_{\Y} \z^{\star} =  (\ystar - \mean{\Y})$\\
    $\scaling{\Z}{\z^{\star}} = (\dof{\Y, \X} + ||\z^{\star}||^2)/(\dof{\Y, \X} + d)$\\
    \tcc{Solve the linear system for the representers $\{ \iup{\BB{b}} \} \in \real{d}$ using the Cholesky factor $\lmap_{\Y}$:}
    \For{i=1:M}{
    $\lmap_{\Y}^\top \iup{\BB{b}} = \iup{\z}$
    }
    
    \tcc{Solve for the linear system for the representer $\BB{b}^\star \in \real{d}$ using the Cholesky factor $\lmap_{\Y}$:}
    $\lmap_{\Y}^\top \BB{b}^\star = \z^\star$\\
    \tcc{Generate the posterior samples}
        \For{i=1:M}{
    $\iup{\x_a} = \meanx + \scale{\X, \Y} \BB{b}^\star + \sqrt{\frac{ \scaling{\Z}{\z^\star}}{ \scaling{\Z}{\iup{\z}} }} \left[(\iup{\x}   - \meanx) - \scale{\X, \Y} \iup{\BB{b}} \right]$
    }
    \vskip 0.2cm
    return $\{\x_a^i\}$
\caption{\code{EnRF}$(\ystar, \pdf{\Y \given \X = \cdot}, \{\x^i \})$ assimilates the data $\ystar$ in the prior samples $\{\x^1, \ldots, \x^M\}$ with the ensemble robust filter for \textbf{a likelihood model $\pdf{\Y \given \X = \cdot}$}}
\label{algo:enrf}
\end{algorithm}

\end{appendices}

\bibliographystyle{unsrt}
\bibliography{mybib.bib}

\end{document}